\documentclass{aastex63}

\usepackage{graphics}

\newcommand{\tbmax}{T$_{B_{\rm max}}$}
\newcommand{\specialnum}{\color{blue}}

\begin{document}

\title{Are Type Ia Supernovae in Restframe $H$ Brighter in More Massive Galaxies?}

\correspondingauthor{Kara Ponder}
\email{kap146@pitt.edu}

\author[0000-0002-8207-3304]{Kara~A.~Ponder}
\affiliation{PITT PACC, Department of Physics and Astronomy,
University of Pittsburgh, Pittsburgh, PA 15260, USA}
\affiliation{
Berkeley Center for Cosmological Physics,
University of California Berkeley,
341 Campbell Hall, Berkeley, CA 94720, USA}
\affiliation{Physics Division, Lawrence Berkeley National Laboratory,
1 Cyclotron Road, Berkeley, CA, 94720, USA}
\affiliation{SLAC National Accelerator Laboratory,
2575 Sand Hill Rd, Menlo Park, CA 94025, USA}

\author{W.~Michael~Wood-Vasey}
\affiliation{PITT PACC, Department of Physics and Astronomy,
University of Pittsburgh, Pittsburgh, PA 15260, USA}

\author{Anja~Weyant}
\affiliation{PITT PACC, Department of Physics and Astronomy,
University of Pittsburgh, Pittsburgh, PA 15260, USA}

\author{Nathan~T.~Barton}
\affiliation{PITT PACC, Department of Physics and Astronomy,
University of Pittsburgh, Pittsburgh, PA 15260, USA}
\affiliation{Department of Mechanical and Civil Engineering,
California Institute of Technology, Pasadena, CA 91126, USA}

\author{Llu\'is~Galbany}
\affiliation{PITT PACC, Department of Physics and Astronomy,
University of Pittsburgh, Pittsburgh, PA 15260, USA}
\affiliation{Departamento de F\'isica Te\'orica y del Cosmos, Universidad de Granada, E-18071 Granada, Spain}

\author{Shu Liu}
\affiliation{PITT PACC, Department of Physics and Astronomy,
University of Pittsburgh, Pittsburgh, PA 15260, USA}

\author{Peter Garnavich}
\affiliation{Physics Department,
University of Notre Dame,
Notre Dame, IN, 46556, USA}

\author[0000-0001-6685-0479]{Thomas Matheson}
\affiliation{NSF's National Optical-Infrared Astronomy Research Laboratory,
950 North Cherry Avenue,
Tucson, AZ 85719, USA}


\newcommand{\numFullHostSample}{220}
\newcommand{\numAll}{144}
\newcommand{\stddevAll}{0.229}
\newcommand{\iqrAll}{0.207}
\newcommand{\numHubbleAll}{80}
\newcommand{\stddevHubbleAll}{0.264}
\newcommand{\numLightHeavy}{143}
\newcommand{\numLight}{59}
\newcommand{\stddevLight}{0.223}
\newcommand{\semLight}{0.029}
\newcommand{\iqrLight}{0.208}
\newcommand{\weightResidualLight}{0.027}
\newcommand{\numHeavy}{84}
\newcommand{\stddevHeavy}{0.231}
\newcommand{\semHeavy}{0.025}
\newcommand{\iqrHeavy}{0.206}
\newcommand{\weightResidualHeavy}{-0.024}
\newcommand{\shiftLightHeavy}{0.051}
\newcommand{\semShiftLightHeavy}{0.038}
\newcommand{\sigmaShiftLightHeavy}{1.34}
\newcommand{\semOutlierLight}{0.025}
\newcommand{\weightResidualOutlierLight}{-0.012}
\newcommand{\semOutlierHeavy}{0.020}
\newcommand{\weightResidualOutlierHeavy}{0.037}
\newcommand{\shiftOutlierLightHeavy}{0.002}
\newcommand{\semShiftOutlierLightHeavy}{0.039}
\newcommand{\sigmaShiftOutlierLightHeavy}{0.005}
\newcommand{\numYesPhotometry}{143}

\begin{abstract}
We analyze \numLightHeavy\ Type Ia supernovae (SNeIa) observed in $H$ band (1.6--1.8~$\mu$m)
and find SNeIa are intrinsically brighter in $H$-band with increasing host galaxy stellar mass.
We find SNeIa in galaxies more massive than $10^{10.43} M_{\odot}$ are $0.13 \pm 0.04$~mag brighter in $H$ than SNeIa in less massive galaxies.
The same set of SNeIa observed at optical wavelengths,
after width-color-luminosity corrections,
exhibit a $0.10 \pm 0.03$~mag offset in the Hubble residuals.
We observe an outlier population ($|\Delta H_{\rm max}| > 0.5$~mag) in the $H$ band and show that removing the outlier population moves the mass threshold to $10^{10.65} M_{\odot}$ and reduces the step in $H$ band to $0.08 \pm 0.04$~mag, but the equivalent optical mass step is increased to $0.13 \pm 0.04$~mag.
We conclude the outliers do not drive the brightness--host-mass correlation.
Less massive galaxies preferentially host more higher-stretch SNeIa,
which are intrinsically brighter and bluer.
It is only after correction for width-luminosity and color-luminosity relationships that SNeIa have brighter optical Hubble residuals in more massive galaxies.
Thus finding SNeIa are intrinsically brighter in $H$ in more massive galaxies
is an opposite correlation to the intrinsic (pre-width-luminosity correction) optical brightness.
If dust and the treatment of intrinsic color variation were the main driver of the host galaxy mass correlation, we would not expect a correlation of brighter $H$-band SNeIa in more massive galaxies.

\end{abstract}

\section{Introduction}\label{sec:introduction} 

Since the late 1990s, Type Ia supernovae (SNeIa) have been used as standard candles to measure the accelerating expansion of the Universe \citep{Riess98, Perlmutter99}.
Much work has gone into further standardizing inferred optical brightness of SNeIa by including corrections based on the stretch~\citep{Phillips93} and color \citep{Riess96, Tripp98} of the lightcurve.
More recent work has started including an additional correction term associated with the stellar mass\footnote{Throughout this paper the term ``mass'' will always refer to the stellar mass of the galaxy.} of the host galaxy of the SNeIa~\citep{Betoule14,Scolnic18,Brout19,Smith20}.

Lightcurves observed at near infrared (NIR) wavelengths ($1~\mu$~m$ < \lambda < 2.5~\mu$~m) are more standard and require no or smaller corrections
to their lightcurves to yield the same precision as optical lightcurves \citep{Kasen06,Wood-Vasey08,Folatelli10,Kattner12,Barone-Nugent12,Dhawan18,Burns18}.
We here compile one of the largest publicly available NIR SN~Ia data sets to further test the standard nature of SNeIa.
We explore different possible correlations between global host galaxy properties and $H$-band luminosity.

The past decade has seen an extensive history of looking for correlations between the standardized optical luminosity of SNeIa and the properties of their host galaxies.
Many papers have studied relationships with global host galaxy properties such as stellar mass, metallicity, star formation rates, and age using galaxy photometry and stellar population synthesis codes \citep{Sullivan06,Gallagher08,Kelly10, Sullivan10,Lampeitl10,Gupta11,DAndrea11, Hayden13, Johansson13, Childress13a,Childress13b,Moreno-Raya16,Campbell16,Wolf16,Roman17,Uddin17a,Jones18,Rose19}.
These papers have found several correlations between standardized brightness and host galaxy properties with the most significant one being host galaxy stellar mass.
Some interpret this as a result of a correlation between galaxy stellar mass and things more physically related to the SN~Ia explosion, such as progenitor metallicity, progenitor age, or dust~\citep{Kelly10,Hayden13, Childress13b,Brout20}.
These analyses show that the standardized brightness of SNeIa hosted in higher-mass galaxies is brighter by $\sim$0.08~mag~\citep{Childress13b} than SNeIa hosted in galaxies with stellar mass less than $10^{10} M_{\odot}$.
The mass ``step'' was also implemented in one of the recent studies to produce cosmological constraints: the Joint Lightcurve Analysis~\citep[JLA;][]{Betoule14}, where they independently measured a correlation with host galaxy stellar mass and implemented a step function to account for it.
Others have focused on local properties of host galaxies such as recent star formation rates within 1--5 kpc of the supernova position using spectroscopy or ultraviolet (UV) photometry \citep{Rigault13,Rigault15,Rigault18,Kelly15,Uddin17b}.
These local property studies find that the standardized brightness of SNeIa in locally passive regions is $\sim$0.094~mag \citep{Rigault15} brighter than those in locally star forming regions.
Furthermore, \cite{Kelly15} showed that SNeIa in locally star forming regions were more standard than those in non-star forming ones.

However, not every analysis suggests that there is a correlation with host galaxy properties.
\cite{Kim14} used an updated lightcurve analysis that is more flexible to intrinsic variations in SNeIa \citep[introduced in][]{Kim13} and found any potential correlations with host galaxy stellar mass, specific star formation rates, and metallicity to be consistent with zero.
\cite{Jones15} found no evidence of a correlation between host galaxy local star formation rates derived from UV photometry by using a larger sample size than previous studies and using different selection criteria.
\cite{Scolnic14} described the systematics utilized in the Pan-STARRS SN~Ia cosmology analysis \citep{Rest14} and found a correlation with host galaxy stellar mass with a step size of $0.037 \pm 0.032$~mag, which is consistent with 0 and is 2-$\sigma$ inconsistent with the previously reported sizes in the literature of $\sim 0.1$~mag.
With twice as many SNeIa, the subsequent Pan-STARRS analysis~\citep{Scolnic18} recovered a very similar small step size of $0.039 \pm 0.016$~mag, but now with a clear deviation from 0.
\citet{Scolnic18} noted that if they did not apply the BEAMS with Bias Correction~\citep[BBC; ][]{Kessler17} method in their analysis they would have found a mass step of $0.064 \pm 0.018$~mag.
The Dark Energy Survey \citep[DES;][]{DES16, DES19, Brout19} originally found no evidence of a host galaxy stellar mass step for 329 SNeIa with 207 observed in the first three years of DES and the rest from low-redshift samples.
However, \citet{Smith20} showed that the DES data do exhibit a mass step if a JLA-like analysis is run, which strongly suggests that such a mass step was being corrected for by the BBC method used in the DES SN cosmology papers to date.

We see much evidence to warrant continued exploration of this parameter space to understand whether we are searching for a real correlation or if we need to improve the analysis of SNeIa lightcurves.

The majority of the previous analyses of host galaxy properties versus SN~Ia corrected brightness have examined correlations using only optical lightcurves.
Doing a similar analysis using NIR lightcurves will help shed light on physical mechanisms and color-dependent intrinsic dispersions.
For example, the NIR is less sensitive to dust such that if there is no correlation found in the NIR, but a correlation found at optical wavelengths, it would provide evidence that the host mass correlation is actually an unaccounted for dust correlation.

In the restframe NIR there have been far fewer studies of host galaxy correlations.
{\cite{Dhawan18} looked at the $J$-band and, with a small sample of 30 SNeIa, found low dispersion ($\sim0.10$~mag) and no obvious trend with host galaxy morphology.
A more in-depth study was done using the Carnegie Supernova Project (CSP) sample by \cite{Burns18}, which compared $H$-band brightnesses to host galaxy stellar mass estimates from $K$-band photometry following the mass-to-light ratio method of \citet{McGaugh14}.
\citet{Uddin20} expanded this analysis to examine correlations from optical to NIR lightcurves with new observations of each host galaxy.
\citet{Uddin20} found a small $\sim 1\sigma$ linear correlation and $\sim 2\sigma$ step function ($0.093 \pm 0.043$~mag) between the restframe $H$-band and host galaxy stellar mass for their sample of $113$ SNeIa. We here analyze a data set with significantly more SN~Ia in low-mass host galaxies, $M<10^{10} M_\Sun$, the canonical break point for the mass step.

SNeIa in the $H$-band have been shown to be standard to $0.15$--$0.2$~mag without lightcurve corrections~\citep{Wood-Vasey08,Folatelli10,Barone-Nugent12,Kattner12,Weyant14,Stanishev15,Avelino19} whereas optical lightcurves before brightness standardization have significantly larger scatter of $\sim 0.8$~mag~\citep{Hamuy95}.
However, there are only $\sim$\numFullHostSample~NIR lightcurves publicly available compared to the over $>1,000$ available for optically observed SNeIa.

We use SNooPy~\citep{Burns11,Burns14} for lightcurve fits as it is has the most developed treatment of NIR templates.
We combine optical and NIR lightcurves to improve fits with the $s_{BV}$ parameter from \citet{Burns14}.
Most previous analyses have explored host galaxy correlations with standardized brightnesses calculated from SALT2~\citep{Guy07} and/or MLCS2k2~\citep{Jha07} fitters~\citep[e.g.,][]{Kelly10}.

This paper is organized as follows:
Section~\ref{sec:data_sample} explains the supernova sample we use and how we collected optical, UV, and NIR photometry of the host galaxies.
Section~\ref{sec:distances} details how we fit lightcurves and created the restframe $H$-band and optical Hubble diagrams.
Section~\ref{sec:analysis} examines the host galaxy stellar mass correlation and shows that the $H$-band Hubble residuals and the optical width-luminosity corrected Hubble residuals are both more negative in higher-mass galaxies.
Section~\ref{sec:discussion} explores the statistical significance of these correlations.
We present our conclusions and recommendations for future work in Section~\ref{sec:conclusion}.

\section{SN~I\lowercase{a} and Host Galaxy Sample}\label{sec:data_sample}

The improved ability to determine standard distances, together with the reduced sensitivity to dust extinction, have motivated several recent projects to pursue larger samples of SNeIa observed in the restframe NIR:
CSP-I, II~\citep{Contreras10,Stritzinger11,Kattner12,Krisciunas17,Phillips19};
CfA~\citep{Wood-Vasey08,Friedman15};
RAISINS~\citep{Kirshner12};
SweetSpot~\citep{Weyant14, Weyant18};
and SIRAH~\citep{Jha19}.
Section~\ref{sec:sn} overviews the NIR sample currently available and used in this analysis.

To gather host galaxy properties,
we used publicly available galaxy catalogs from Sloan Digital Sky Survey (SDSS),
Panoramic Survey Telescope and Rapid Response System (Pan-STARRS),
DECam Legacy Survey (DECaLS),
Galaxy Evolution Explorer (GALEX),
and Two Micron All-Sky Survey (2MASS).
We used \texttt{kcorrect}~\citep{Blanton07} to estimate estimate galaxy properties including the stellar mass of the host galaxies.
We detail this process in Section~\ref{sec:hostgalaxies}.

\subsection{SNeIa} \label{sec:sn}
We start with the compilation of literature SNeIa gathered in \citet{Weyant14}.
% with small tweaks on the naming scheme.
We assigned each SN~Ia to ``belong'' to a given survey to be able to examine properties as a function of survey.
If a SN~Ia was found in multiple surveys, we labeled that object with the survey name containing the most lightcurve points in $H$; however, lightcurve points from all surveys were included when running the lightcurve fits.
We used the following survey codes: K+, CSP, BN12, F15, W18.
\begin{itemize}
\item K+ is the miscellaneous early sample~\citep{Jha99, Hernandez00, Krisciunas00, Krisciunas03, Krisciunas04a, Krisciunas04b, Krisciunas07, Valentini03, Phillips06, Pastorello07b, Pastorello07a, Stanishev07, Pignata08} and named for extensive early work by Kevin Krisciunas and originally defined in  \citet{Weyant14}.
\item CSP refers to the lightcurves released from~\citet[C10]{Contreras10} and~\citet[S11]{Stritzinger11}.
There were NIR observations of 71 SNeIa in the C10+S11 samples.
One of these, SN~2004eo, is placed in the K+ sample and 11 are placed in the CfA sample.
\item BN12 covers the SNeIa from~\citet{Barone-Nugent12}.
% and has one overlap with CfA which is placed in that sample instead of BN12.
\item We renamed the CfA sample from WV08~\citep{Wood-Vasey08} to F15 due to the 74 additional SNeIa from the final data release~\citep{Friedman15}.  We do not use any of the peculiar Iax supernovae \citep{Foley13} from \citet{Friedman15}.
5 of the F15 SNeIa overlap with and are placed in the CSP sample and 1 F15 SN~Ia is placed in the BN12 sample.
\item The SweetSpot W14 sample is replaced by W18 due to the addition of 34\footnote{Though \citet{Weyant18} states that 33 lightcurves were released, 34 lightcurves were actually provided by the authors for download.} SNeIa from SweetSpot's first data release~\citep{Weyant18}.
\end{itemize}
All of the SNeIa from BN12 were discovered in the non-targeted PTF survey.  The other samples here have SNeIa mostly discovered in surveys targeted on known galaxies.

Our full sample of SNeIa $H$-band lightcurves consists of \numFullHostSample\ SNeIa.
Table~\ref{tab:sneia_surveys1} gives the breakdown per survey.
We used the Open Supernova Catalog\footnote{\url{https://sne.space/}}~\citep[OSC;][]{Guillochon17} to retrieve all lightcurve data.
We inspected the filter choices and definitions listed in the OSC compilation and spot-checked data against the published tables in the original papers.

We removed the 10 91bg-like SNeIa, two 02cx-like SNeIa, and two more generically peculiar SNeIa as reported in their classification spectra.
We removed SN~2011aa because its lightcurve in F15 seems to be in error as its almost flat over 40 days in $J$, $H$, and $K$.
We then fit these SNeIa with SNooPy (see Section~\ref{sec:distances}).
In order to be included on the Hubble diagram, a supernova was required to have at least 3 observations with a signal-to-noise greater than 3.5 and have a fit with a chi-square per degree of freedom less than 3.
These cuts removed 41~SNeIa nominally successful SNooPy fits.
Table~\ref{tab:sneia_surveys1} shows how many SNeIa have lightcurve fits that pass the quality cuts (``Pass LC Fit Cuts'').
The left side of Figure~\ref{fig:all_hubble_redshift_mass} shows the redshift distribution of the full sample compared to the distribution of the Hubble diagram sample used for the analysis below.
The full sample has a median redshift of 0.026 while the Hubble diagram sample has a slightly lower median redshift of 0.022.

We were not able to obtain reliable host galaxy photometry for SN~2004S.
Thus out of the \numAll\ SNeIa, \numLightHeavy\ have sufficient host galaxy photometry to derive stellar masses (see Section~\ref{sec:hostgalaxies}).

\begin{figure}
\plottwo{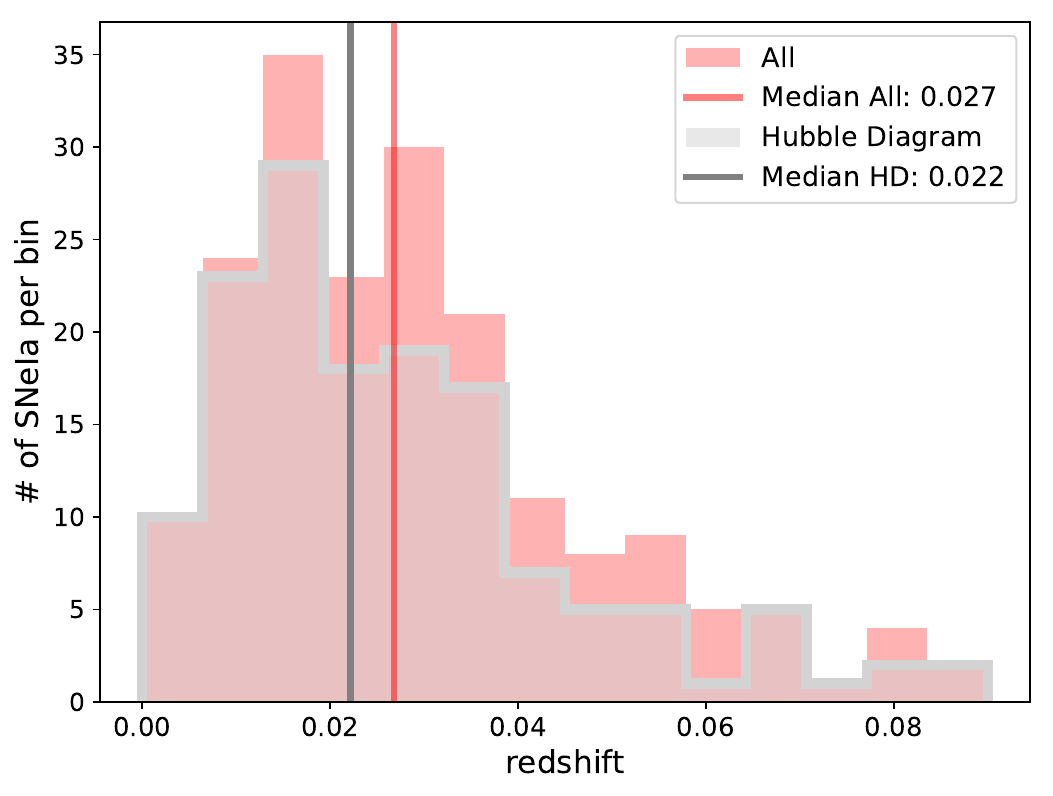}{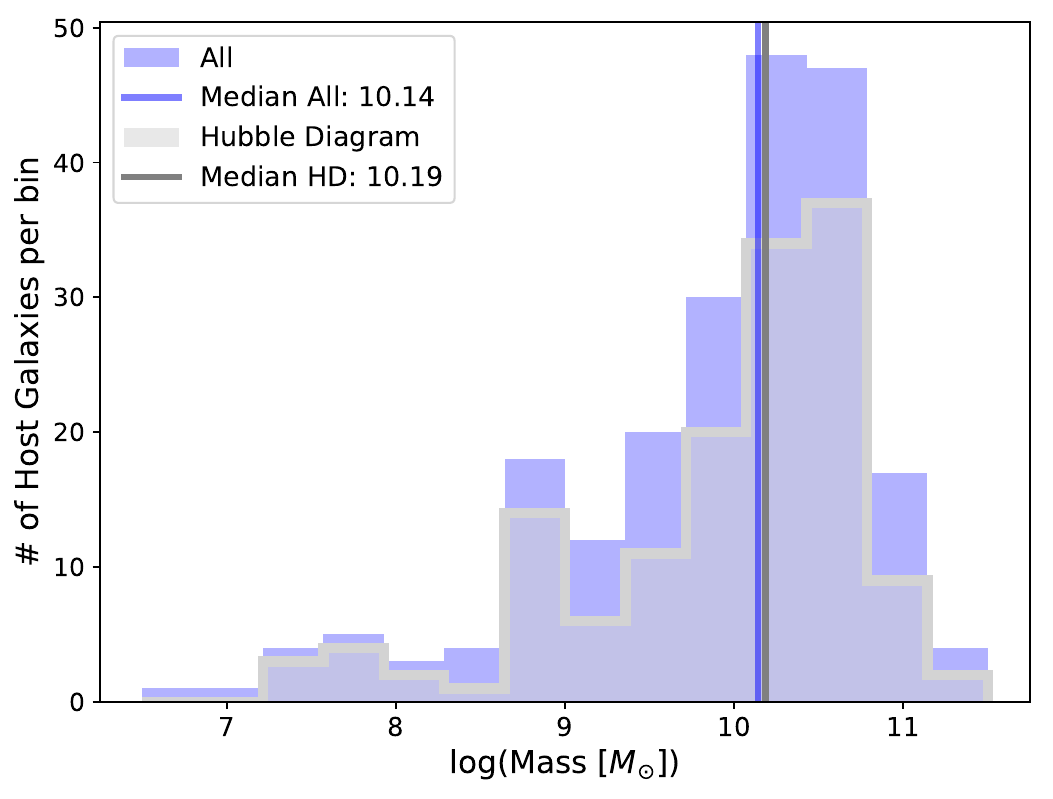}
\caption{\textit{Left:} Redshift distribution of the publicly available NIR sample of \numFullHostSample\ SNeIa and the Hubble Diagram sample of \numAll\ SNeIa.
\textit{Right:} Mass distribution of the host galaxies in log space of the publicly available NIR sample and the Hubble Diagram sample.
}
\label{fig:all_hubble_redshift_mass}
\end{figure}

The right side of Figure~\ref{fig:all_hubble_redshift_mass} compares the mass distribution of the full sample versus those SNe~Ia that are in the Hubble diagram sample.
The medians for both distributions are comparable and the distribution shapes are consistent.
Table~\ref{tab:sneia_surveys1} details how many SNeIa each sample have a successful SNooPy lightcurve fit, pass the quality cuts, and have sufficient information to calculate a host stellar mass (``Pass LC + Host Mass'').

For all \numFullHostSample\ SNeIa with $H$-band lightcurves, we used the OSC to download any available corresponding optical lightcurves.
Some surveys such as CSP and F15 obtained complementary optical lightcurves as a part of their survey; however, other surveys such as BN12 and W18 did not.
 All optical lightcurves go through the same quality cuts as the the NIR sample.
103 SNeIa observed in optical wavelengths passed the LC fit cuts and had usable host galaxy masses.
The optical sample contains 99 SNeIa that are also in the NIR sample.

\begin{deluxetable*}{lrrrr}
\tablewidth{0pt}
\tablecaption{Number of SNeIa from NIR Surveys \label{tab:sneia_surveys1}}
\tablehead{
 \colhead{SN Survey} &  \colhead{Total} &  \colhead{Pass LC Fit Cuts}  & \colhead{Pass LC+Host Mass}}
\startdata
 K+   & 23 & 11 & 10 \\
 CSP  & 59 & 47 & 47 \\
 BN12 & 12 & 12 & 12 \\
 F15  & 92 & 56 & 56 \\
 W18  & 34 & 18 & 18 \\
 \hline
 Total & \numFullHostSample\  & \numAll\  & \numLightHeavy\ \\
\enddata
\end{deluxetable*}

\subsection{Host Galaxies}\label{sec:hostgalaxies}
The host galaxy for all \numFullHostSample\ SNeIa was identified from the IAU list of supernovae\footnote{\url{http://www.cbat.eps.harvard.edu/lists/Supernovae.html}} and the NASA Extragalactic Database (NED)\footnote{\url{http://ned.ipac.caltech.edu/}}.
The procedure to confirm a host galaxy started with either the suggested host from the transient announcement or a distance search in NED.
We confirmed each host redshift matched the supernova redshift and visually examined other potential hosts in the vicinity.
We used NED to collect the host galaxy name, coordinates, and the heliocentric redshift for each galaxy.
If NED did not have a spectroscopic redshift, we recorded the redshift of the hosted supernova from the classification spectrum.

We obtained optical photometry from the SDSS Data Release 13~\citep{DR13}, the DECam Legacy Survey~\citep[DECaLS;][]{Dey19}, and the Pan-STARRS1 Data Release 2~\citep[PS1;][]{Chambers16,Flewelling16,Magnier16}.
The SDSS and PS1 photometry were downloaded using their respective CasJobs\footnote{\url{http://skyserver.sdss.org/CasJobs/}, \url{http://mastweb.stsci.edu/ps1casjobs/}} websites.
For SDSS photometry, we used the $ugriz$ ``modelMag'' magnitudes, which are based on the best fit de Vaucouleurs or Exponential profile in the $r$-band.
Though ``cmodelMag'' magnitudes give a more accurate description of the total flux in each filter, ``modelMag'' magnitudes are better for color studies because the flux is measured consistently across all filters~\citep{Stoughton02}.

We also obtained the PS1 stacked Kron~\citep{Kron80} magnitudes\footnote{We tested a procedure to recreate a ``modelMag'' with PS1 de Vaucouleurs and Exponential profile fits. We measured these masses to have the largest bias compared to SDSS. There was enough coverage from the Kron magnitudes that we did not need to use the PS1 modelMags.}, which uses the first moment of an image to determine the radius out to which flux should be integrated. This photometry and the masses derived using them are referred to as ``PS1''.
PS1 does not always have all five $grizy$ magnitudes for all of our objects.
If $gi$~magnitudes were not available in PS1, we did not use that host galaxy photometry as we could not calculate extinction coefficients~\citep{Tonry12}.

We used the Astro Data Lab at NSF's National Optical-Infrared Astronomy Research Laboratory cross-match service\footnote{\url{https://datalab.noirlab.edu/xmatch.php}} to query the DECaLS catalog.
DECaLS uses Tractor~\citep{Lang} to fit a morphological type to each source and then extracts the photometry measured in AB magnitudes
This method is conceptually similar to what is done in SDSS.
We only use the optical filters available ($grz$) from DECaLS.

We obtained GALEX GR6/GR7\footnote{\url{http://galex.stsci.edu/GR6/}}~\citep{Bianchi14} far ultraviolet (FUV/$F$) and near ultraviolet (NUV/$N$) information where available from the MAST data archive.\footnote{\url{https://galex.stsci.edu/casjobs/}}
We used the photometry that is the result of the elliptical aperture method ``MAG$\_$AUTO'', which is similar to the Kron radius calculation, in Source Extractor~\citep{Bertin96}.
GALEX often reported detections in only one of FUV or NUV magnitudes, but we only required one of these to mark an object as having UV data.

We also gathered $JHK_s$ magnitudes from the 2MASS All-Sky Extended Source Catalog~\citep[XSC;][]{TMASS} using the NASA/IPAC Infrared Science Archive (IRSA).\footnote{\url{http://irsa.ipac.caltech.edu/frontpage/}}
We used the total magnitude calculated from the extrapolated radial surface brightness profile.
One object (PGC~1361264, host of SN~2010ho) had an $H$-band uncertainty of zero and a magnitude significantly inconsistent with its $JK_s$ magnitudes, so that $H$-band photometric point was not used to determine the mass of PGC~1361264.

There were 5 galaxies that did not have optical photometry in any of these three surveys or GALEX UV photometry and measurements only from 2MASS were not sufficient to get a reliable host galaxy mass.
We supplemented those 5 galaxies with optical photometry from \citet[][U20 hereafter]{Uddin20}.
U20 compared their galaxy photometry with SDSS and derived the best fit linear offsets (U20 Table 1).
We applied these offsets to their photometry and treated them as SDSS measurements to determine host galaxy mass.
This set of host galaxies is labeled as ``CSP Host Phot'' to distinguish it from the CSP subsample of SNeIa.

We use \texttt{kcorrect} \citep{Blanton03, Blanton07}
to transform the photometry to the restframe and infer physical parameters\footnote{\texttt{kcorrect} does not return uncertainties on the physical parameters.} such as stellar mass.
\texttt{kcorrect} fits galaxy spectral energy distributions from the UV to NIR and relies on \cite{BC03} stellar evolution synthesis models using the \citet{Chabrier03} stellar initial mass function (IMF).
The physical parameters that \texttt{kcorrect} reports are based on those of the galaxy templates from these models.
Adding the UV and NIR photometry to the optical photometry gives sharper constraints on dust absorption and thus help distinguish the different galaxy models that overlap at optical wavelengths.
Figure~\ref{fig:transmission} shows an example where a spiral and elliptical galaxy largely agree in optical wavelengths, but are clearly distinguished with the addition of UV and NIR measurements.
All magnitudes are converted to the AB magnitude system and are extinction corrected for Milky Way dust before being input into \texttt{kcorrect}.
We derive $K$-corrections and host galaxy stellar mass
by combining optical photometry plus GALEX and 2MASS for each host galaxy.
If there was no optical photometry, we required the galaxy to have four observations between GALEX and 2MASS to ensure at least one point in the UV and in the NIR.
Table~\ref{tab:surveys} lists how many SN~Ia host galaxies have photometry for each of the surveys that are in our analysis.

\begin{figure*}
\plotone{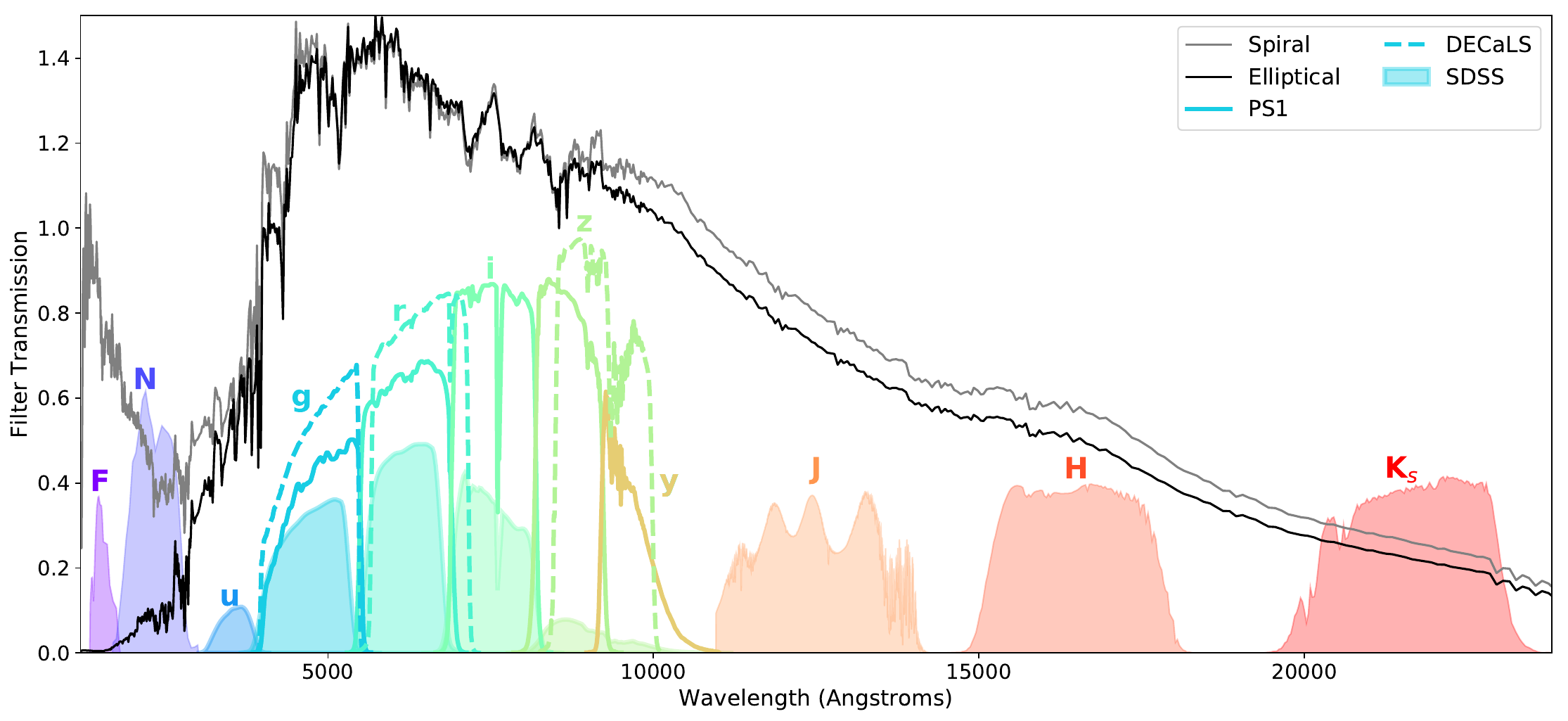}
\caption{System transmission functions
for GALEX ($F$, $N$), SDSS ($ugriz$), DECaLS ($grz$), PS1 ($grizy$), and 2MASS ($JHK_s$).
These transmission functions accounts for the detector, optics, filter, and atmosphere.
Over-plotted is the normalized spectral energy distribution (SED) for a Sc spiral galaxy in grey and an elliptical galaxy in black from the SWIRE Template Library~\citep{Polletta07}.
For both spiral and elliptical galaxies, the majority of their flux is emitted at restframe optical wavelengths peaking in $gri$ bands.
The Sc spiral galaxies have a small bulge and obvious spiral arms containing young stars emitting heavily in the UV.
Though having optical photometry observes most of the flux, observing UV and NIR can help constrain the spectrum.
}
\label{fig:transmission}
\end{figure*}

Figure~\ref{fig:transmission} illustrates the wavelength coverage from these surveys.
All \numAll\ lightcurves have host galaxy photometry available in at least one of these catalogs but only 143 galaxies meet our requirement for a robust host galaxy stellar mass measurement.
The redshift and host galaxy mass distributions of our final sample versus the full sample is presented in of Figure~\ref{fig:all_hubble_redshift_mass}.

Figure~\ref{fig:redshift_mass_all_surveys_hubble} shows a histogram of the galaxies in the Hubble diagram sample that have photometry in each survey.
There is a large overlap between the galaxies observed by SDSS, DECaLS, PS1, and CSP Host Phot.
Figure~\ref{fig:redshift_mass_per_survey} shows the final breakdown of photometry used in this analysis.
We preferentially chose SDSS over DECaLS over PS1 (See Section~\ref{sec:OpticalPhotComp}).
We summarize the photometric data in Table~\ref{tab:snia_gal} for all \numFullHostSample\ SNeIa.

\begin{deluxetable*}{llrrrr|r}
\tablewidth{0pt}
\tablecaption{Number of Host Galaxies Observed in Our Sample per Galaxy Catalog.\label{tab:surveys}}
\tablehead{
\colhead{Sample} & \colhead{Survey} &  \colhead{Optical} &  \colhead{GALEX} & \colhead{2MASS} & \colhead{GALEX+2MASS} & \colhead{Total}
}
\startdata
All & SDSS & 14 & 14 & 14 & 55 & 97 \\
& DECaLS & 7 & 2 & 18 & 32 & 59 \\
&PS1 & 3 & 1 & 16 & 21 & 41 \\
&CSP Host Phot & 0 & 0 & 5 & 0 & 5 \\
&None & \nodata & \nodata & {\specialnum 2}\tablenotemark{a} & 16 & 18 \\
\hline
&Total & 24 & 17 & 55 & 124 & 220 \\
\hline
  & \quad &  \quad & \quad  & \quad  &  \\
    & \quad &  \quad & \quad  & \quad  &  \\
 \hline
Hubble & SDSS & 9 & 8 & 10 & 32 & 59 \\
Diagram &DECaLS & 4 & 2 & 14 & 19 & 39 \\
&PS1 & 3 & 0 & 13 & 15 & 31 \\
&CSP Host Phot & 0 & 0 & 5 & 0 & 5 \\
&None & \nodata & \nodata & {\specialnum 1}\tablenotemark{a} & 9 & 10 \\
\hline
&Total & 16 & 10 & 43 & 75 & 144 \\
\enddata
\tablenotetext{a}{Included for completeness but not used to determine galaxy mass.}
\tablecomments{The Survey column denotes in which survey the optical photometry was used.
The Optical column presents how many objects had \textit{only} optical data.
The GALEX column are galaxies with optical and UV observations.
The 2MASS column are galaxies with optical and NIR observations.
The GALEX+2MASS column presents how many galaxies have observations in all three optical, UV, and NIR catalogs.}
\end{deluxetable*}

\begin{figure}
\epsscale{0.75}
\plotone{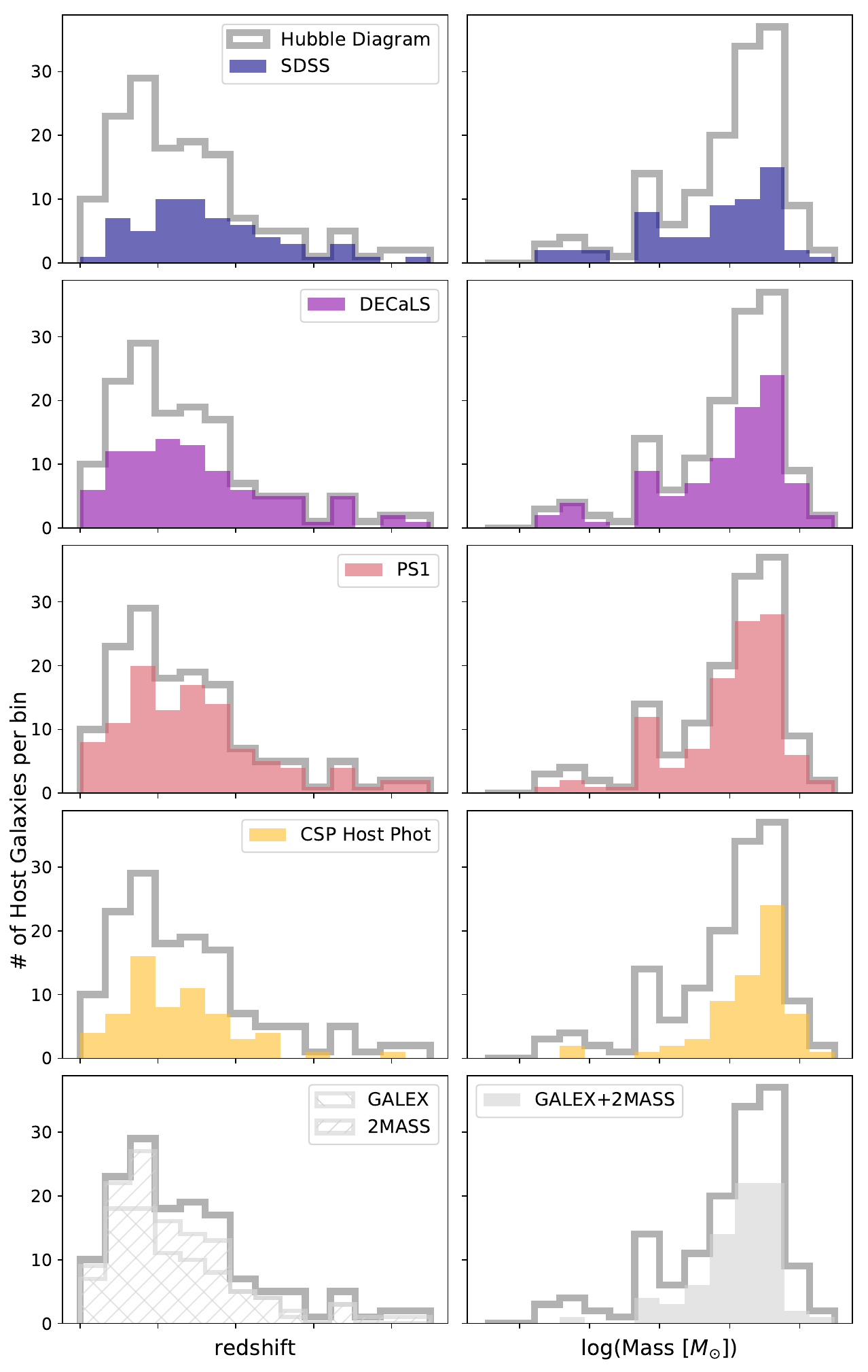}
\caption{The redshift (left) and host galaxy mass (right) histograms of the \textit{total available} host galaxy photometry in SDSS, DECaLS, PS1, and CSP Host Phot for the 144 SNeIa used in the Hubble diagram analysis, which is outlined in grey for all plots.
The bottom left histogram shows the total coverage in redshift for GALEX and 2MASS separately; however, in the bottom right mass histogram we required both GALEX and 2MASS measurements to ensure more reliable mass measurements.
}
\label{fig:redshift_mass_all_surveys_hubble}
\end{figure}

\begin{figure}
\plottwo{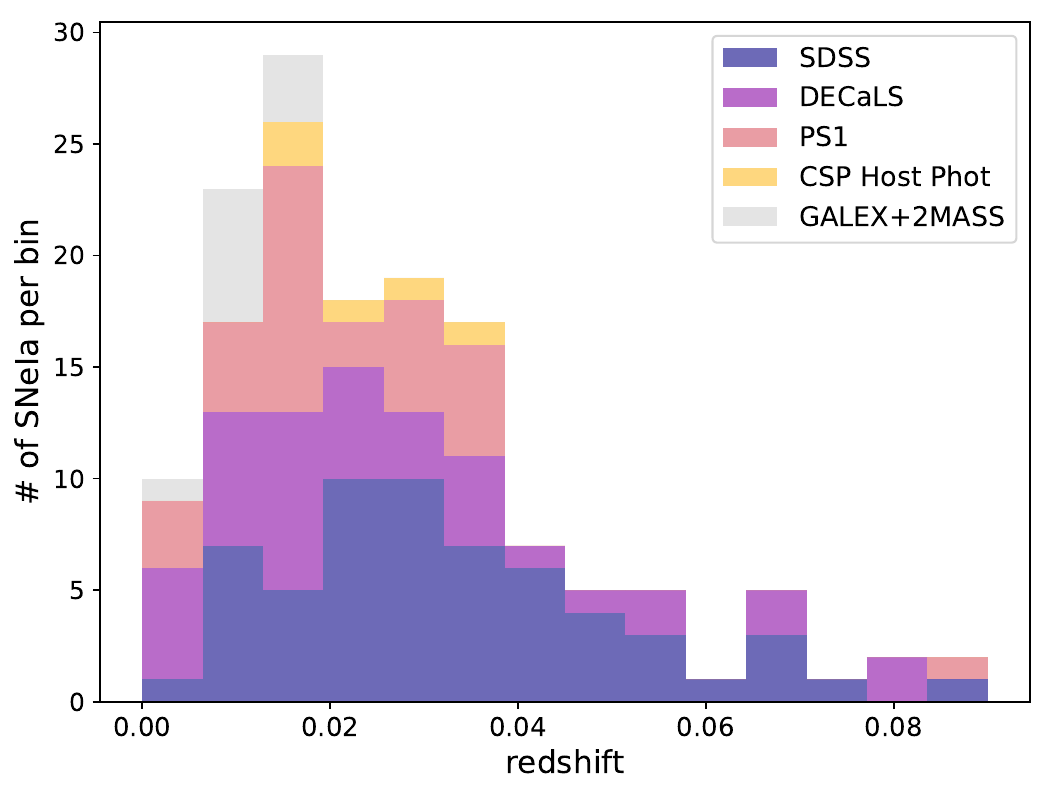}{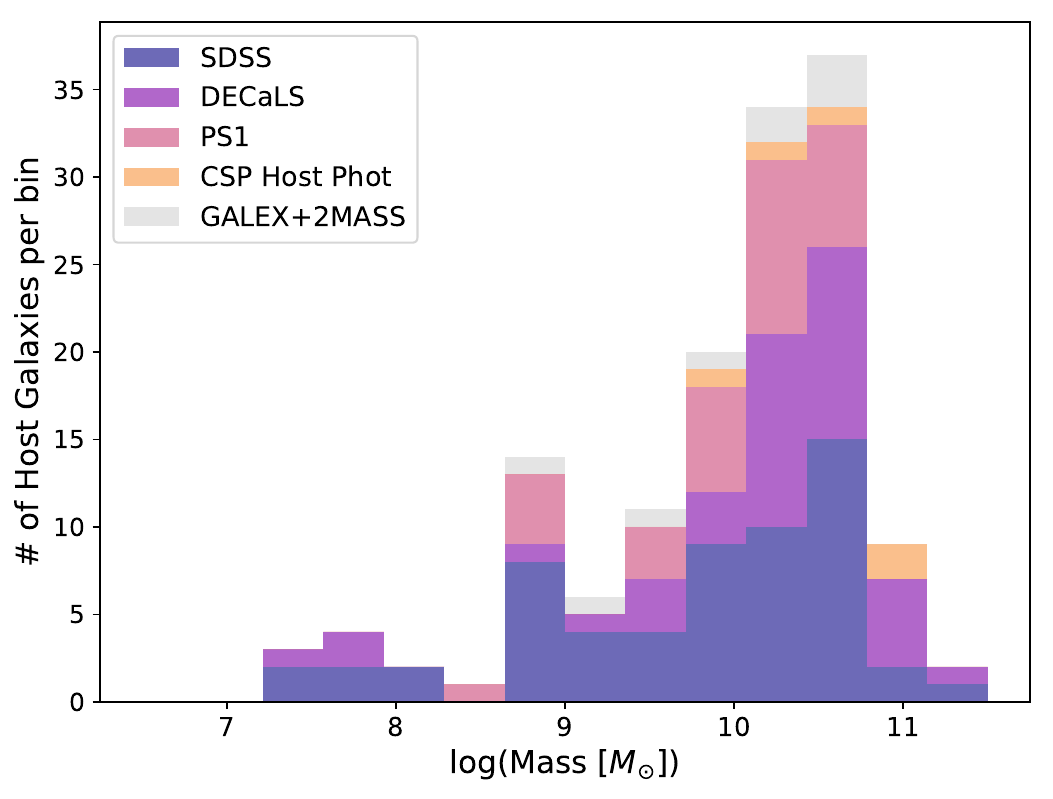}
\caption{\textit{Left:} Redshift distribution identifying which survey's photometry was used to determine the host galaxy mass the final sample of \numAll\ SNeIa used in the Hubble residual analysis.
In addition to optical photometry, most host galaxies also have photometry in GALEX and/or 2MASS but the grey ``GALEX+2MASS'' indicates that no optical photometry was used to determine the mass.
\textit{Right:} Mass distribution of the 143 host galaxies in log space separated by which survey was used to determine the mass.
}
\label{fig:redshift_mass_per_survey}
\end{figure}

\subsubsection{Special Cases for Optical Galaxy Photometry}
A few galaxies observed by SDSS and PS1 could not be handled using the normal catalog searches.
In SDSS, the host galaxy for SN~2011fe is M101.
SN~2011fe was not used in the Hubble diagram analysis, but the mass was used in Figure~\ref{fig:all_hubble_redshift_mass}.
We had to use special large galaxy catalogs for SDSS~\citep[DR7; ][]{DR7}, the GALEX Ultraviolet Atlas of Nearby Galaxies~\citep{Paz07}, the 2MASS Large Galaxy Atlas~\citep{Jarrett03}}.

In PS1, there are two objects that are in galaxy pairs: SN~2007sr located in the Antennae Galaxies (NGC~4038/NGC~4039) and SN~2011aa located in UGC 3906.
For both of these, we kept photometry for each galaxy and averaged the host galaxy mass from \texttt{kcorrect}.
SN~2011aa is not used in the Hubble diagram analysis.

Three host galaxies had observations in PS1 but no associated catalog photometry.
Here we list the galaxy, associated supernova, and probable cause for the failure of a catalog measurement:
\begin{enumerate}
\item  UGC~3329, SN~1999ek, Two bright stars in the foreground;
\item Unnamed host of PTF13dad, Galaxy is faint;
\item Unnamed host of PS1-13dkh, There is a live SN in the images and has a bright star nearby.
\end{enumerate}
After masking the bright stars and SN, we performed aperture photometry from a derived Kron radius using the calibration information provided by PS1\footnote{\url{https://outerspace.stsci.edu/display/PANSTARRS/PS1+Stack+images}}.
We counted these as PS1 magnitudes.

The host galaxy for SN~2004S (ESO 427-G6) did have some imaging from PS1, but it was at a low declination causing the survey to cut off close to the observation.
No catalog information was available and we were unable to use these images with our own algorithms.
ESO 427-G6 was also missing information from GALEX and 2MASS-only photometry is not sufficient to measure a mass.
This is the only object missing a mass estimate in the Hubble sample.
NGC~4679, the host of SN~2001cz, was not used in the Hubble diagram analysis but should have been included in Figure~\ref{fig:all_hubble_redshift_mass}; however, it only has measurements from 2MASS and was excluded.

If any of the surveys observed a host galaxy when the respective SN~Ia was active, the SN~Ia could contaminate the measured flux.
We cross matched the years that SDSS, PS1, DECaLS, GALEX, and 2MASS were active with the time of maximum light of our supernovae and examined the respective galaxy images if there was an overlap in time.
The only contamination (PS1-13dkh in PS1) we discovered was already accounted for in this section.

\subsubsection{Comparing Masses Derived using Different Surveys}
\label{sec:OpticalPhotComp}

We used galaxies in Stripe 82~\citep{Abazajian09} at redshifts between 0.001 and 0.1 to compare the masses derived with photometry from different optical surveys.
We downloaded the galaxies from SDSS and then matched on their coordinates to catalogs from DECaLS, PS1, GALEX, and 2MASS.
The different sets of optical photometry were used in \texttt{kcorrect} to derive masses, as is done for the SNeIa host galaxies.
We also compared different combinations of data including only optical data, optical plus GALEX plus 2MASS, optical plus GALEX, and optical plus 2MASS.

\begin{figure}
\gridline{\fig{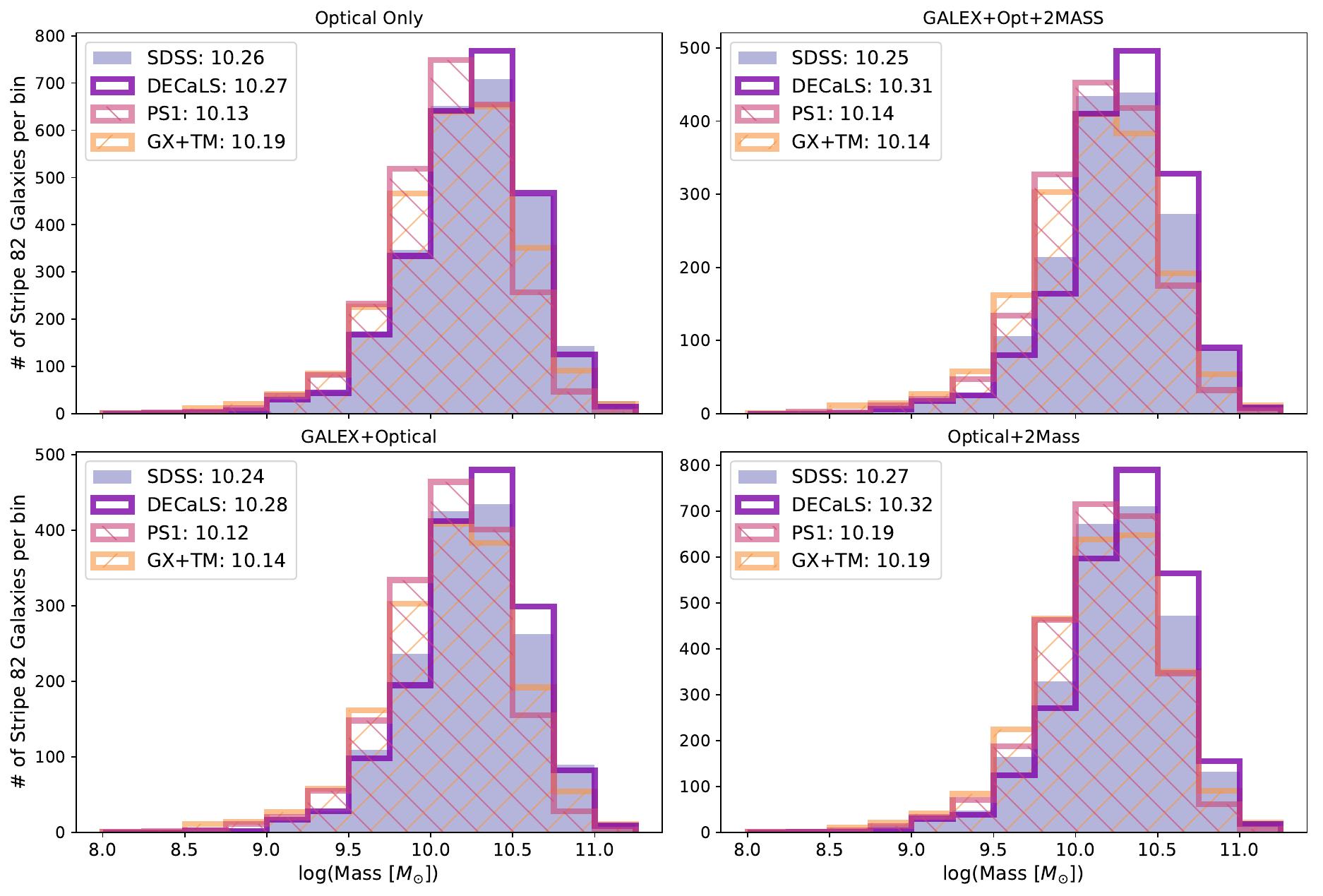}{0.665\textwidth}{}
              \fig{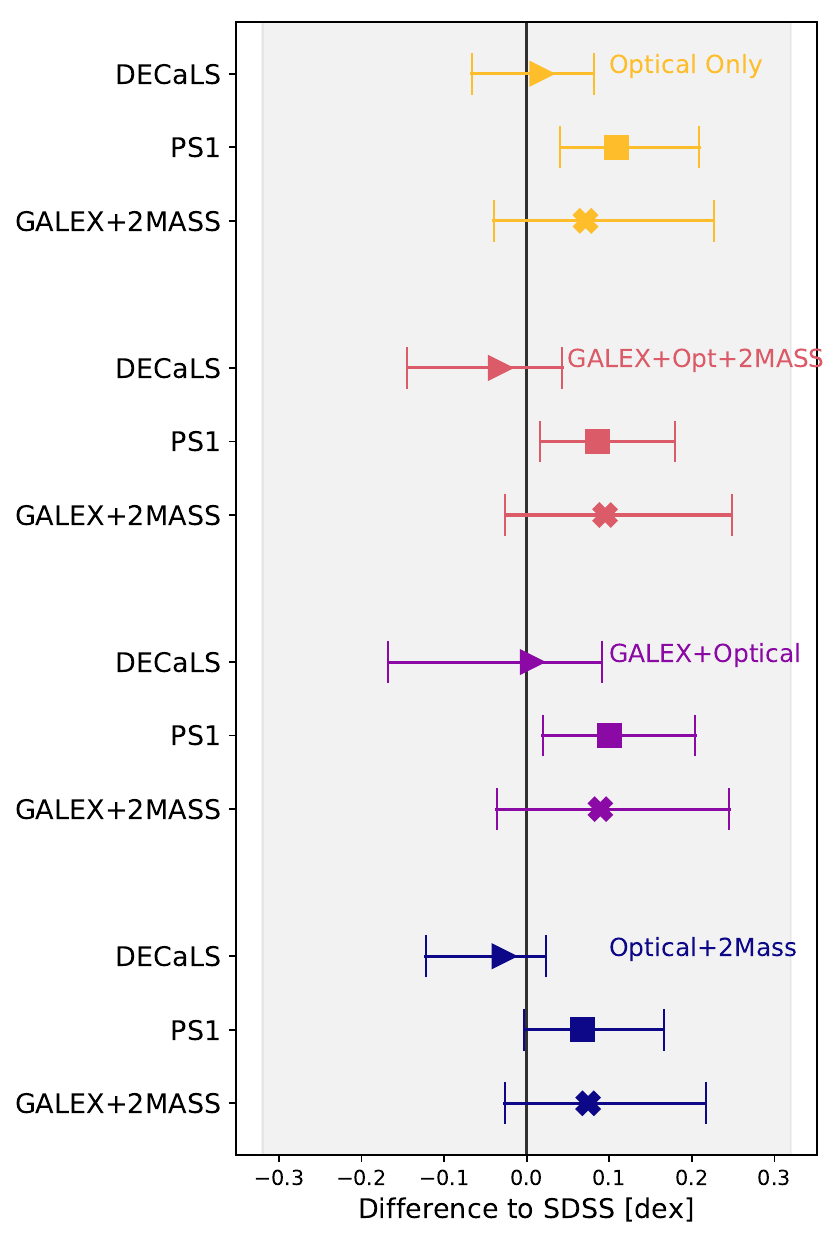}{0.3\textwidth}{}}
\caption{\textit{Left:}  Masses derived for galaxies in Stripe 82.
Top left is using only optical photometry.
Top right uses optical photometry and GALEX and 2MASS.
Bottom left uses only optical and GALEX photometry.
Bottom right uses only optical and 2MASS photometry.
For comparison, we have also included the histogram for the GALEX+2MASS (``GX+2M'') derived masses where no optical photometry is used.
The legends contain the median mass in log solar masses.
\textit{Right:} The difference between the median mass derived with photometry from DECaLS, PS1, and GALEX+2MASS (no optical) compared to the median mass from SDSS.
Each set of comparisons corresponds to a frame on the left side of this figure.
The error bars are the 16th and 84th ($1~\sigma$) percentiles.
The grey area corresponds to the estimated error on galaxy mass at 0.32 dex (Section~\ref{sec:mass_bias}).
}
\label{fig:stripe82_hist_diff}
\end{figure}

Figure~\ref{fig:stripe82_hist_diff} shows the mass histograms for the Stripe 82 galaxies for the different combinations of multi-wavelength photometry.
Each photometry combination in the different frames contains the same galaxies such that each frame is comparing the same set of objects.
Table~\ref{tab:stripe82_rmse} shows the number of galaxies used in each frame.
Overlaid on each frame is are the galaxy mass measurements for objects with no optical photometry.
These mass measurements are required to have both GALEX and 2MASS information; however, this excludes many low mass objects from our comparison as they are too dim to be detected by 2MASS.
For the optical photometry-only mass measurements, the low mass galaxies observed with DECaLS and PS1 follow the same distribution relative to SDSS as the higher mass galaxies.

The right side of Figure~\ref{fig:stripe82_hist_diff} shows the difference of each optical photometry source compared to SDSS with the estimated error on SDSS mass in grey (see Section~\ref{sec:mass_bias}).
It further shows how each survey difference changes depending on which supplementary photometry is included.
In all cases, DECaLS has a nearly zero median offset from SDSS indicating that it is the most similar to \texttt{kcorrect} masses derived with SDSS photometry.
PS1 is systematically offset to lower masses.

To determine how the surveys differ in a one-to-one galaxy mass comparison, we calculated the root mean square (RMS) error while taking the measurement from SDSS to be the truth.
Table~\ref{tab:stripe82_rmse} summarizes these results and shows that DECaLS is consistently more aligned with SDSS than PS1.
Though the distribution of masses for GALEX+2MASS aligns slightly better with SDSS than PS1 (Figure~\ref{fig:stripe82_hist_diff}), PS1 does better when comparing individual galaxies.

\begin{deluxetable}{lccccc}
\tablewidth{0pt}
\tablecaption{RMS Difference in Mass Compared to SDSS for Stripe 82 Galaxies. }
\tablehead{
 \colhead{Photometry Combination} &  \colhead{$\#$}  & {PS1} & {DECaLS} & {GALEX+2MASS}
}
\startdata
Optical Only  & 2603 &  0.18 & 0.14 & 0.17\\
GALEX + Optical + 2MASS    & 1627 &  0.16 & 0.13 & 0.18  \\
GALEX + Optical       & 1627   &  0.18 & 0.16 & 0.19 \\
Optical + 2MASS & 2601  &  0.16 & 0.13 & 0.17 \\
\enddata
\tablecomments{RMS units in $\log_{10}(M_{\odot})$   }
\label{tab:stripe82_rmse}
\end{deluxetable}

Since \texttt{kcorrect} was intended to be used for SDSS photometric observations, we gave priority to SDSS photometry first.
Comparing other optical photometry to SDSS, we preferentially used the DECaLS optical photometry for objects without SDSS photometry.
We then used PS1 if neither SDSS nor DECaLS photometry were available.
We give the lowest priority to observations without optical photometry.
Table~\ref{tab:surveys} gives the final number breakdown for which optical photometry was used.

\begin{figure}
\plotone{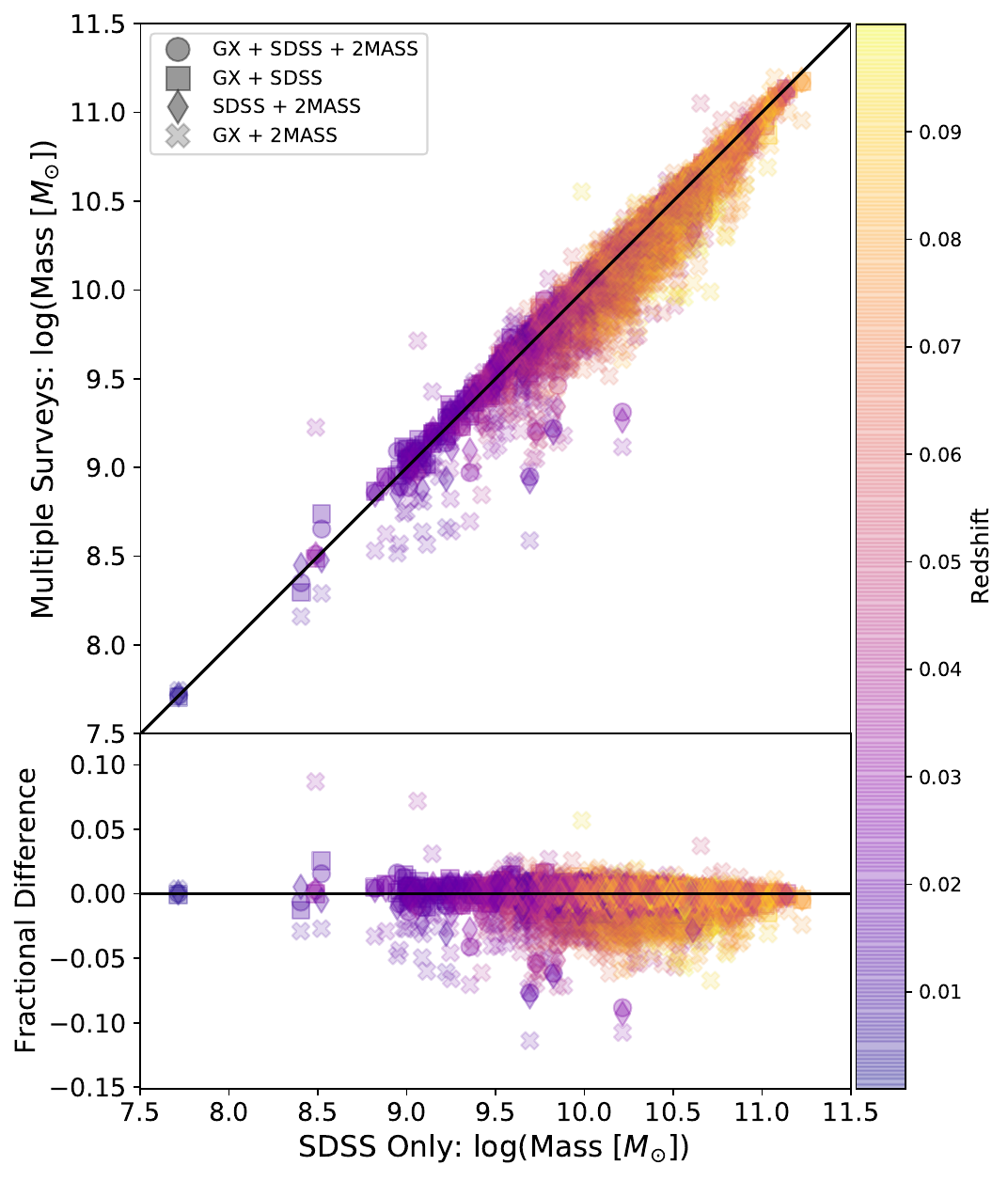}
\caption{Comparison of the derived mass from \texttt{kcorrect} for Stripe 82 galaxies when only optical SDSS data is used versus GALEX+SDSS+2MASS (squares), GALEX+SDSS (circles), SDSS+2MASS (diamonds), and GALEX+2MASS (x).
\textit{Top:} One-to-one mass comparisons.
\textit{Bottom:} Fractional difference between the SDSS-only derived masses and the masses derived from other or additional sources.
The color map indicates the redshift of the host galaxy.
Most high mass galaxies are in agreement with the optical only measurements; however, low mass galaxies show more variation.
}
\label{fig:optical_v_opticalplus}
\end{figure}

Figure~\ref{fig:optical_v_opticalplus} compares \texttt{kcorrect}-derived properties for SDSS versus SDSS+UV, SDSS+NIR, or SDSS+UV+NIR for the 1627 Stripe 82 galaxies with optical, GALEX, and 2MASS photometry.
All the optical surveys show similar correlations as SDSS does in this figure.
The high-mass galaxies agree with the optical-only measurements, because they have less dust and star formation such that the mass-age degeneracy that is broken by adding UV and/or NIR information is less relevant.
At masses $<10^{10} M_{\odot}$, there are differences between the SDSS-only and SDSS+ results with additional discrepancies between SDSS+UV versus SDSS+NIR in the derived mass.
Adding UV and NIR wavelength coverage improves estimates of low mass galaxies as long as the templates cover the same parameter space.
GALEX+2MASS is the most discrepant, but as shown in Figure~\ref{fig:stripe82_hist_diff}, the overall distribution is in agreement with SDSS and the errors are within the estimated mass error.

\begin{figure}
\plotone{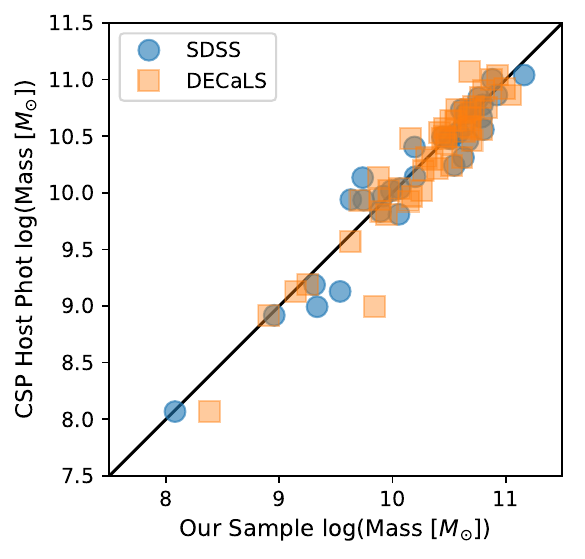}
\caption{Comparison of masses derived from kcorrect using SDSS (circles) and DECaLS (squares) photometry versus CSP photometry.
Median offset of 0.022-0.035 dex. Standard deviation of 0.18-0.34 dex. These offsets are well within the estimated errors that we derive for \texttt{kcorrect} in Section~\ref{sec:mass_bias}.
The black line is a one-to-one comparison to guide the eye.
}
\label{fig:csp_mass_comparison}
\end{figure}

Though the photometry should be similar to SDSS, we compared the CSP host galaxy photometry masses derived using \texttt{kcorrect} to those calculated using SDSS and DECaLS photometry in Figure~\ref{fig:csp_mass_comparison}.
We cannot use Stripe 82 galaxies in this comparison, so we used the 29 host galaxies that overlap between U20 and this paper's sample.
The median offset between SDSS (DECaLS) photometry-derived masses and CSP photometry-derived masses is 0.022 (0.035) dex.
These offsets are well within the estimated errors that we derive for \texttt{kcorrect} in Section~\ref{sec:mass_bias}.

Of the 29 host galaxies in common with the U20 derived masses, one host mass derived using CSP photometry APMUKS(BJ) B051529.79-235009.8 for SN~2006is had a discrepancy of almost $2~\log(M_{\odot})$ compared to the mass determined using DECaLS.
The galaxy is very faint with only $griJ$ observations from U20.
No other object in the CSP sample is missing both $u$ and $H$ measurements.
The DECaLS derived mass matches within the error bar of the U20 mass measurement.
All 5 galaxies from U20 used in this analysis have a mass equal to or greater than $10^{10}$ and $ugriJH$ (one is missing $u$) and should be consistent with the other surveys.

\subsubsection{Aperture Photometry from Different Surveys}
When using photometry from the different surveys, we did not extract new photometry with equivalent effective radii to ensure that each survey was measuring the same area of flux.

\begin{figure}
\plotone{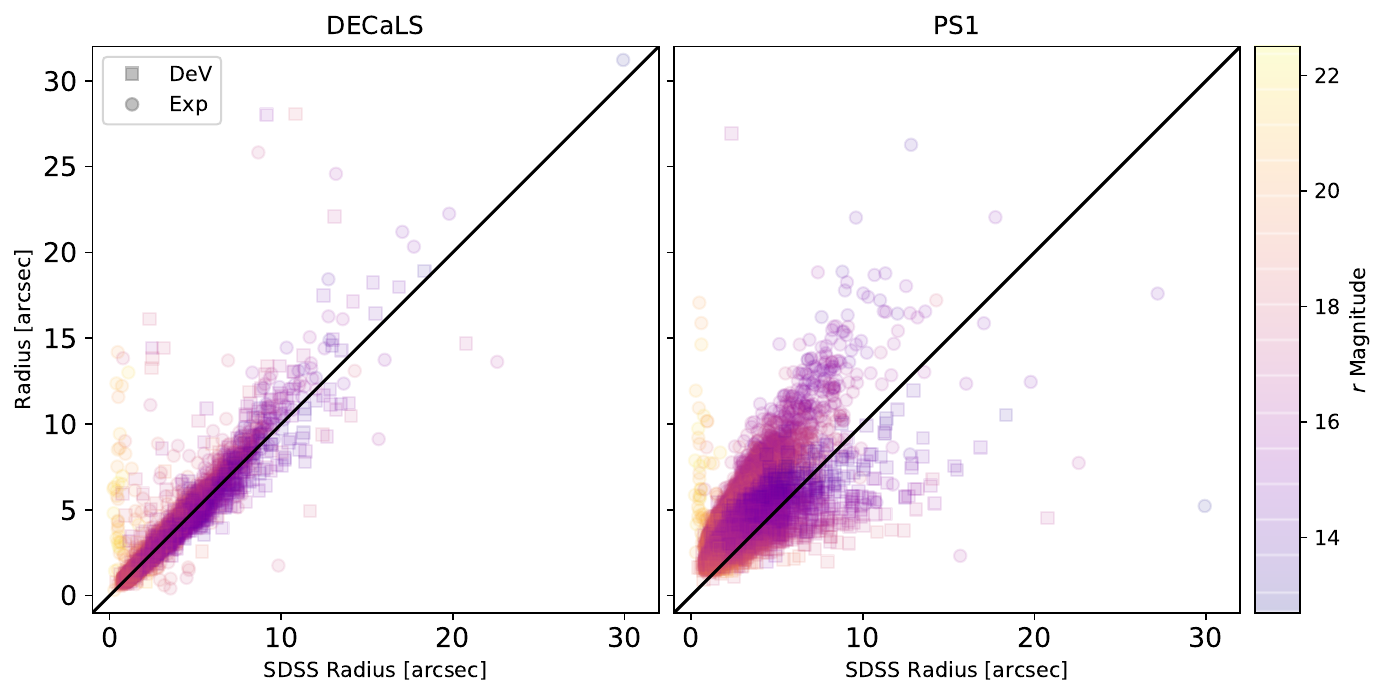}
\caption{One-to-one comparison of the effective radii compared to SDSS for DECaLS and PS1.
The squares correspond to a de Vaucouleurs (DeV) profile fit and the circules correspond to an Exponential (Exp) profile fit.
The color scheme corresponds to the SDSS $r$ model magnitude.
The black line is a one-to-one comparison to guide the eye.
The dim objects with small radii in SDSS and large radii in the other surveys are mismatches between the surveys.
The SDSS objects are typically small galaxies that are too faint for DECaLS or PS1.
The PS1 Kron radii have been adjusted to account for the DeV or Exp scaling factor provided in~\citet{Graham05}.
}
\label{fig:stripe82_aperture_onetoone_comparison}
\end{figure}

SDSS and DECaLS are calculated in similar ways such that each object is fit with an exponential profile and a de Vaucouleurs profile.
We downloaded all of the measured effective radii for the Stripe 82 galaxies.
We then sorted the effective radii into groups based on whether they were classified as having an exponential profile or a de Vaucouleurs profile.
Figure~\ref{fig:stripe82_aperture_onetoone_comparison} shows a one-to-one comparison of effective radii for DECaLS versus SDSS effective radii.
There is a tight correlation with the SDSS radius with a median difference of approximately zero, specifically $-0.02 \pm 0.01''$.
The outliers at low magnitudes ($\sim22$~mag) correspond to very small and dim galaxies in SDSS that were not detected in DECaLS but did have a brighter object nearby.
In some cases the dim objects are subsections of larger galaxies in SDSS that are matched the whole galaxy in DECaLS.
Large ($>10$\arcsec) outliers are due to crowded fields, mix-ups in object identification, extremely faint objects, clumpy/highly structured objects, or irregular shaped galaxies that confuse the algorithms.
We visually inspected images for all of the galaxies in our sample to check for possible erroneous magnitude measurements.
In general, these two surveys are using similar effective radii to SDSS to do their aperture photometry.

The PS1 sample uses Kron magnitudes that are measured using a different algorithm than the other optical photometry.
They are measured using the first radial moment and contains $~90\%$ of the total flux.
We expect the total flux to be underestimated and we see from Figure~\ref{fig:stripe82_hist_diff} that the PS1 mass measurements are systematically offset by about 0.1~dex.
To compare the Kron radii to the effective radii for the profile fitting from the other surveys, we use the relationships from \citet{Graham05} for the exponential ($r_{\rm Kron} = 1.19 r_{\rm effective}$) and de Vaucouleurs ($r_{\rm Kron} = 2.29 r_{\rm effective}$) profile.
We applied these two corrections and compared the two profiles in Figure~\ref{fig:stripe82_aperture_onetoone_comparison}.
There is a much larger scatter and offset compared to SDSS than with DECaLS with a median and standard deviation of $-2.32 \pm 0.41''$ for the exponential profile and $-0.62 \pm 0.24''$ for the de Vaucouleurs profile.

Though there are differences in the photometry from the different surveys, the measurement of galaxy mass has wide error bars which keeps the masses derived from different optical photometry consistent.

For GALEX and 2MASS, \texttt{kcorrect} was written to work with SDSS-matched GALEX and 2MASS catalog data as presented in~\citet{Blanton07} so we will not present an analysis of their different apertures.

\subsubsection{Bias in Calculated Host Galaxy Mass}\label{sec:mass_bias}
Twelve of our SNeIa with SDSS photometry overlapped with those used in the \citet{Kelly10} analysis.
\citet{Kelly10} fit {\it ugriz} photometry to different spectral energy distributions from PEGASE2~\citep{Fioc97, Fioc99} stellar population synthesis models using LePhare~\citep{Arnouts99,Ilbert06} using the IMF from \citet{Rana92}.
We found that our host galaxy masses using SDSS-only photometry are consistently lower than those reported in \citet{Kelly10} by a median value of 0.35~dex.
However, with the large uncertainties on host mass, we are consistent within $1$--$3~\sigma$.

The \texttt{kcorrect} approach derives lower masses because it calculates the current mass of the stars in a galaxy instead of the mass from integrating the total star formation rate over time which includes stars that died before we observed the galaxy.
To explore the bias in our data, we compared our \texttt{kcorrect}-derived masses in Stripe 82 to the photometric mass estimates from the MPA/JHU\footnote{\url{http://home.strw.leidenuniv.nl/~jarle/SDSS/}} originally presented in \citet{Kauffmann03} for SDSS DR4~\citep{Adelman06} and updated for SDSS DR7~\citep{Abazajian09}.
The original  \citet{Kauffmann03} analysis used the \citet{Kroupa01} IMF, but the updated version used the \citet{Chabrier03} IMF which matches the IMF used in  \texttt{kcorrect}.
\citet{Kelly10} compared their derived masses with~\citet{Kauffmann03} as well and found a mean bias of 0.033~dex with a dispersion of 0.15~dex, which is consistent with the \citet{Kauffmann03} data.

\begin{figure}
\epsscale{1.0}
\plotone{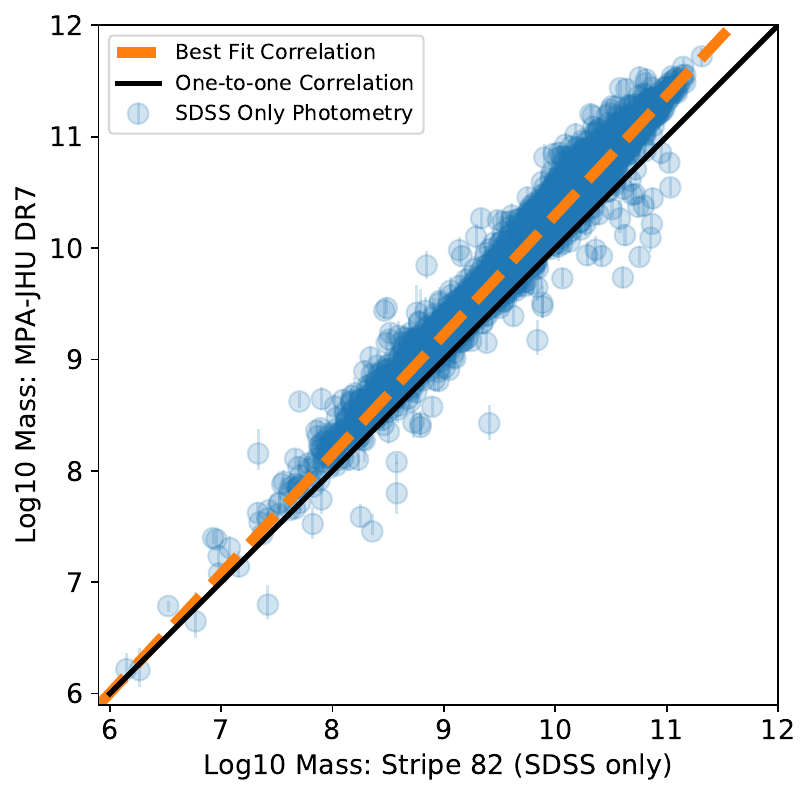}
\caption{Comparison of \texttt{kcorrect}-derived Stripe 82 galaxy masses and the masses from the MPA-JHU sample originally presented in~\citet{Kauffmann03} with DR4 data and updated for DR7.
The fit slope is 1.07 with offset -0.43 dex.
}
\label{fig:mpa_jhu}
\end{figure}

In our Stripe 82 sample of SDSS photometry, 8707 have overlapping information in MPA/JHU.
Figure~\ref{fig:mpa_jhu} plots the MPA/JHU DR7 masses versus our \texttt{kcorrect} masses and it is clear \texttt{kcorrect} systematically underestimates masses.
This offset is linear in log mass with a slope of 1.07 and an intercept of $-0.43~\log(M_{\odot})$ such that the effect increases as mass increases.
Both \citet{Bernardi10} and \citet{Moustakas13} have previously reported that \texttt{kcorrect} produces lower masses for high mass, elliptical galaxies.
\citet{Blanton07} compared their \texttt{kcorrect}-derived masses to those calculated in \citet{Kauffmann03} (on which MPA/JHU DR7 is based) and showed that the results agreed to within 0.2~dex with a 0.1~dex scatter, which roughly agrees with our findings with a mean bias of 0.29~dex and a dispersion of 0.12~dex.

If we assume that the error in \texttt{kcorrect} can be estimated by the root-mean-square of the difference between \texttt{kcorrect} and MPA/JHU, then the error is $\sim 0.32$~dex.
This error estimate is also consistent with the error determined in~\citet{Rose19}.
With this estimate of the mass error, we can confirm that our derived masses are systematically lower than those see in \citet{Kelly10}.
But these differences are not significant on the scale of the mass range of the host galaxies, and most importantly, do not preferentially change the ordering of galaxies in mass.

\section{Hubble Diagram}\label{sec:distances}

We here present the NIR and optical Hubble diagram from the current global collection of literature data on SNeIa observed in restframe $H$.

\subsection{Lightcurves}\label{sec:lightcurves}

We used the SNooPy\footnote{Version 2.0, \url{https://github.com/obscode/snpy}} fitter of \citet{Burns11} to estimate maximum magnitudes in $H$ with the ``max\_model'' for the collected sample of supernovae.
We fit the optical lightcurves with the SNooPy ``EBV\_model2''.
For both models, we use the parameterization based on the updated $s_{BV}$ width parameter introduced in \citet{Burns14}.

We adopted the same approach as in \citet{Weyant14} of fitting separately in each band using the ``max\_model'' SNooPy model.
Unlike in \citet{Weyant14}, where we held $\Delta m_{15}=1.1$ fixed, we here fit for the width parameter $s_{BV}$.
We first fit with the reported time of maximum B-band light, \tbmax, from the original spectroscopic confirmation announcement (generally ATel or CBET).
For most of the SNeIa, we had constraining lightcurve information in the optical or NIR that started before peak brightness, we generated an updated \tbmax\ from a lightcurve fit to all available data.
We then recorded these updated \tbmax\ values along with the original estimates for those not updated and ran the final lightcurve fits with \tbmax\ fixed.

In total, there are 36 objects without optical or NIR observations before \tbmax; however, only 16 of these objects pass the cuts to be in either the NIR or optical Hubble diagrams.
For these 16 SNeIa that had no optical or NIR lightcurve points before \tbmax, we used the spectroscopic original \tbmax.
Two of them, PTF11qri and PTF13ddg, did have raw lightcurves of the transient as published by PTF DR3.\footnote{\url{https://www.ptf.caltech.edu/page/DR3}}$^{,}$\footnote{\url{https://irsa.ipac.caltech.edu/Missions/ptf.html}}
These lightcurves did not include explicit host-galaxy subtraction and so are unreliable for determining accurate brightness, but the relative magnitudes of the lightcurves are useful to determine a time of \tbmax.
We used these lightcurves to confirm that the time of maximum light from the optical photometry was consistent with the spectroscopic estimate.

The sample of SNeIa came from several surveys and the different transmission curves were accounted for in SNooPy using the corresponding CSP transmission curves, WHIRC transmission curves, and 2MASS (for PAIRITEL) transmission curves.

We used the default SNooPy $K$-corrections using the \citet{Hsiao07} spectral templates, but we did not warp the spectral templates to match the observed color (``mangle=False'').
We do not apply any color-luminosity correction as we do not assume a relationship between the different filters in our ``max\_model'' fitting.

We did not use lightcurves that were observed before 1990,
had no known optical \tbmax,
or were known to be SN~1991bg-like or other peculiar types (although we include SN~1991T-like events).
We excluded from the Hubble residual analysis any SN~Ia that had fewer than three lightcurve points in the $H$-band.
After these quality cuts, we have a sample of \numAll\ SNeIa.

\begin{figure*}
\epsscale{1.0}
\plotone{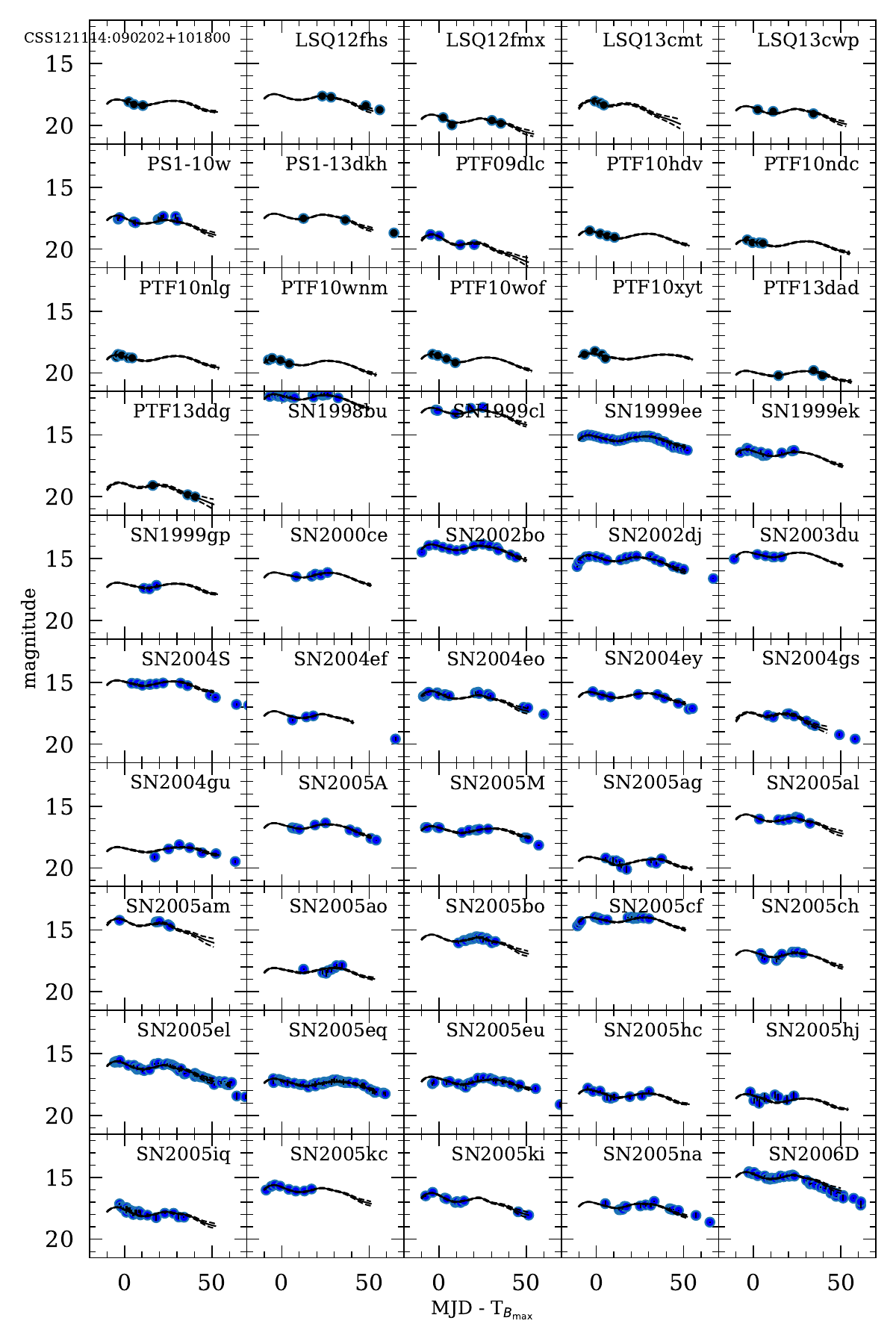}
\caption{The SNooPy $H_{\rm max}$ lightcurve fits to \numAll\ SNeIa.
Error bars are smaller than the markers.
}
\label{fig:lightcurvefitsa}
\end{figure*}

\begin{figure*}
\epsscale{1.0}
\figurenum{10.1}
\plotone{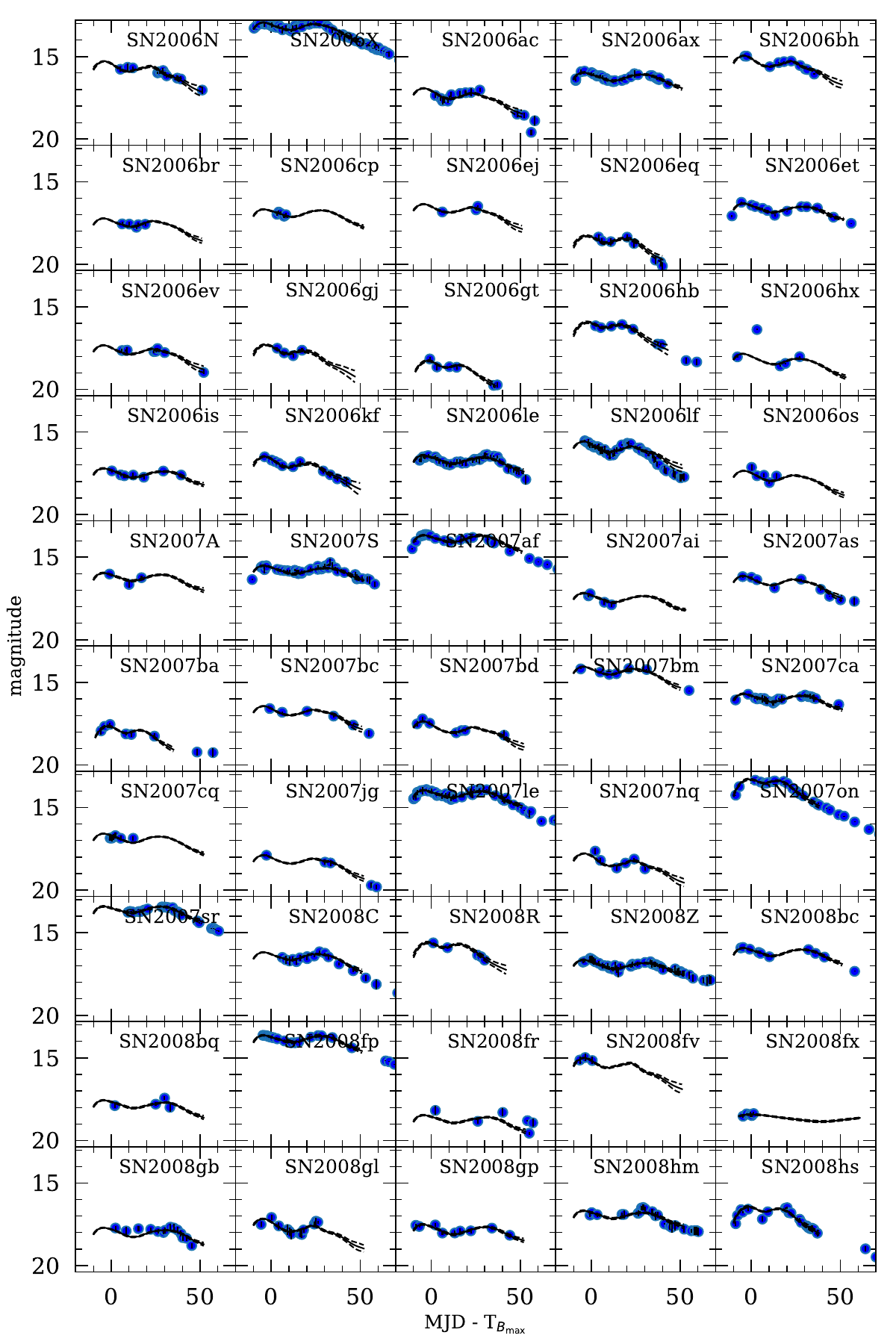}
\caption{The SNooPy $H_{\rm max}$ lightcurve fits to \numAll\ SNeIa.
Error bars are smaller than the markers.
}
\label{fig:lightcurvefitsb}
\end{figure*}

\begin{figure*}
\epsscale{1.0}
\figurenum{10.2}
\plotone{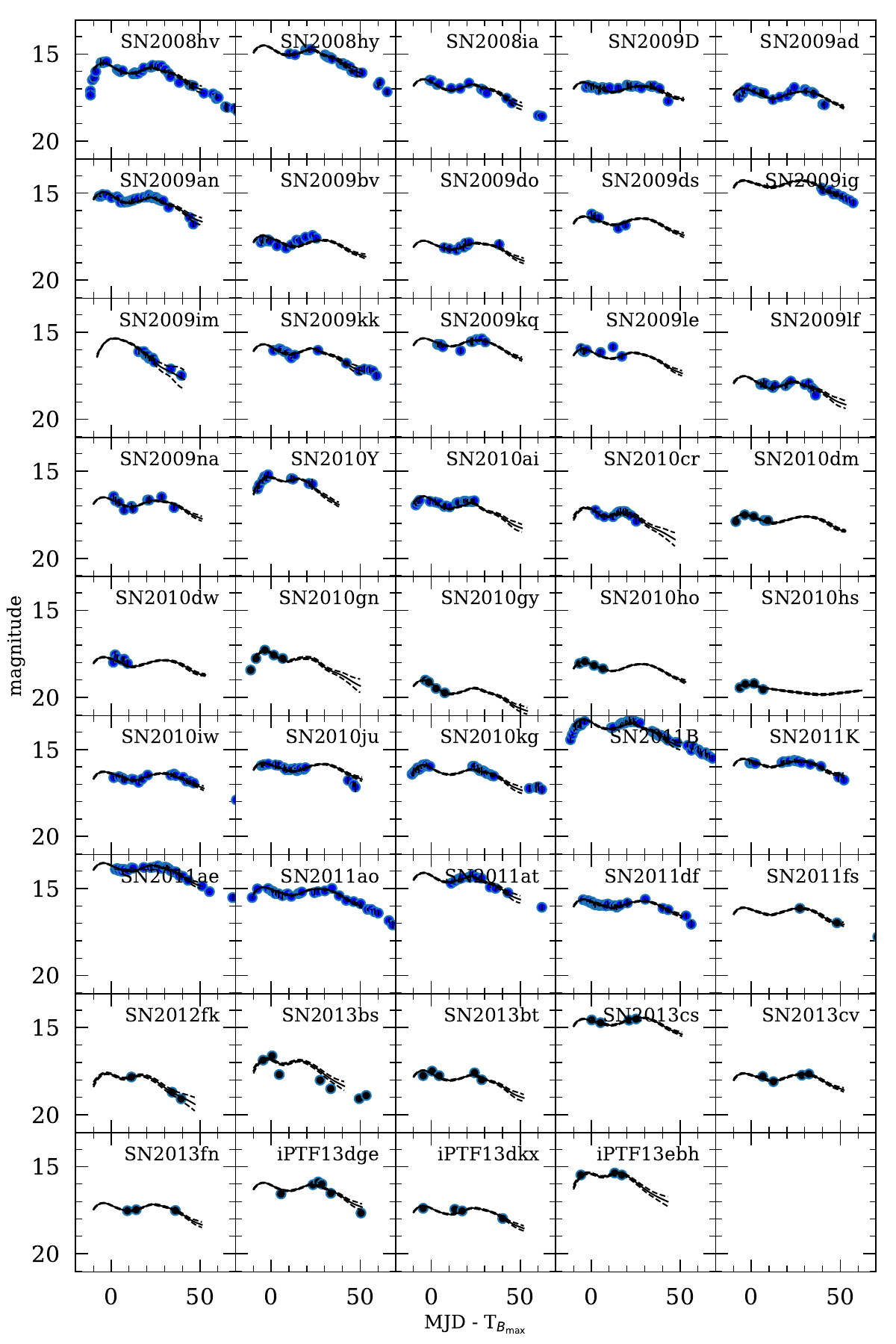}
\caption{The SNooPy $H_{\rm max}$ lightcurve fits to \numAll\ SNeIa.
Error bars are smaller than the markers.
}
\label{fig:lightcurvefitsc}
\end{figure*}

The lightcurve fits to SNe~Ia presented here are shown in Figures~\ref{fig:lightcurvefitsa}--\ref{fig:lightcurvefitsc}.
Errorbars are included on the plot but are smaller than the markers.
We did not use non-detections in these fits.

The $H$-band magnitudes are reported as the fit apparent magnitude based on the ``max\_model'' template for the given $s_{BV}$.
The SNooPy ``max\_model'' templates are normalized to a magnitude of 0 at maximum light individually for each filter and for all values of $s_{BV}$.
We apply no correction to the $H$-band apparent magnitude based on $s_{BV}$.
For the optical fits, we do include the $s_{BV}$ correction to the apparent brightness through the default use of the SNooPy ``EBV\_model2'' which includes the stretch-luminosity correction as part of the fit and estimation of distance modulus.
The ``max\_model'' and ``EBV\_model2'' do not account explicitly for intrinsic color variations, but both models do remove Milky Way reddening~\citep{Schlafly11}. ``EBV\_model2'' also accounts for host galaxy extinction using all of the optical filters by fitting for E(B-V) while holding $R_V$ constant.
To emphasize this distinction we quote the $H$-band fits in terms of apparent magnitude and the optical fit results in terms of distance modulus ($\mu$).

\subsection{Hubble Diagram}

We compare our measured SN~Ia apparent brightness to that predicted by a flat LCDM model with $H_0=72$~km~s$^{-1}$~Mpc$^{-1}$ and $\Omega_M=0.28$ \citep{Perlmutter99,Freedman01}.
We calculated the weighted best fit value of the absolute magnitude, after adding both an intrinsic dispersion of 0.08~mag~\citep[as reported in][]{Barone-Nugent12} and the equivalent magnitude uncertainty from a peculiar velocity of 300~km~s$^{-1}$
in quadrature to the reported statistical fit uncertainty from SNooPy.
We redid the full analysis at 150~km s$^{-1}$ and found minimal differences.
These additions to the uncertainty were used in computing the weighted average, but are not included in the errors plotted on the residual plots or reported in Table~\ref{tab:sn_hubble}.
While SNooPy ``max\_model'' reports apparent brightness and ``EBV\_model2'' returns distance modulus, the actual calculation of residuals follows the same process.
The absolute magnitude is entirely degenerate with the chosen value for $H_0$.  As we are here looking at residual relative brightness, the absolute brightness and value of $H_0$ are not directly relevant.
This model was then subtracted from the data points to yield the residuals that were used to compare against properties of the host galaxies.

\begin{figure}
\epsscale{1.0}
\plotone{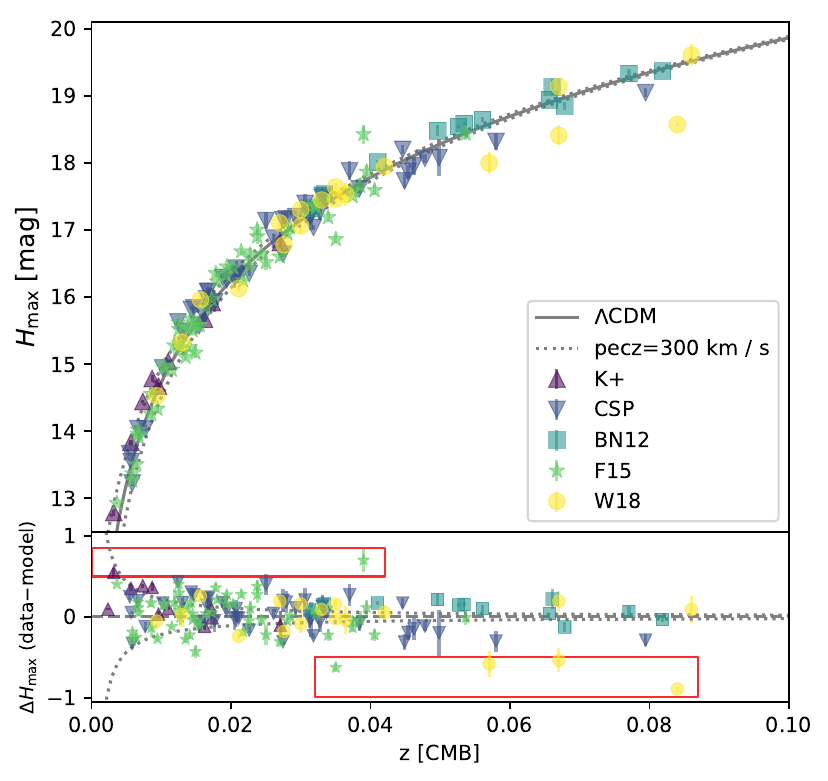}
\caption{\textit{Top:} SN~Ia $H$-band Hubble diagram for the sample considered in this paper.  \textit{Bottom:} The residuals from the apparent $H$ magnitude at maximum light (data$-$model) for the best-fit $\Lambda$CDM cosmology.  The points are coded in different shapes to indicate the source of the SN~Ia lightcurve data.
The 6 outliers referred to in Section~\ref{sec:remove_outlier} are highlighted in the red boxes.
The additional intrinsic scatter (0.08~mag) and the error from peculiar velocities are not included in the error bars.
}
\label{fig:SNIa_NIR_hubble}
\end{figure}

The results from these fits are tabulated in Table~\ref{tab:sn_hubble} and the resulting Hubble diagram with residuals is shown in Figure~\ref{fig:SNIa_NIR_hubble}.

\section{Analysis}\label{sec:analysis}
In this section, we examine the host galaxy stellar mass correlations with the restframe $H$-band residuals and the optical width-luminosity-corrected distance modulus residuals.
Though we present an in-depth study of host galaxy stellar mass since it is the largest trend seen in the literature with optical lightcurves, we have done the same studies examining restframe $K$-corrected absolute $r$-band magnitude as well as briefly exploring other properties of the supernova environment ($g-r$ color, Hubble flow, NUV colors, and distance from center of host galaxy).
These studies are summarized in Appendix~\ref{sec:other}.

\begin{figure}
\plotone{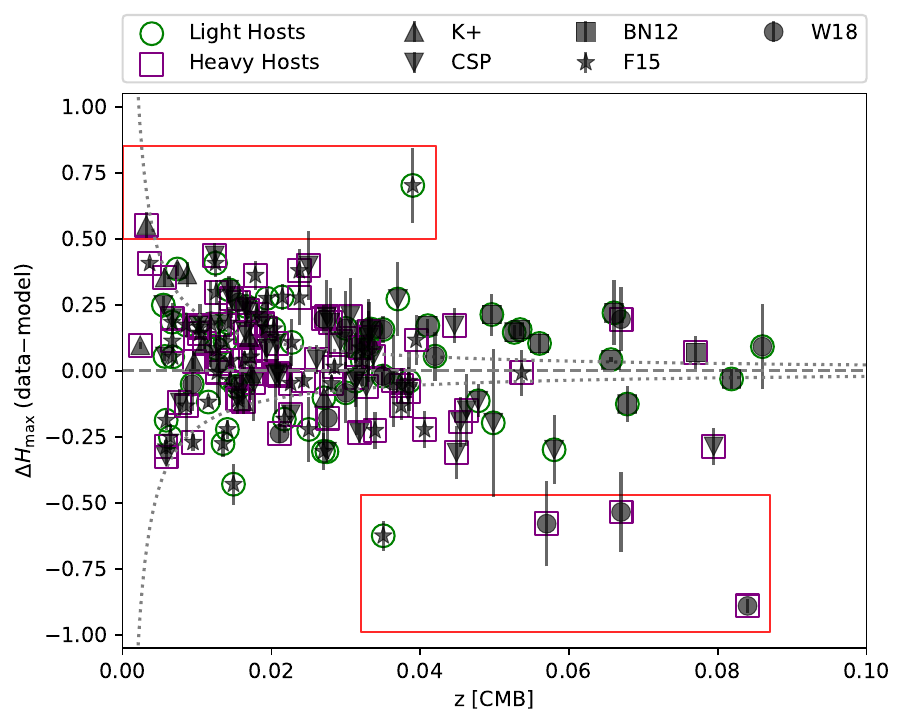}
\caption{The residuals from the apparent $H$ magnitude at maximum light (data$-$model) for the best-fit $\Lambda$CDM cosmology.  The points are coded in different shapes to indicate the source of the SN~Ia lightcurve data.
If no host galaxy mass was calculated, the point does not have a circle or square around it.
Overlaid on the points are the classification of their host galaxy: green circles are galaxies with mass $<10^{10} M_{\odot}$ and purple squares are galaxies with mass $>10^{10} M_{\odot}$.
The 6 outliers referred to in Section~\ref{sec:remove_outlier} are highlighted in the red boxes.
The additional intrinsic scatter (0.08~mag) and the error from peculiar velocities are not included in the error bars.
}
\label{fig:SNIa_NIR_residual_color}
\end{figure}

\subsection{Statistical Properties of the Distributions}\label{sec:distributions}
Having collected UV, optical, and/or NIR data allows us to estimate stellar masses for \numLightHeavy\ out of \numAll\ host galaxies.
We separate this sample by mass where the ``Light'' population corresponds to galaxies with masses less than $10^{10}~M_{\odot}$ and the ``Heavy'' population corresponds to galaxies with masses greater than  $10^{10}~M_{\odot}$.
Figure~\ref{fig:SNIa_NIR_residual_color} shows the Hubble residuals as a function of redshift with Light and Heavy galaxies highlighted.
Those with no indicator do not have sufficient host galaxy photometry to estimate mass.
We observe a population of 4 bright (residual $< -0.5$~mag) SNeIa at $z > 0.03$ plus 2 additional dim outlier (residuals $>0.5$~mag).
We see no clear trend in host galaxy mass versus redshift.

\begin{figure*}
\gridline{\fig{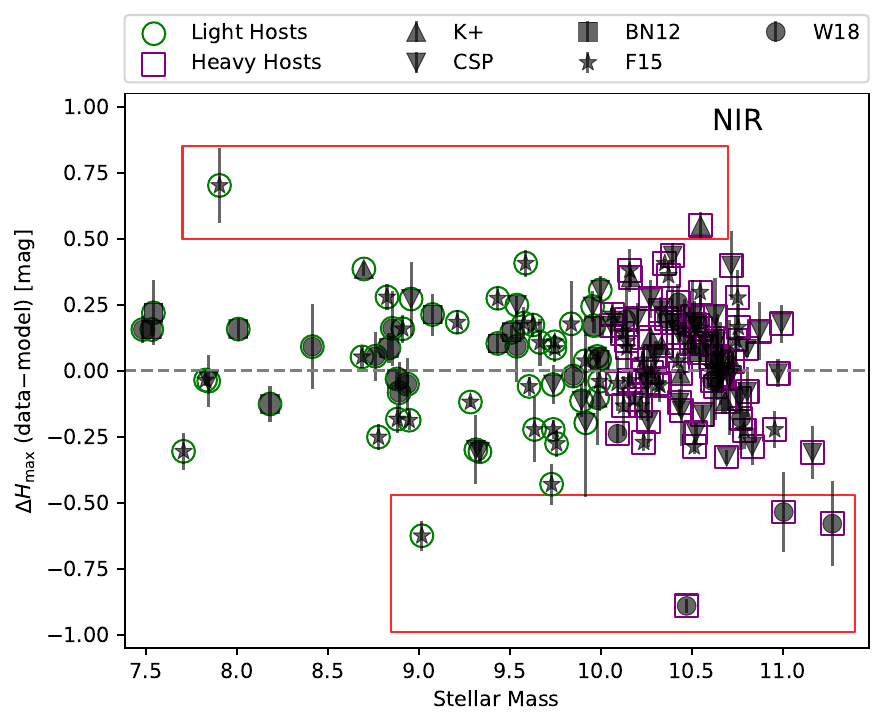}{0.52\textwidth}{}
          \fig{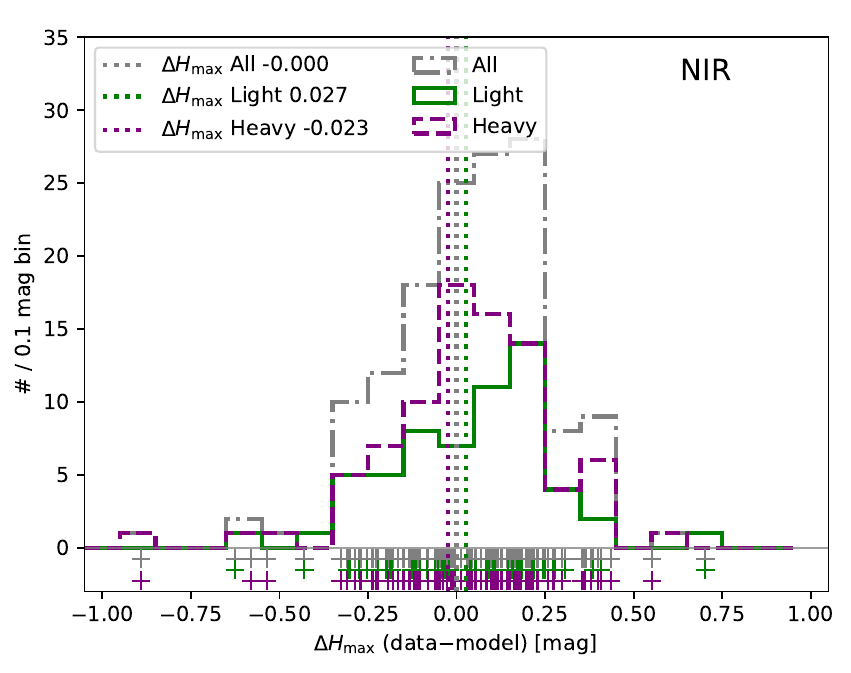}{0.4935\textwidth}{}}
\vspace{-11mm}
\gridline{\fig{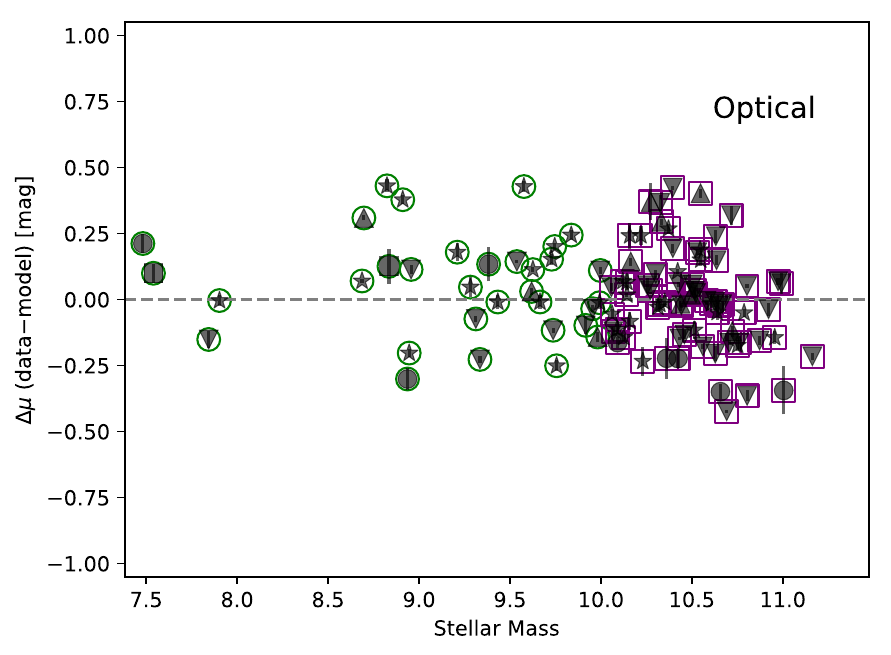}{0.52\textwidth}{}
          \fig{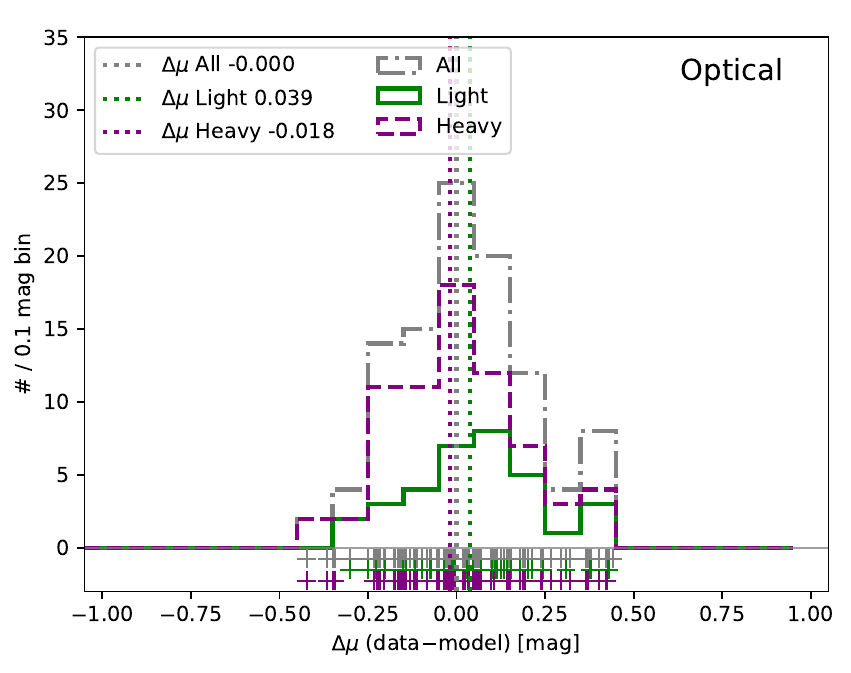}{0.4935\textwidth}{}}
\caption{
\textit{Top Left:} SN~Ia Hubble residuals vs host galaxy mass.
The points are coded in different shapes to indicate the source of the SN~Ia lightcurve data.
Overlaid on the points are the classification of their host galaxy: green circles are galaxies with mass $<10^{10} M_{\odot}$ and purple squares are galaxies with mass $>10^{10} M_{\odot}$.
\textit{Bottom Left:} Same as top left but for the distance modulus from optical lightcurves.
\textit{Top Right:} Histogram of Hubble residuals (data$-$model) for the SNeIa of the full sample (grey dashdotted).
The dotted vertical lines correspond to the weighted averages of the full (grey), light (green), and heavy (purple) samples.
\textit{Bottom Right:} Same as top right but for the distance modulus from optical lightcurves.
The additional intrinsic scatter (0.08~mag) and the error from peculiar velocities are not included in the error bars.
}
\label{fig:host_mass}
\label{fig:dm_mass}
\end{figure*}

The top left plot of Figure~\ref{fig:host_mass} shows the $H_{\rm max}$ Hubble residuals ($\Delta H_{\rm max}$) versus host galaxy mass and the top right plot shows a histogram of the Hubble residuals grouped by mass with the full sample included in grey for comparison.
Table~\ref{tab:stddev} shows the full details of the fits for the different populations including their peak residual magnitude, weighted peak residual magnitude, $\chi^2$, $\chi^2$/DoF, standard deviation, interquartile range (IQR), the standard error on the mean (SEM), and the intrinsic standard deviation that would result in a reduced $\chi^2 = 1$.

We find that the measured unweighted standard deviation of the whole sample is \stddevAll\ mag and the IQR equivalent to $1~\sigma$ assuming a Gaussian distribution is \iqrAll\ mag.
The standard deviation (IQR) of SN~Ia residuals in Light hosts is \stddevLight\ (\iqrLight)~mag, while the standard deviation (IQR) of SNeIa residuals in Heavy hosts is \stddevHeavy\ (\iqrHeavy)~mag.

The weighted average residual of the Light population is \weightResidualLight\ $\pm$ \semLight\ mag and the weighted average residual of the Heavy population is \weightResidualHeavy\ $\pm$ \semHeavy\ mag.
The difference in average weighted residuals is \shiftLightHeavy\ $\pm$ \semShiftLightHeavy\ mag with more massive galaxies hosting brighter SNeIa, which is not a detection at \sigmaShiftLightHeavy-$\sigma$ but has an amplitude in agreement with the literature.

If we remove the outlier population at $|\Delta H_{\rm max}| \geq 0.5$~mag, the separation between the peaks drops to \shiftOutlierLightHeavy\ $\pm$ \semShiftOutlierLightHeavy\ mag, a \sigmaShiftOutlierLightHeavy-$\sigma$ significance (Table~\ref{tab:stddev}) indicating these outliers are driving the $\sim1$-$\sigma$ shift seen in the full sample.
We will explore this ``outlier population'' further in Section~\ref{sec:remove_outlier}.

\begin{deluxetable*}{lrrrrrrrrcl}
\tablecaption{NIR SN Sample Mean and Std Deviations \label{tab:stddev}}
\tablewidth{0pt}
\tabletypesize{\footnotesize}
\tablehead{\colhead{Sample} & \colhead{SNeIa} & \colhead{residual} & \colhead{wgt residual} & \colhead{$\chi^2$} & \colhead{$\chi^2$/DoF} & \colhead{stddev} & \colhead{IQR} & \colhead{SEM} & \colhead{Implied $\sigma_H^{int}$}  & \colhead{Notes}\\ \colhead{ } & \colhead{ } & \colhead{$\mathrm{mag}$} & \colhead{$\mathrm{mag}$} & \colhead{ } & \colhead{ } & \colhead{mag} & \colhead{$\mathrm{mag}$} & \colhead{mag} & \colhead{$\mathrm{mag}$}}
\startdata
All & 144 & 0.031 & -0.000 & 346.2 & 2.40 & 0.229 & 0.207 & 0.019 & 0.174 &  \\
\\
Light   & 59 & 0.032 &  0.027 & 125.2 & 2.12 & 0.223 & 0.208 & 0.029 & 0.171 & $M <$  1e+10 $M_{\odot}$ \\
Heavy & 84 & 0.026 & -0.024 & 219.1 & 2.61 & 0.231 & 0.206 & 0.025 & 0.175 & $M \geq$ 1e+10 $M_{\odot}$ \\
\\
Light & 57 & 0.032 & 0.037 & 79.6 & 1.40 & 0.190 & 0.202 & 0.025 & 0.120 & $M <$  1e+10 $M_{\odot}$, \\
&  &  &  & &  &  & &  &  & $| \Delta H_{\rm max}|<0.5$~mag \\
Heavy & 80 & 0.046 & 0.022 & 97.3 & 1.22 & 0.182 & 0.177 & 0.020 & 0.105 & $M \geq$ 1e+10 $M_{\odot}$,  \\
&  &  &  & &  &  & &  &  & $|\Delta H_{\rm max}|<0.5$~mag \\
\\
Hubble Flow & 80 & -0.013 & -0.028 & 282.0 & 3.53 & 0.236 & 0.214 & 0.026 & 0.208 & $z>0.02$ \\
\\
Hubble Light & 35 & 0.020 & 0.022 & 97.0 & 2.77 & 0.226 & 0.196 & 0.038 & 0.193 &  $z>0.02$, $M <$  1e+10 $M_{\odot}$ \\
Hubble Heavy & 45 & -0.038 & -0.073 & 185.1 & 4.11 & 0.240 & 0.208 & 0.036 & 0.218 &  $z>0.02$, $M \geq$ 1e+10 $M_{\odot}$ \\
\\
Hubble Light & 33 & 0.019 & 0.034 & 51.3 & 1.55 & 0.166 & 0.191 & 0.029 & 0.127 &  $z>0.02$, $M <$  1e+10 $M_{\odot}$ \\
&  &  &  & &  &  & &  &  & $| \Delta H_{\rm max}|<0.5$~mag \\
Hubble Heavy & 42 & 0.007 & -0.012 &   63.9 & 1.52 & 0.172 & 0.180 & 0.027 & 0.126 &  $z>0.02$, $M \geq$ 1e+10 $M_{\odot}$ \\
&  &  &  & &  &  & &  &  & $|\Delta H_{\rm max}|<0.5$~mag \\
\enddata
\end{deluxetable*}

\subsubsection{Correlations with Corresponding Optical Lightcurves}
Host galaxy correlations have been well studied in the optical wavelengths.
To compare our results to these studies, we repeated the analysis with optical lightcurves of SNeIa observed in the $H$-band.
The optical data set is only 104 SNeIa in total and 103 with host galaxy mass estimates.
The bottom panels in Figure~\ref{fig:dm_mass} shows the distributions from host galaxy stellar mass compared with the optical distance modulus ($\mu$) residuals.
Table~\ref{tab:stddev_optical} presents the resulting weighted residuals and standard deviations.
Here we see no difference in the Light and Heavy host galaxies with a difference in average weighted residuals $0.058 \pm 0.039$~mag, which is $\sim1.5~\sigma$.
The $\Delta H_{\rm max}$ outliers are not outliers in this sample and no objects have $|\Delta \mu| \geq 0.5$~mag.

Comparing the NIR and optical data sets, the NIR sample is $41\%$ low mass galaxies while the optical sample consists of only $32\%$ low mass host galaxies.
The low mass galaxies are more represented in the NIR than in the optical, but from Figure~\ref{fig:dm_mass} we can see that the low mass distributions have similar shapes.
We used the Z-test statistic to compare how similar the NIR and optical low mass galaxy residual distributions were and found a value of 0.07, which corresponds to a p-value of 0.47, indicating they are from the same distribution.
Though the low mass galaxies may be a smaller percentage of the total population, they are still representative of the full distribution.

In this histogram analysis, we found no statistically significant trends between restframe $H$ or optical SN~Ia brightness and host galaxy mass.

\begin{deluxetable*}{lrrrrrrrrcl}
\tablecaption{Optical SN Sample Mean and Std Deviations \label{tab:stddev_optical}}
\tablewidth{0pt}
\tabletypesize{\footnotesize}
\tablehead{\colhead{Sample} & \colhead{SNeIa} & \colhead{residual} & \colhead{wgt residual} & \colhead{$\chi^2$} & \colhead{$\chi^2$/DoF} & \colhead{stddev} & \colhead{IQR} & \colhead{SEM} & \colhead{Implied $\sigma_\mu^{int}$}  & \colhead{Notes}\\ \colhead{ } & \colhead{ } & \colhead{$\mathrm{mag}$} & \colhead{$\mathrm{mag}$} & \colhead{ } & \colhead{ } & \colhead{mag} & \colhead{$\mathrm{mag}$} & \colhead{mag}}
\startdata
All & 104 & 0.022 & -0.000 & 165.4 & 1.59 & 0.190 & 0.196 & 0.019 & 0.128 &   \\
\\
Light & 33 &  0.057 &  0.039 & 52.5 & 1.59 & 0.185 & 0.168 & 0.032 & 0.132 &  $M <$  1e+10 $M_{\odot}$  \\
Heavy & 70 & -0.000 & -0.019 & 110.1 & 1.57 & 0.185 & 0.171 & 0.022 & 0.124 & $M \geq$ 1e+10 $M_{\odot}$  \\
\\
Hubble Flow & 52 & -0.009 & -0.017 & 112.8 & 2.17 & 0.170 & 0.178 & 0.024 & 0.147 & $z>0.02$ \\
\\
Hubble Light & 15 & 0.052 & 0.033 & 36.4 & 2.43 & 0.183 & 0.160 & 0.047 & 0.165 &  $z>0.02$, $M <$  1e+10 $M_{\odot}$ \\
Hubble Heavy & 37 & -0.034 & -0.038 & 76.3 & 2.06 & 0.158 & 0.157 & 0.026 & 0.140 &  $z>0.02$, $M \geq$ 1e+10 $M_{\odot}$ \\
\enddata
\end{deluxetable*}

\subsection{Functional Form of Correlation}\label{sec:functionalform}
In the previous section, we compared the weighted mean residuals of SNeIa separated by host galaxy mass.
However, there is no strong reason to model any host-galaxy brightness dependence by a simple step function.
To further test the significance of this correlation, we explore different function forms for to relationship between SN~Ia Hubble diagram residuals and the host galaxy stellar mass.

\subsubsection{Different Models to Fit}
We fit 7 different models using \texttt{scipy.optimize.curve$\_$fit}:
a constant function corresponding to a single population and no correlation, a linear function, a step function with a break corresponding to the threshold used in the previous section ($10^{10} M_{\odot}$), a step function that fits for the location of the break as well as the amplitude and y-intercept, and several logistic functions.
We fit three logistic functions: one where the threshold was held constant at $10^{10}~M_{\odot}$, one where it was allowed to float, and the generalized logistic equation.
The error on the fitted model parameters corresponds to the diagonal elements of the resulting covariance matrix.

After fitting the different functions to our data, we compare which model describes the data better using two different information criteria (ICs): the Akaike Information Criterion \citep[AIC; ][]{Akaike74} and the Bayesian/Schwartz Information Criterion \citep[BIC; ][]{Schwarz1978}.
We use the updated AIC$_c$~\citep{Sugiura78}, which is more suitable for smaller samples.

Information criteria allow for a comparison of different models by balancing an improved $\chi^2$ versus an increase in the number of fit parameters.
However, AIC$_c$ and BIC cannot be used to determine the absolute goodness-of-fit of the model; they can only establish which model the data favor compared to another model.
We calculate $\Delta$AIC$_c$ and $\Delta$BIC relative to the constant model.
If the difference in IC is $>2$, a constant model is preferred; $>5$,  a constant model is strongly preferred;  $<-2$, the compared model is preferred; and $<-5$  the compared model is strongly preferred.
When $0 < $ IC $< 2$, there is a preference for a constant model, but not a statistically significant one.
Likewise, an IC between $-2$ and $0$ shows a preference for the compared model, but it is not significant.

\begin{figure*}
\gridline{\fig{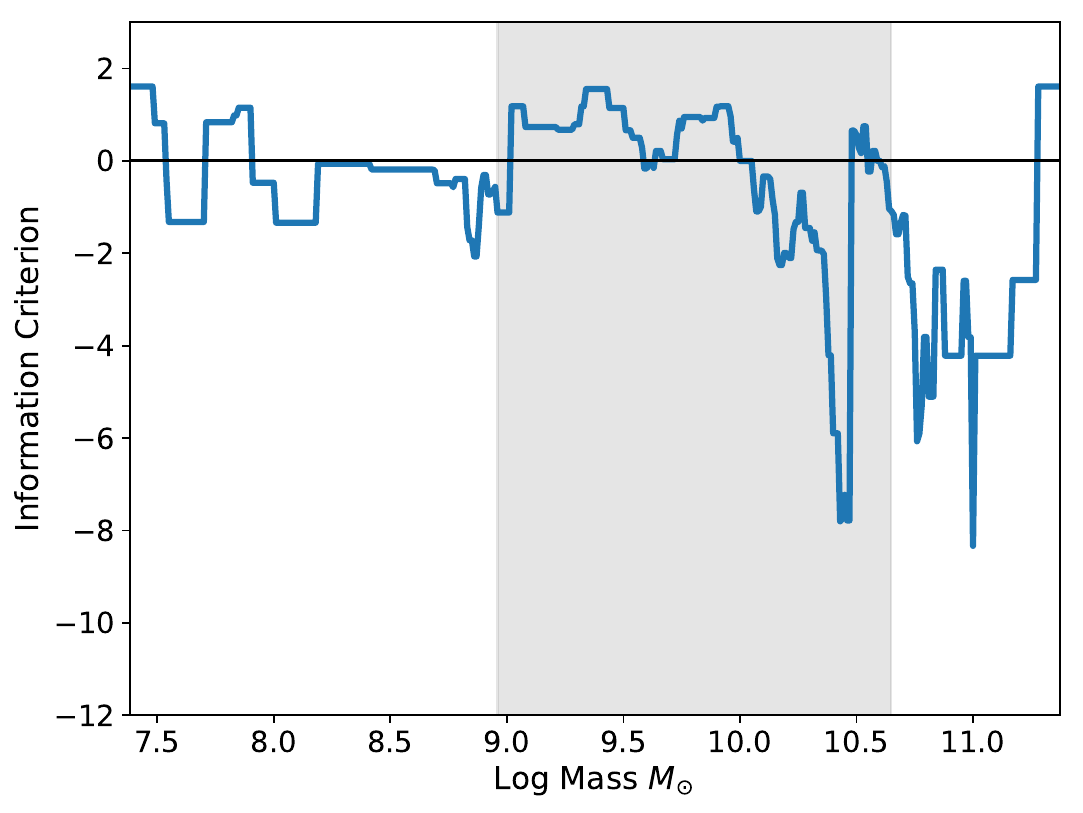}{0.5\textwidth}{}
          \fig{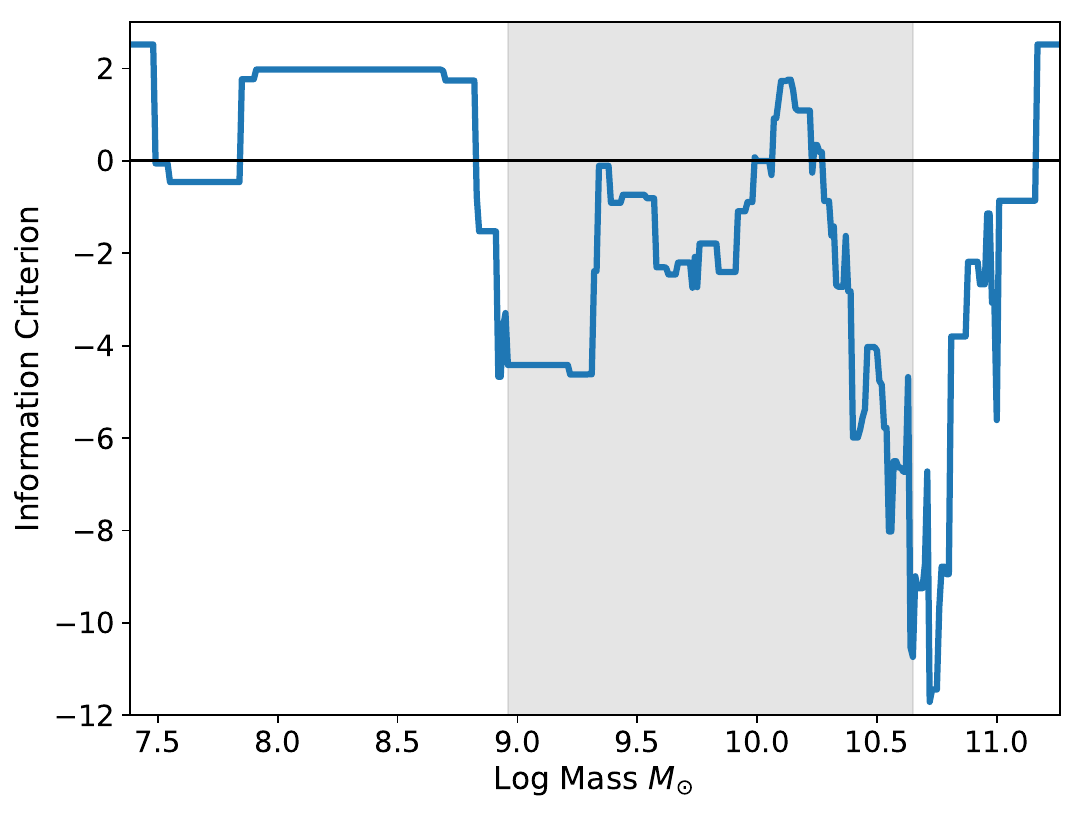}{0.5\textwidth}{}}
\vspace{-11mm}
\gridline{\fig{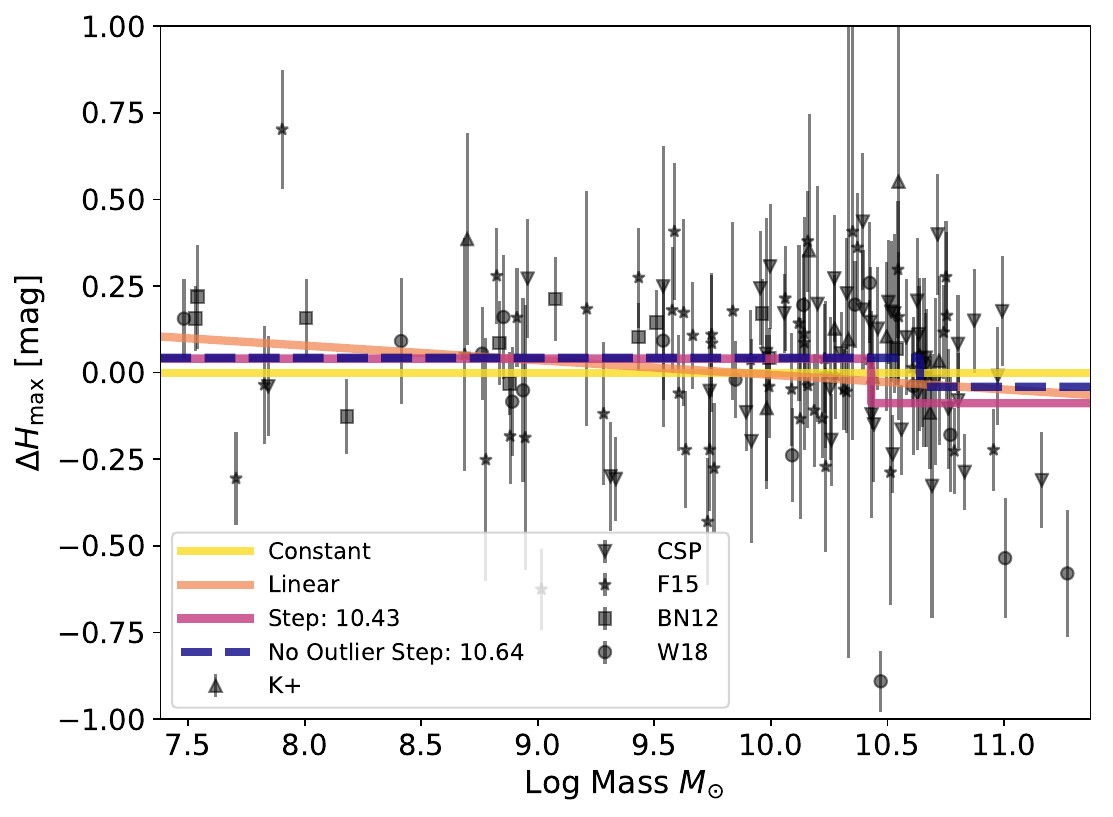}{0.5\textwidth}{}
          \fig{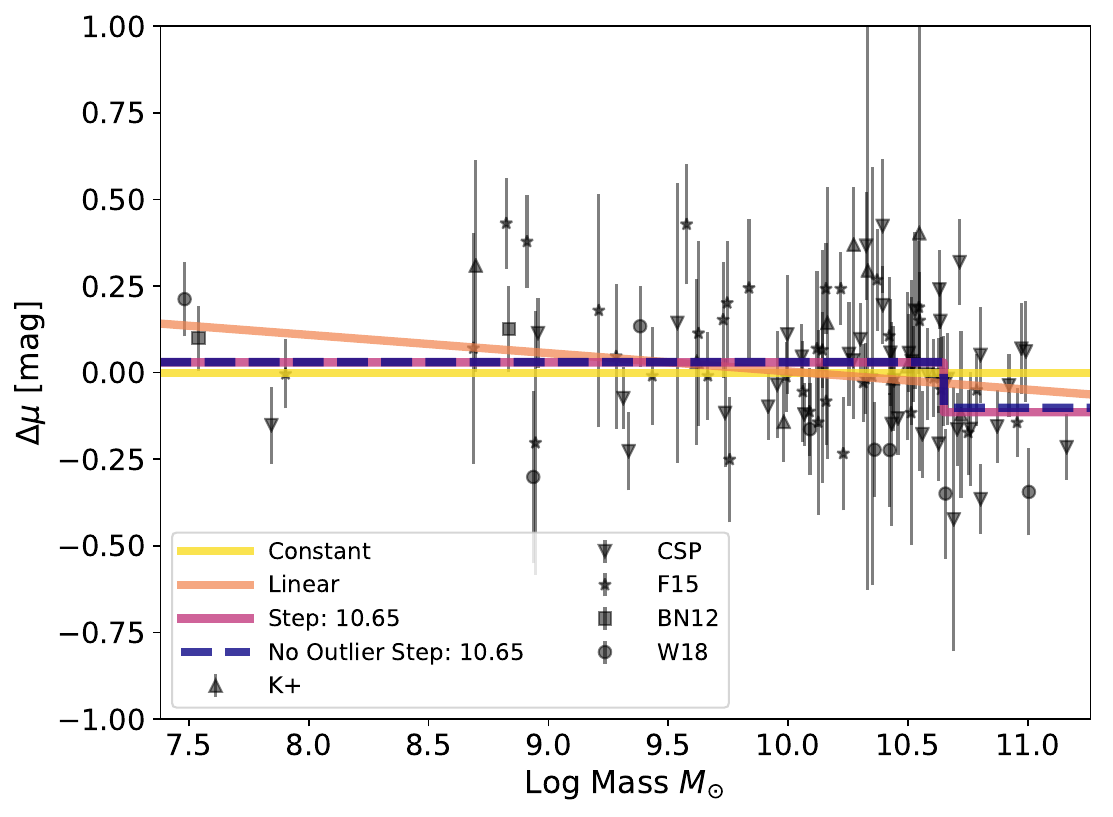}{0.5\textwidth}{}}
\caption{
Results from finding the best fit step location and fitting several functions to the residuals versus host galaxy mass.
\textit{Top:} Best fit location of the step function threshold shown using the AIC$_c$. 
The grey band highlights the area that we consider for the step function break such that the lower or higher mass bin has at least $20\%$ of the total SNeIa.
\textit{Bottom:} Various functions fit to the host galaxy mass versus Hubble residuals.
The blue, dashed line shows the best fit step function is the outliers are removed (Section~\ref{sec:remove_outlier}).
These plots do show the additional 0.08~mag intrinsic scatter that was introduced as a lower bound and the peculiar velocity errors.
\textit{Left: } Results from using the $H_{\rm max}$ Hubble residuals.
\textit{Right: } Results from using the distance modulus ($\mu$) optical lightcurve Hubble residuals.
The additional intrinsic scatter (0.08~mag) and the error from peculiar velocities \textbf{are} included in the error bars for both $H_{\rm max}$ and $\mu$.
}
\centering
\label{fig:model_fits_mass}
\end{figure*}

\subsubsection{$H$-band and Optical Results}\label{section:mass_fit}
To estimate the best site of the break (step function) or midpoint (logistic function), we fixed the position at a range of values between $7 < \log_{10}{({\rm mass}/M_{\odot})} < 12$ and fit for the other parameters in the respective models.
We then use the ICs to compare the model at each transition location versus the model with the step or midpoint located at the original threshold of $10^{10}~M_{\odot}$ and chose the location with the lowest IC.
The top panels of Figure~\ref{fig:model_fits_mass} show the results from doing this procedure for the step function for $\Delta H_{\rm max}$ and the optical distance modulus ($\Delta \mu$).

The top left panel is the result of fitting \numLightHeavy\ $H_{\rm max}$ residuals and has a minimum at $10^{10.43}~M_{\odot}$. Below a mass of $10^{8.96}~M_{\odot}$, the lower mass bin has less than $20\%$ of the total number of SNeIa making it more susceptible to edge effects during fitting.
The same is true for the higher mass bin above a mass of $10^{10.65}~M_{\odot}$.
Therefore, we only consider breaks in the step function between $10^{8.96}~M_{\odot}$ and  $10^{10.65}~M_{\odot}$, which is indicated in the grey band in the top panels in Figure~\ref{fig:model_fits_mass}.
The top right panel finds the best fit location for the 103 optical lightcurves favors a threshold at $10^{10.65}~M_{\odot}$.

The ICs strongly prefer a break at $10^{10.43}~M_{\odot}$ over $10^{10}~M_{\odot}$ for $H_{\rm max}$ residuals but prefer a larger mass  $10^{10.65}~M_{\odot}$ for the break in $\mu$ residuals.
Both the $H_{\rm max}$ and $\mu$ residuals favor a mass step that is in between the typical number found at $10^{10}~M_{\odot}$~\citep[e.g., ][]{Sullivan10,Lampeitl10, Gupta11,Childress13b} and $10^{10.8}~M_{\odot}$ found in \citet{Kelly10}.

The bottom panels of Figure~\ref{fig:model_fits_mass} show the models from the best fits: constant, linear, and the best-fit step function.
Table~\ref{tab:IC_mass} summarizes the best fit models using ICs and Table~\ref{tab:significance_mass} outlines the significance in the slope of the linear function, the step size of the best-fit step function, and the step size of the step function with a break at the original threshold.
We recover a $2$-$\sigma$ detection of a small slope but the ICs do not have a significant preference for a linear or constant model ($-2 <$ ICs $< 2$).
However, the AIC$_{c}$ strongly prefers a step function at the best-fit break over a constant model while the BIC favors a step function without being conclusive.
The best-fit step at $10^{10.43}~M_{\odot}$ finds a $0.13 \pm 0.04$~mag step at $3.25~\sigma$.

\begin{deluxetable}{llrr}
\tablewidth{0pt}
\tabletypesize{\normalsize}
\tablecaption{Information Criteria Results for Different Models \label{tab:IC_mass}}
\tablehead{\colhead{Residual} & \colhead{Fit\tablenotemark{a}} & \colhead{$\Delta$ AIC$_c$} & \colhead{$\Delta$ BIC}}
\startdata
$H_{\rm max}$  & Constant & 0.00 & 0.00 \\
 & Linear & -1.68 & 1.23 \\
 & Step: 10.00 & 0.45 & 3.35 \\
 & Step: 10.43 & -5.27 & -0.51 \\
 & Modified Logistic: 10.65 & 0.62 & 9.25 \\
 & Modified Logistic: 10.00 & 0.41 & 6.19 \\
 & Generalized Logistic & 10.59 & 24.81 \\
\hline
$\mu$ & Constant  & 0.00 & 0.00 \\
 & Linear & -5.22 & -2.66 \\
 & Step: 10.00 & -0.44 & 2.11 \\
 & Step: 10.65 & -9.06 & -3.99  \\
 & Modified Logistic: 10.65 & -6.58 & 0.96 \\
 & Modified Logistic: 10.00 & -3.75 & 1.32  \\
 & Generalized Logistic & 10.84 & 23.17 \\
\enddata
\tablenotetext{a}{If the fit is followed by a number, the number is the location of either the best fit break (step function) or the midpoint (logistic function) in units of log~$M_{\odot}$.}
\end{deluxetable}

\begin{deluxetable*}{llrrrrl}
\tablecaption{Significance of Linear and Step Function Fits \label{tab:significance_mass}}
\tablewidth{0pt}
\tabletypesize{\normalsize}
\tablehead{\colhead{Residual} & \colhead{Fit} & \colhead{Constant} & \colhead{$\sigma_{\rm Constant}$}  & \colhead{Slope${\rm |}$Step} & \colhead{$\sigma _{\rm Slope|Step}$} & \colhead{Units}}
\startdata
$H_{\rm max}$  & Constant & 0.00 & 0.02 &  & & mag  \\
 & Linear &  0.42 & 0.22 & -0.04 & 0.02  & mag (log~$M_{\odot}$)$^{-1}$ \\
 & Step: 10.00 & 0.03 & 0.03 & -0.05  & 0.04 & mag \\
 & Step: 10.43 & 0.04  & 0.02 & -0.13 & 0.04  & mag \\
\hline
$\mu$ & Constant & 0.00 & 0.02 &  & & mag  \\
 & Linear &  0.53 & 0.20 & -0.05 & 0.03 & mag (log~$M_{\odot}$)$^{-1}$ \\
 & Step: 10.00 & 0.04 & 0.04 & -0.06  & 0.04  & mag\\
 & Step: 10.65 & 0.03  & 0.02 & -0.14 & 0.04  & mag  \\
\enddata
\end{deluxetable*}

The ICs from distance modulus residuals favor a non-constant model more frequently than the $H_{\rm max}$ residuals.
A linear correlation is found at a $1.67$-$\sigma$ significance level and the ICs favor/strongly favor a linear model over the constant model.
The best-fit step function was found at a 3.5-$\sigma$ significance level and the ICs favor to strongly favor this model over the constant model.
The best-fit step at $10^{10.65}~M_{\odot}$ finds a $0.14 \pm 0.04$~mag step, but if we move the step to match the $H_{\rm max}$ residuals, the step size is reduced in size to $0.10 \pm 0.03$~mag with similar significance.
We do not measure a significant step at $10^{10}~M_{\odot}$ as found previously in the literature.

The modified logistic function provides a smooth transition between two populations unlike a step function which is an abrupt change; however, this model introduces an additional free parameter.
For both residuals, the best fit midpoint is at the highest allowed mass.
With the additional free parameter, the ICs more clearly favor a constant model with the AIC$_c$ preferring no model and the BIC strongly preferring a constant function.
No transition between populations was found when using a modified logistic function at the $10^{10}~M_{\odot}$ midpoint and the ICs prefer a constant model.
Given the ICs favor the step function more when compared to a constant model, we do not show the curves in Figure~\ref{fig:model_fits_mass} or include any of the fit parameters.

For the $H_{\rm max}$ and $\mu$ residuals, the generalized logistic function returned a straight line which completely overlaps with the constant model; however, the ICs do not prefer this model due to the additional parameters it introduced.
Since the ICs were strongly against these models in every scenario, we do not include the fit on the plots or the fit parameters.

We showed here that there is evidence of a trend between host galaxy mass and the $H_{\rm max}$ NIR lightcurves in which more massive galaxies host SNeIa that are brighter than those hosted in lower mass galaxies by $0.13 \pm 0.04$~mag.
We also measured a trend between host galaxy mass and optical lightcurves in which more massive galaxies host SNeIa that have more negative width-luminosity corrected optical brightnesses by $0.14 \pm 0.04$~mag.
Our results also agree with the literature~\citep{Childress13b} in that a step function is more preferred over a linear function to describe the correlation between residuals and host galaxy mass.
Though we found some evidence for a trend, the information criteria failed to provide strong, conclusive support in $H_{\rm max}$ lightcurves.
However, the ICs do enforce the trend measured with the optical lightcurves.

\section{Discussion}\label{sec:discussion}
In this section we will further explore the statistical significance of our analysis by studying
the dependence on the number of pre-maximum lightcurve points,
effects from using a heterogeneous set of SNeIa,
the outlier population,
modeling the underlying distribution with a Gaussian Mixture Model,
joint data samples,
the dependence on the location of the step,
and finally whether we are adding new information by including the NIR.
We also compare our result to U20 and discuss the physical interpretation of the results.

\begin{figure}
\gridline{\fig{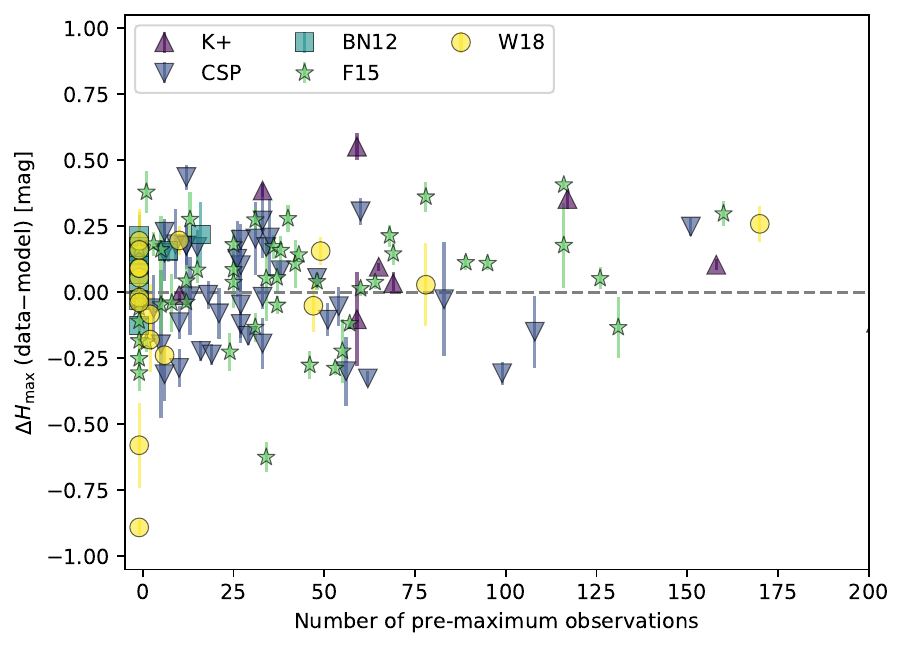}{0.52\textwidth}{}
          \fig{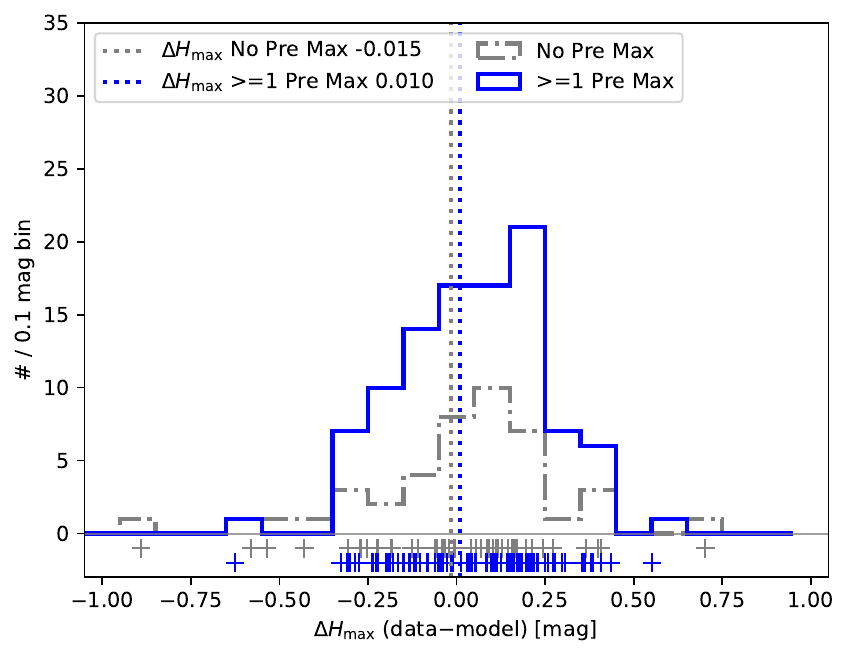}{0.4935\textwidth}{}}
\caption{
\textit{Left: }  SN~Ia Hubble residuals vs the of number of lightcurve points before \tbmax.
\textit{Right: }  Histogram of $H$-band residuals with 0 pre-max points in any band (grey dashed) and those with at least one pre-max point (blue, solid).
There is no dependence of the residuals on the number of pre-maximum points.
The brightest outlier, PTF13ddg, has no pre-maximum published lightcurve observations.  It does, however, have unsubtracted photometry from the PTF lightcurve database that shows behavior consistent with its quoted \tbmax.
}
\label{fig:SNIa_pre_max_residuals}
\end{figure}

\subsection{Residuals are Not Dependent on Number of Pre-Maximum Lightcurve Points}

One reasonable concern is that the number of lightcurve points observed before \tbmax could affect the reliability of inferred maximum brightness.
As discussed in Section~\ref{sec:lightcurves}, there were 14 SNeIa with no pre-maximum lightcurve points.
Figure~\ref{fig:SNIa_pre_max_residuals} shows that the Hubble diagram residuals were not dependent on the number of pre-maximum lightcurve points.
We thus conclude that our brightness measurements are robust to the number of pre-maximum lightcurve points.

\subsection{Comparison of Residuals per Sample}
\begin{figure}
\centering
\epsscale{1.0}
\plotone{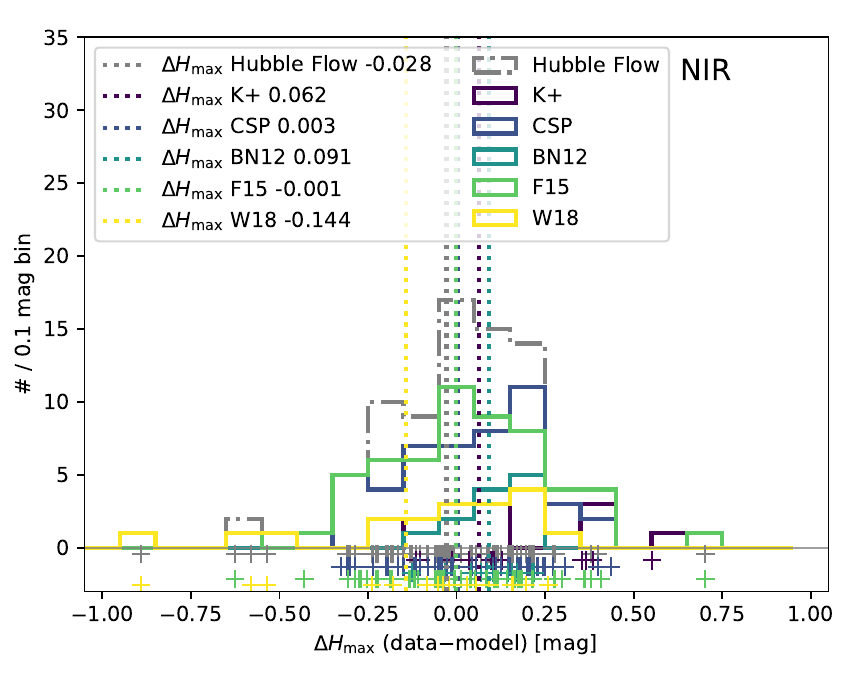}
\caption{
The $H$-band residuals for all Hubble flow SNeIa (grey dashed) and for each sample.
The individual points are shown below the $y=0$ axis for ease of reference to the original data.
The dotted vertical lines correspond to the weighted averages of the individual samples.
}
\label{fig:hubble_samples_residuals}
\end{figure}

Here we look at the statistical properties of the residuals if we separate them per sample, which are summarized in Table~\ref{tab:stddev_surveys}.
Figure~\ref{fig:hubble_samples_residuals} shows the $H_{\rm max}$ residuals colored by SN lightcurve source (Sample).
The difference in weighted mean residuals between the brightest (W18) and dimmest (BN12) samples is 0.24~mag and 0.20~mag for the $H_{\rm max}$ and $\mu$ residuals, respectively (see Table~\ref{tab:stddev_surveys}).
This difference between surveys is larger than any step size we see based on any host-galaxy feature.
However, the brightest population comes from W18 which features 3 of the bright outlier SNeIa.
These 3 SNeIa also factor into the larger standard deviation and intrinsic dispersion seen in W18.
BN12, the dimmest sample, has the tightest standard deviation.
We note that BN12 reported a small range in $B$-band stretch for their lightcurves indicating a data set lacking in intrinsic variation of SNeIa and 8 out of 9 BN12 SNeIa with host galaxy photometry are in blue galaxies.

Table~\ref{tab:stddev_surveys} has an additional column denoted ``Step'' which is the size and direction of the step function assuming a mass step break at $10^{10} M_{\odot}$.
The $H_{\rm max}$ residuals per sample are all consistent with zero except the W18 sample.
W18 contains most of the NIR outliers (Section~\ref{sec:remove_outlier}).
If these objects are removed, the size of the step, $-0.06 \pm 0.09$~mag and $-0.08 \pm 0.09$~mag for peculiar velocities of 300~km~s$^{-1}$ and 150~km~s$^{-1}$, respectively, is consistent with zero and the other samples.
The $\mu$ residuals are more complicated with respect to the step.
F15 shows $\sim2$-$\sigma$ step inline with the literature.
W18 exhibits a large, significant step that is not affected by the $H_{\rm max}$ outliers; however, there are only 8 objects in this sample and small number statistics is the likely driver of this result.

While the surveys have different mean properties in their residuals, they overall appear to form a continuous distribution.
We thus assert that using SNeIa from different samples is not greatly biasing our results.
A possible exception is BN12, which shows little variation in host galaxy type and may contain an intrinsically different distribution of SNeIa.

In Table~\ref{tab:stddev_surveys}, the K+ sample does not have a reported intrinsic dispersion for the $H_{\rm max}$ residuals.
To determine the intrinsic dispersion, we set the $\chi^2$/DoF equal to one and solve for the intrinsic dispersion.
For K+, this dispersion would have to be imaginary since the $\chi^2$/DoF is less than one.
The K+ sample has a lower redshift distribution than the other surveys (see the purple triangles in Figure~\ref{fig:SNIa_NIR_hubble}).
The imaginary implied intrinsic dispersion is a manifestation of the high peculiar velocity choice of 300 km~s$^{-1}$.
The K+ sample is the lowest redshift collection (you can piece this out of Figure~\ref{fig:SNIa_NIR_hubble}) and is thus most sensitive to the peculiar velocity assumed.
If the assumed peculiar velocity is reduced to 150~km~s$^{-1}$, the implied intrinsic dispersion for $H_{\rm max}$ residuals is $0.159$~mag.
For all other surveys, the intrinsic dispersion changes by $\sim$-0.005$\pm$0.029~mag if the peculiar velocity is 150~km~s$^{-1}$.
We see a similar response to the change in peculiar velocities from the $\mu$ residuals.
Though we do see small differences noted in Table~\ref{tab:stddev_surveys}, reducing the peculiar velocity has no affect on the outcome of this analysis.

The intrinsic dispersion assumed in the fitting analysis (0.08~mag) is clearly underestimating the intrinsic dispersion as reported in Table~\ref{tab:stddev_surveys}.
We continue to use 0.08~mag as it represents a lower limit on what the intrinsic dispersion could be for a single, well-sampled survey.
However, we reran the analysis while increasing the intrinsic dispersion to match the full sample implied $\sigma^{\rm int}$ from Table~\ref{tab:stddev_surveys}.
For $\Delta H_{\rm max}$, the size of the step is reduced to  $0.10 \pm 0.04$~mag at $10^{10.43} M_{\odot}$, but $10^{10.43} M_{\odot}$ is still the location of the best fit step.
The AIC$_c$ is negative but favors no model and the BIC favors a constant model.
The slope of the linear function is consistent with zero.
For $\Delta \mu_{\rm max}$, the slope and step results from Table~\ref{tab:significance_mass} are still valid but the ICs are reduced slightly.
The step function at $10^{10.65} M_{\odot}$ is still strongly preferred over a constant model.
Increasing the intrinsic dispersion degraded the step between host galaxy mass and $H$ residuals to 2.5-$\sigma$ but had no effect on the step for $\mu$ residuals.

\begin{deluxetable}{cclrrrrrrrrcr}
\tablecaption{SN Sample Mean and Std Deviations - Surveys \label{tab:stddev_surveys}}
\tablewidth{0pt}
\tabletypesize{\scriptsize}
\tablehead{\colhead{Residual} & \colhead{Pec. Vel.} & \colhead{Sample} & \colhead{SNeIa} & \colhead{residual} & \colhead{wgt residual} & \colhead{$\chi^2$} & \colhead{$\chi^2$/DoF} & \colhead{stddev} & \colhead{IQR} & \colhead{SEM} & \colhead{Implied $\sigma^{\rm int}$} & \colhead{Step\tablenotemark{a}} \\ \colhead{ } & \colhead{km~s$^{-1}$} & \colhead{ } & \colhead{ } & \colhead{$\mathrm{mag}$} & \colhead{$\mathrm{mag}$} & \colhead{ } & \colhead{ } & \colhead{mag} & \colhead{$\mathrm{mag}$} & \colhead{mag} & \colhead{mag} & \colhead{mag}}
\startdata
$H_{\rm max}$ & 300 & All & 144 & 0.031 & 0.000 & 346.2 & 2.40 & 0.229 & 0.207 & 0.019 & 0.174 & $-0.05 \pm 0.04$ \\
 & & K+ & 11 & 0.162 & 0.063 & 6.6 & 0.60 & 0.209 & 0.258 & 0.063 &\nodata\tablenotemark{b} & $-0.01 \pm 0.14$ \\
 & & W18 & 18 & -0.077 & -0.144 & 135.6 & 7.53 & 0.301 & 0.219 & 0.071 & 0.291  & $-0.36 \pm 0.17$ \\
 & & F15 & 56 & 0.022 & -0.001 & 115.6 & 2.06 & 0.233 & 0.222 & 0.031 & 0.174 & $0.05 \pm 0.06$ \\
 & & CSP & 47 & 0.034 & 0.003 & 69.5 & 1.48 & 0.195 & 0.214 & 0.029 & 0.130 & $0.06 \pm 0.06$ \\
 & & BN12 & 12 & 0.100 & 0.091 & 19.0 & 1.59 & 0.097 & 0.073 & 0.028 & 0.115 & $-0.02 \pm 0.11$ \\
\hline
$H_{\rm max}$ & 150  & All & 144 & 0.017 & 0.000 & 474.0 & 3.29 & 0.229 & 0.207 & 0.019 & 0.197 & $-0.03 \pm 0.04$ \\
 & & K+ & 11 & 0.148 & 0.080 & 18.0 & 1.64 & 0.209 & 0.258 & 0.063 & 0.159  & $-0.13 \pm 0.16$ \\
 & & W18 & 18 & -0.091 & -0.131 & 153.4 & 8.52 & 0.301 & 0.219 & 0.071 & 0.297 & $-0.30 \pm 0.17$ \\
 & & F15 & 56 & 0.008 & -0.008 & 180.4 & 3.22 & 0.233 & 0.222 & 0.031 & 0.203  & $0.05 \pm 0.06$ \\
 & & CSP & 47 & 0.020 & 0.007 & 103.3 & 2.20 & 0.195 & 0.214 & 0.029 & 0.160  & $0.04 \pm 0.07$ \\
 & & BN12 & 12 & 0.086 & 0.081 & 19.0 & 1.59 & 0.097 & 0.073 & 0.028 & 0.110 & $-0.03 \pm 0.11$ \\
 \hline
 & &  &  & & & &  &  & &  &   \\
 \hline
 $\mu$ & 300 & All & 104 & 0.022 & 0.000 & 165.4 & 1.59 & 0.190 & 0.196 & 0.019 & 0.128 & $-0.05 \pm 0.06$ \\
 & & K+ & 11 & 0.157 & 0.050 & 11.1 & 1.01 & 0.205 & 0.251 & 0.062 & 0.082 & $0.17 \pm 0.12$  \\
 & & W18 & 8 & -0.157 & -0.091 & 23.7 & 2.96 & 0.201 & 0.165 & 0.071 & 0.214  & $-0.39 \pm 0.09$ \\
 & & F15 & 39 & 0.048 &  0.042 &   54.4 & 1.40 & 0.171 & 0.162 & 0.027 & 0.118  & $-0.10 \pm 0.05$ \\
 & & CSP & 44 & -0.006 & -0.024 & 74.1 & 1.68 & 0.173 & 0.157 & 0.026 & 0.126 & $0.05 \pm 0.06$ \\
 & & BN12 & 2 & 0.112 & 0.109 & 2.1 & 1.06 & 0.013 & 0.010 & 0.009 & 0.084 &  \nodata\tablenotemark{c} \\
\hline
$\mu$ & 150 & All & 104 & 0.012 & -0.000 & 292.0 & 2.81 & 0.190 & 0.196 & 0.019 & 0.165 & $-0.06 \pm 0.04$ \\
 & & K+ & 11 & 0.147 & 0.061 & 25.0 & 2.28 & 0.205 & 0.251 & 0.062 & 0.182 & $0.11 \pm 0.13$ \\
 & & W18 & 8 & -0.167 & -0.128 & 39.4 & 4.92 & 0.201 & 0.165 & 0.071 & 0.245 & $-0.34 \pm 0.11$ \\
 & & F15 & 39 & 0.038 & 0.041 & 102.4 & 2.63 & 0.171 & 0.162 & 0.027 & 0.157 & $-0.11 \pm 0.06$ \\
 & & CSP & 44 & -0.016 & -0.023 & 123.1 & 2.80 & 0.173 & 0.157 & 0.026 & 0.155 & $0.06 \pm 0.06$ \\
 & & BN12 & 2 & 0.103 & 0.100 & 2.1 & 1.05 & 0.013 & 0.010 & 0.009 & 0.083 & \nodata\tablenotemark{c} \\
\enddata
\tablenotetext{a}{Size and direction of step assuming a break at $10^{10} M_{\odot}$.}
\tablenotetext{b}{The K+ sample does not return an implied $\sigma_H^{\rm int}$. $\sigma^{\rm int}$ is determined by setting $\chi^2$/DoF equal to one and solving for the intrinsic dispersion. The K+ $\chi^2$/DoF is less than one which would result in an imaginary intrinsic dispersion.}
\tablenotetext{c}{With only 2 objects for BN12 in the optical, we did not fit for a step function.}
\end{deluxetable}

\subsection{Impact of the NIR Outlier Population}\label{sec:remove_outlier}
Out of \numAll\ SNeIa there are 6 SNeIa with $|\Delta H_{\rm max}| > 0.5$~mag as listed in Table~\ref{tab:outlier}.
The four bright outliers SNeIa are LSQ13cmt, LSQ13cwp, PTF13ddg, and SN~2005eu.
The two faint outliers are  SN~1999cl and SN~2008fr.

\begin{deluxetable*}{lccccrcrccrrrrc}
\tablewidth{0pt}
\tablecaption{Outlier SNeIa \label{tab:outlier}}
\tablehead{
 \colhead{SN} &   \colhead{Sample} & \colhead{$z$} & \colhead{$\sigma_{{\rm out}, H}$} &\colhead{$\Delta H_{\rm max}$} &  \colhead{$\Delta \mu$} & \colhead{$s_{BV, H}$}  & \colhead{$N_H$} &\colhead{Phot\tablenotemark{a}} &  \colhead{Profile} &  \colhead{Mass} &  \colhead{$g-r$} &  \colhead{$M_r$} & \colhead{PGCD}  \\
  \colhead{} &  \colhead{}  & \colhead{} & \colhead{}  & \colhead{mag} &  \colhead{mag} & \colhead{} & \colhead{} & \colhead{} &  \colhead{} &  \colhead{log$_{10}(M_{\odot})$} &  \colhead{mag} &  \colhead{mag} & \colhead{Mpc}
}
\startdata
LSQ13cmt & W18 & 0.057 & 2.49 & -0.57 & -- & 0.68 & 3 &  DT & DeV & 11.27 &  0.81 &  -23.56 & 0.0383  \\
LSQ13cwp  & W18 & 0.067& 2.31 & -0.53 & -0.32 & 0.91 & 3 & DT  & DeV & 11.00 & 0.78 & -22.87  & 0.0153 \\
PTF13ddg	 & W18 & 0.084 & 3.84 & -0.88 & -- &  0.70  & 3 & ST & DeV & 10.47 & 0.72 & -21.50  & 0.0728  \\
SN~2005eu & F15 & 0.035  & 2.71 & -0.62 & -- & 1.11 & 23 & D & REX\tablenotemark{b} & 9.02 & 0.49 & -18.71 & 0.0009  \\
SN~2008fr & F15 & 0.039 & 3.05 & 0.70 & -0.00 &1.10 & 6 & S & DeV &  7.90  & 0.38 & -17.11 &  0.0008 \\
SN~1999cl & K+ & 0.003 & 2.40 & 0.55 & 0.40 &  0.93 & 5& K\tablenotemark{c}T & EXP & 10.55 & -- & -- & 0.0034  \\
\enddata
\tablenotetext{a}{Galaxy photometry source where ``D'' is DECaLS only, ``S'' is SDSS only, ``DT'' is DECaLS plus 2MASS, ``KT'' is Kron plus 2MASS, and ``ST'' is SDSS plus 2MASS.}
\tablenotetext{b}{Round exponential galaxy, which is an extended but low signal-to-noise galaxy.}
\tablenotetext{c}{Only $giy$.}
\end{deluxetable*}

We excluded the 6 outliers and repeated the $\Delta H_{\rm max}$ fits versus host galaxy mass from Section~\ref{sec:functionalform}.
Table~\ref{tab:summary_discussion} presents the number of supernovae, the step size, and the best fit (BF) step location for the original sample and for the sample without the large outliers.
The size of the step drops from $0.13 \pm 0.04$~mag to $0.08 \pm 0.04$~mag and the location of the best-fit step moved to $10^{10.64} M_{\odot}$, which is more similar to the optical residuals' mass split location.
The step function with outliers and without the outliers is directly comparable in Figure~\ref{fig:model_fits_mass}.
The slope from the linear model was consistent with zero.
All of the ICs either favor no model or the constant model.

3 out of 6 outlier SNeIa were also present in the optical data set with host galaxy mass, but none are also an outlier in that sample.
The size of the step decreased by 0.01~mag which is a $3$-$\sigma$ detection and the best fit break stays the same, see Table~\ref{tab:summary_discussion}.
The ICs are now more conflicted with the the step function at $10^{10.65} M_{\odot}$ being strongly preferred by the AIC$_c$ but the BIC has no model preference.

In the NIR, removing these large $H$-band outliers reduced the significance of the step reported above to $2$-$\sigma$ and moves the location of the step to $10^{10.65}~M_{\odot}$, which is the edge of the allowed range.
The global minimum is closer to  $10^{10.8}~M_{\odot}$ where there are fewer objects.
The correlation is no longer partially preferred by the ICs over a constant model.
If we force the step to be at $10^{10.43}~M_{\odot}$, the step size is $0.06 \pm 0.03$~mag and the ICs do not prefer a step function over a constant model ($\Delta$ AIC$_c$=-1.53, $\Delta$ BIC=1.33).
Part of the trend in the NIR was driven by the outlier population, mostly at  $\Delta H_{\rm max} \leq -0.5$~mag, but a $2$-$\sigma$ detection still remains.

The outlier population impacts but does not fully determine the correlation of brighter SNeIa in more massive galaxies.

\begin{deluxetable}{llcccrr|cccrr}
\tablewidth{0pt}
\tablecaption{Number of SNeIa for Different Sections \label{tab:summary_discussion}}
\tablehead{
\colhead{Section} &  \colhead{Type} &  \colhead{\# $H_{\rm max}$}  & \colhead{Step $H_{\rm max}$}   & \colhead{BF Step Loc $H_{\rm max}$} & \colhead{AIC$_c$} &\colhead{BIC} & \colhead{\# $\mu$}  & \colhead{Step $\mu$} & \colhead{BF Step Loc $\mu$ } & \colhead{AIC$_c$}&\colhead{BIC}\\ \colhead{} & \colhead{} &  \colhead{} & \colhead{mag}   & \colhead{log10 $M_{\odot}$}  &\colhead{}& \colhead{}   & \colhead{mag} & \colhead{log10 $M_{\odot}$} & \colhead{} & \colhead{} }
\startdata
\ref{section:mass_fit} & Original & 143 & $0.13 \pm 0.04$ & $10.43$ & -5.3&-0.5 & 103 & $0.14 \pm 0.04$ & $10.65$ & -9.1&-4.0 \\
\ref{sec:remove_outlier} & No Outlier & 137 & $0.08 \pm 0.04$ &  $10.64$& -0.3&5.4 & 100 &  $0.13 \pm 0.04$ & $10.65$& -6.5&-1.5 \\
\ref{sec:both} & Joint & 99 & $0.12 \pm 0.06$ &$9.22$ & 0.1&5.0  & 99 & $0.12 \pm 0.04$ & $10.55$& -7.4&-2.5 \\
\enddata
\tablecomments{The ICs are relative to a constant model.}
\end{deluxetable}

\subsubsection{A Closer Look at the Outlier Population}

We examined each outlier lightcurve more closely and found nothing unusual for LSQ13cmt or SN~2005eu.
Though LSQ13cwp helped to discover a lensed galaxy system due to its proximity to it, this supernova is removed enough from the system to not be affected by the lens \citep{Galbany18}.
PTF13ddg is notable because it is the largest outlier but it is at a redshift of 0.084 and only has 3 data points in the $H$-band.
SN~2005eu from F15 has 23 lightcurve points in the $H$-band while the W18 SNeIa all have only 3 data points per lightcurve.

There are also two dim outliers.
SN~1999cl is at a very low redshift of 0.003.
While it is $>0.5$~mag dimmer, its significantly greater uncertainty from peculiar velocity makes it a non-significant outlier.
The other dim outlier, SN~2008fr, is notable because it has no pre-max optical data (See Figure~\ref{fig:SNIa_pre_max_residuals}) -- but it does have good sampling after \tbmax\ such that we are confident that its time of maximum is reasonably accurate.

\begin{figure}
\centering
\epsscale{1.15}
\plottwo{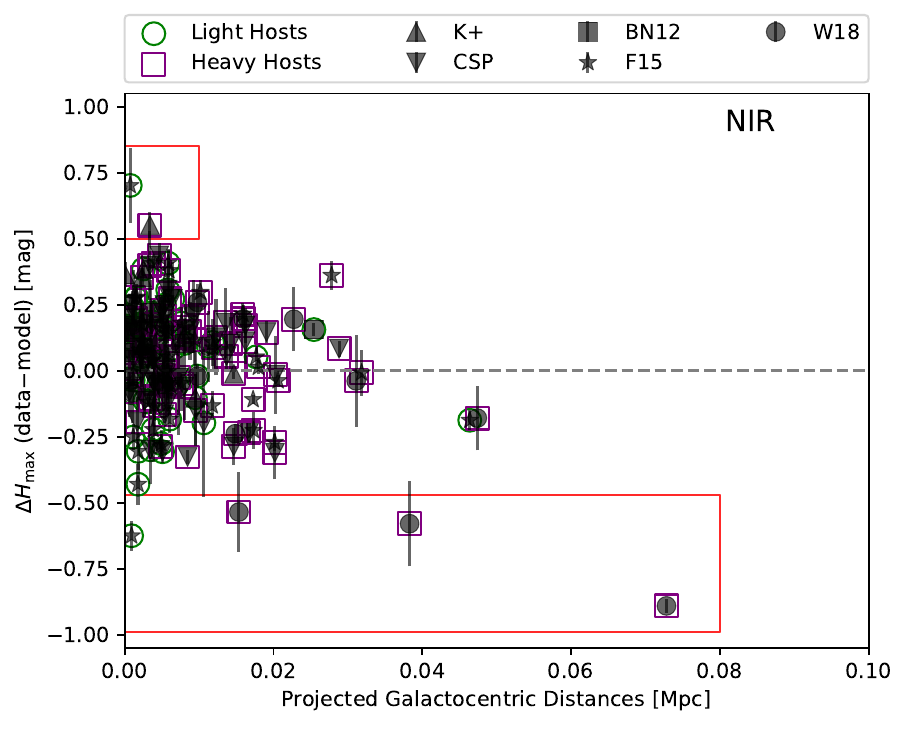}{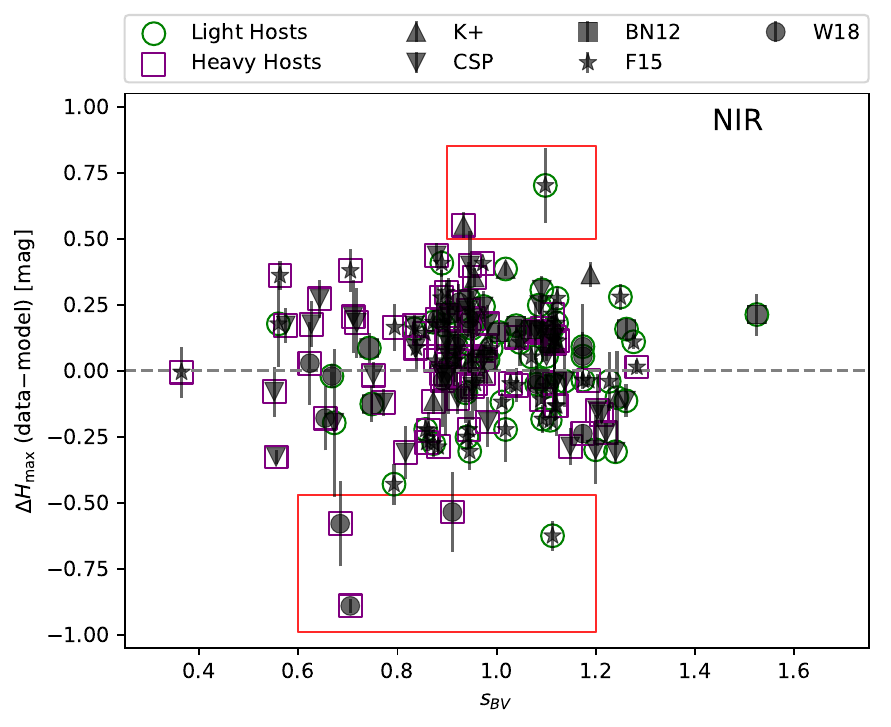}
\caption{
\textit{Left: } SN~Ia Hubble residuals vs projected galactocentric distances between the supernova and its host galaxy.
\textit{Right: } SN~Ia Hubble residuals vs $s_{BV}$ stretch factor.
For both figures, the points are coded in different shapes to indicate the source of the SN~Ia lightcurve data.
Overlaid on the points are the classification of their host galaxy: green circles are galaxies with mass $<10^{10} M_{\odot}$ and purple squares are galaxies with mass $>10^{10} M_{\odot}$.
The six outliers referred to in Section~\ref{sec:remove_outlier} are highlighted in the red boxes.
The additional intrinsic scatter (0.08~mag) and the error from peculiar velocities are not included in the error bars.
}
\label{fig:pgcd}
\end{figure}

The W18 outlier population are located far from the center of the galaxy when compared to the bulk of the sample, see Figure~\ref{fig:pgcd}.
The median projected galactocentric distance (PGCD) for our full sample is $\sim0.006$~Mpc and these SNeIa have values $0.015-0.073$~Mpc.
The outliers from F15 and K+ are much closer to their host galaxies.
The one feature (other than $z>0.03$) that is common amongst the 5 outliers not from peculiar velocities is that they are all in elliptical-like galaxies.
The W18 outliers are from red ellipticals while the F15 galaxies are from small, blue, and round galaxies.

The right side of Figure~\ref{fig:pgcd} presents the distribution of the $s_{BV}$ values for the $H_{\rm max}$ model.
These objects fall well within the distribution of the wider sample.

The column $\sigma_{{\rm out}, H}$ in Table~\ref{tab:outlier} presents how many sigma outside of the distribution these objects are.
Only PTF13ddg and SN~2008fr are more than $3$-$\sigma$ outliers but all objects are within $4$~$\sigma$ of the distribution.
For any cosmological-based analysis with this data set, we suggest removing this outlier population based on its large Hubble residual, but it is unclear that these are true outliers and not statistical fluctuations.

\subsection{Gaussian Mixture Model}
If a step function is an appropriate parameterization of the residual data, then a Gaussian Mixture Model (GMM) with two subpopulations should model the underlying distribution.
\citet{Ponder16} fit a GMM model assuming an evolution in the relative weights of two populations as a function of redshift.
Here, we similarly fit for an evolution in population as function of host galaxy mass.

The \texttt{scipy.optimize.curve$\_$fit} function used for the different parameterizations cannot handle fitting for GMM parameters.
Instead, we implemented the GMM likelihood in a Stan \citep{Carpenter17} model using PyStan \citep{Riddell18} which creates the full posterior distribution using a Hamiltonian Monte Carlo method.
We fit a GMM with a mass evolution and without a mass evolution for the full sample and with the outliers removed.
For the GMM with evolution, we normalized the $\log_{10}$ mass data by subtracting the mean and dividing by the standard deviation so that the fits were more stable.
To compare how well this model performed, we used the AIC$_{c}$ and BIC where the comparison is made to a single Gaussian model, which we took as the mean and standard deviation results from Tables~\ref{tab:stddev} and~\ref{tab:stddev_optical}.

\begin{deluxetable}{llrr}
\tablewidth{0pt}
\tabletypesize{\normalsize}
\tablecaption{Information Criteria Results for Gaussian Mixture Models versus Single Gaussian Models \label{tab:IC_gmm}}
\tablehead{\colhead{Residual} & \colhead{Fit} & \colhead{$\Delta$ AIC$_c$} & \colhead{$\Delta$ BIC}}
\startdata
$H_{\rm max}$  & Single Gaussian & 0.00 & 0.00 \\
 & GMM & -4.75 & -13.02 \\
 & GMM: No Evolution & -7.27 & -15.54 \\
 & GMM: No Outliers & -2.16 & 14.96 \\
 & GMM: No Outliers/Evolution & -4.65 & 12.47 \\
\hline
$\mu$ &  Single Gaussian & 0.00 & 0.00 \\
 & GMM & 1.58 & 13.97 \\
 & GMM: No Evolution & -0.57 & 11.82 \\
 & GMM: No Outliers & -6.64 & 4.26 \\
 & GMM: No Outliers/Evolution & -8.49 & 2.41 \\
\enddata
\end{deluxetable}

\begin{deluxetable*}{llrrrrrr}
\tablecaption{GMM Fits \label{tab:gmm_fits}}
\tablewidth{0pt}
\tabletypesize{\normalsize}
\tablehead{\colhead{Residual} & \colhead{Fit} & \colhead{Mean$_{\rm A}$} & \colhead{$\sigma_{\rm intrinsic, A}$}  & \colhead{Mean$_{\rm B}$} & \colhead{$\sigma _{\rm intrinsic, B}$} & \colhead{w$_{\rm A}$/slope} & \colhead{intercept}}
\startdata
$H_{\rm max}$  & GMM &  -0.187 & 0.313 & 0.054 & 0.087  & 0.001 & 0.217 \\
 & GMM: No Evolution & -0.110 & 0.272 & 0.062  & 0.052 & 0.304 &\\
 & GMM: No Outliers & -0.055  & 0.080 & 0.086 & 0.050  & 0.023 & 0.435 \\
 & GMM: No Outliers/Evolution & -0.054  & 0.079 & 0.089 & 0.043  & 0.437 &\\
\hline
$\mu$ & GMM & -0.054 & 0.062 & 0.125 & 0.190 & 0.084 & 0.629 \\
 & GMM: No Evolution & -0.054 & 0.065 & 0.082  & 0.100 & 0.532 & \\
 & GMM: No Outliers & -0.051 & 0.057 & 0.133 & 0.086  & 0.085  & 0.639 \\
 & GMM: No Outliers/Evolution & -0.051  & 0.060 & 0.089 & 0.098  & 0.537 & \\
\enddata
\end{deluxetable*}

\begin{figure*}
\centering
\epsscale{1.0}
\plotone{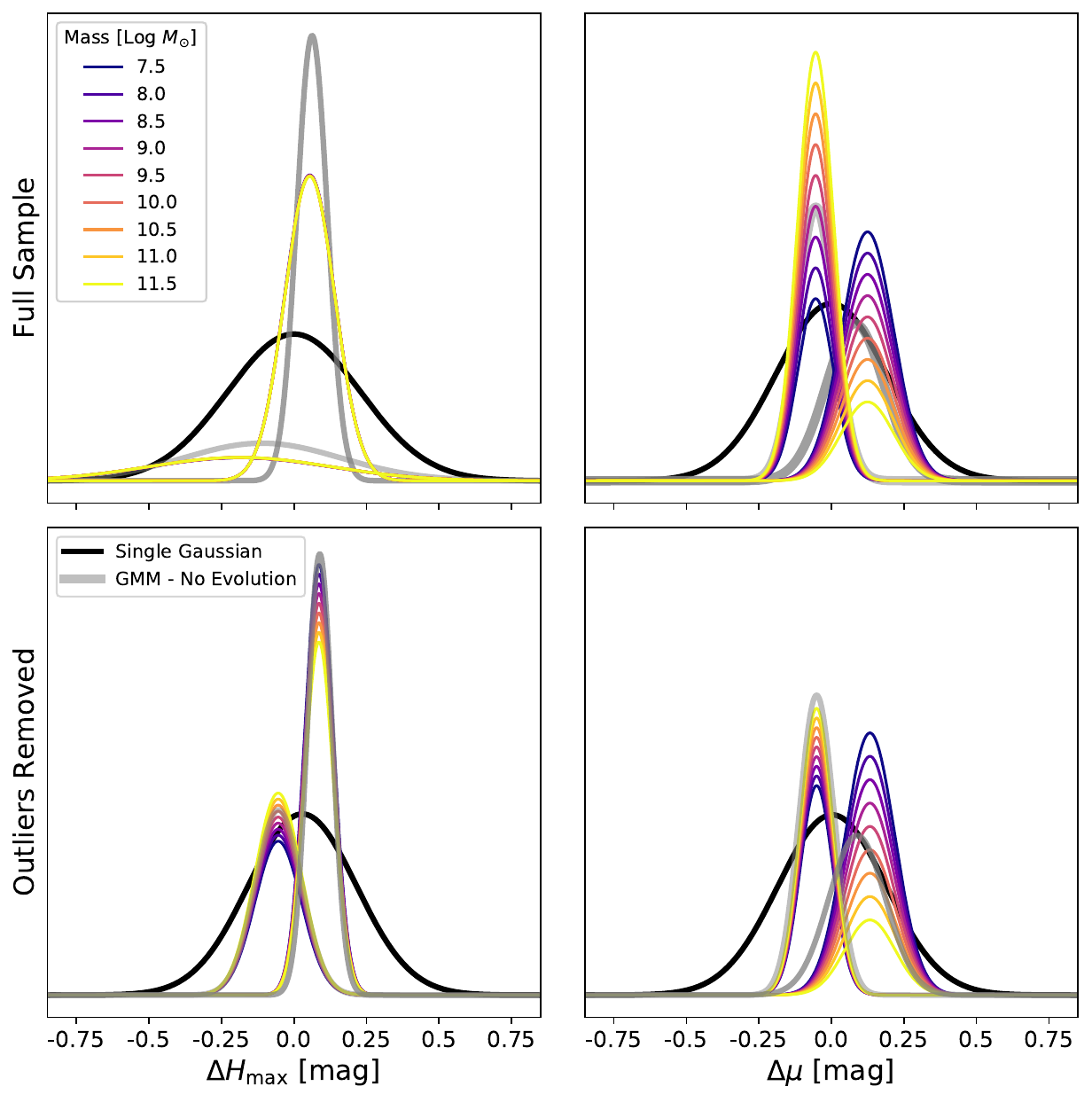}
\caption{
Fits of different Gaussian Mixture Models (GMMs) to $H$-band residuals (\textit{Left}) and the optical residuals (\textit{Right}) for the full samples (\textit{Top}) and with the $H_{\rm max}$ outliers removed (\textit{Bottom}).
The multi-colored curves correspond to a GMM that evolves as a function of host galaxy mass.
The grey curves are a GMM with no evolution and the black curves is a single Gaussian fit to the data.
}
\label{fig:gmm}
\end{figure*}

The PDFs of the results of these fits can be examined in Figure~\ref{fig:gmm} with the IC information is in Table~\ref{tab:IC_gmm} and the parameter values given in Table~\ref{tab:gmm_fits}.
The GMM with $H_{\rm max}$ is finding a second population mostly driven by the bright outliers.
The ICs strongly favor a GMM with the outliers but are split between a single Gaussian model and GMM once they are removed.
The $\mu$ residuals strongly favor a single Gaussian model with the outliers, but have some preference for a GMM once they are removed.

The slope of the evolving GMM is larger for $\Delta \mu$ residuals than $H_{\rm max}$ indicating that the optical residuals may be more sensitive to changes in host galaxy mass.

A non-evolving GMM is more favored in the ICs than the evolving GMM partially because there is one less parameter to fit.
Typically, the evolving GMM yields a larger difference in means except for $H_{\rm max}$ without outliers.

The bright $H_{\rm max}$ outliers are partially driving two populations in the $H$-band data but the optical data favors a GMM more when they are removed.

\subsection{SNeIa with Both $H$-band and Optical Lightcurves}\label{sec:both}
We here explore the results from limiting the data set to only the SNeIa that have both $H$-band and optical lightcurves.
A summary of the results is presented in Table~\ref{tab:summary_discussion} under ``Joint''.

99 SNeIa with host galaxy stellar masses measured have both NIR and optical lightcurves that satisfy our quality cuts for inclusion in the Hubble analysis.
The $H$-band brightness residuals favor a step function at $10^{9.22} M_{\odot}$ with a decreased step of $0.12 \pm 0.06$~mag and the BIC strong prefers a constant model while the AIC$_c$ prefers no model.
If we measure the step at  $10^{10.43} M_{\odot}$, it is reduced to $0.07 \pm 0.04$~mag.
The optical correlation step size decreased by 0.02~mag but the location of the step moved to $10^{10.55} M_{\odot}$.
The ICs are equivalent to those for the full sample.
If we hold the step at $10^{10.65} M_{\odot}$, the step size is $0.07 \pm 0.04$~mag.

The joint sample produced very different best fit step locations but neither had a strong preference with the ICs.
If we measure both the optical and NIR step at the best fit location for the full sample, they both produce the same step amplitude at a $\sim2$-$\sigma$ correlation.

\subsection{Importance of the Location of the Step}
\begin{figure}
\centering
\epsscale{1.15}
\plottwo{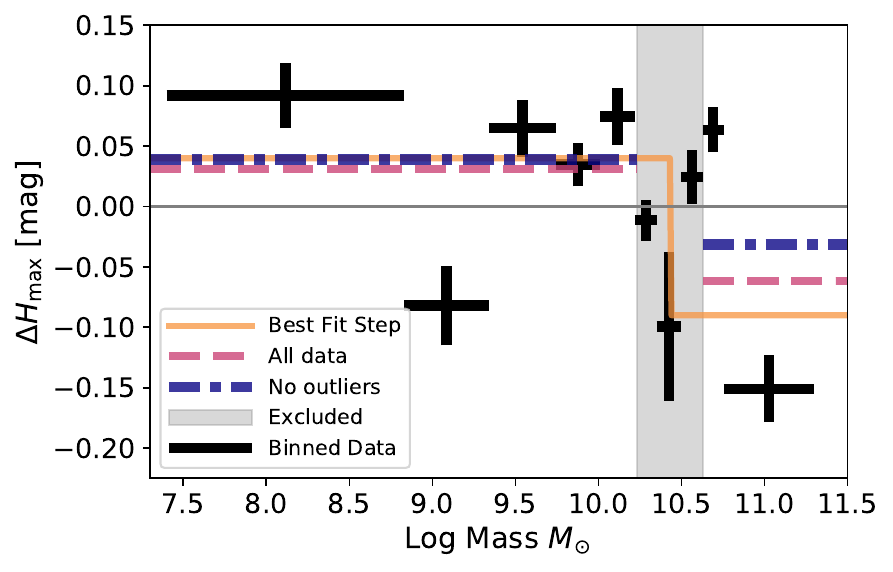}{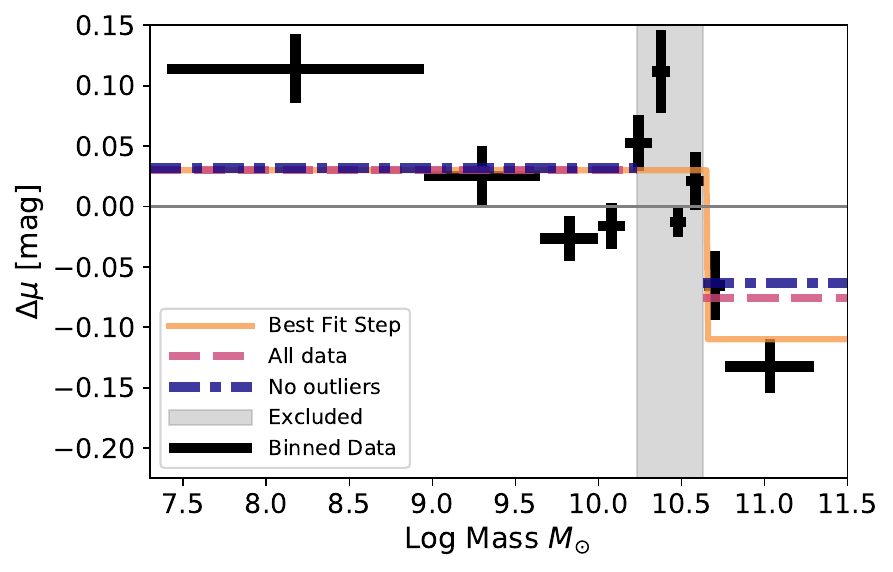}
\caption{
\textit{Left:} The $H$-band residuals in 10 mass bins where the three lowest bins have 15 objects and the rest have 14 objects.
The highest bin has a large offset from the others.
The best fit break is at 10.43 log Mass $M_{\odot}$.
The orange solid line is the best fit step function from Figure~\ref{fig:model_fits_mass}.
The greyed area is the $\pm~0.2$~dex surrounded that break.
The pink dashed line are the fits to the low and high mass objects on either side of the excluded area.
The blue dash dotted line is the same as the pink line but with the outliers removed.
The difference in the high and low mass fits are 0.09~mag and 0.07~mag for the full sample and the sample without outliers, respectively.
\textit{Right:} Same as left except for the $\mu$ residuals in 10 mass bins where the three lowest bins have 11 objects and the rest have 10 objects.
The different in the high and low mass fits are 0.10~mag and 0.10~mag for the full sample and the sample without outliers, respectively.
}
\label{fig:binned}
\end{figure}

To explore what is happening at low and high mass, we fit a constant model excluding the $\pm 0.2$~dex region surrounding the best fit break of $10^{10.43} M_{\odot}$.
The constant fit model is consistent with the weighted average.
Figure~\ref{fig:binned} shows the results of this fit with the best fit step function for the full sample as reference.
The data in Figure~\ref{fig:binned} is binned with evenly distributed number of objects.
It is easier to see the evolution and offsets between high and low mass in this reduced form.
The difference between the low and high mass samples is $0.09 \pm 0.05$~mag for $H_{\rm max}$ and $0.11 \pm  0.05$~mag for $\mu$.
Without the outliers, we measure a difference between the two is $0.07 \pm 0.04 $~mag for $H_{\rm max}$ and $0.10 \pm 0.05$~mag for $\mu$.

\subsection{Location of the Step Compared to Other Analyses}
The best fit step locations ($10^{10.43} M_{\odot}$/$10^{10.65} M_{\odot}$) are higher than typically reported in other analyses.
Since the masses here may be underestimated by 0.3~dex, the mass values are closer to $10^{11} M_{\odot}$.
We examined 13 of the papers measuring a host galaxy mass step in the last 10 years and found that the majority of them assigned the value of $10^{10} M_{\odot}$ because it was used in \citet{Sullivan10}.
\citet{Sullivan10} used $10^{10} M_{\odot}$ because it was the median of the sample; however, they did also fit for a step function at $10^{10.5} M_{\odot}$ and found a $\sim4$-$\sigma$ step.
If an analysis did not use $10^{10} M_{\odot}$, then they used the median of their own sample which was typically much larger than $10^{10} M_{\odot}$ ($10^{10.5} M_{\odot}$-$10^{10.8} M_{\odot}$).
It is of note that \citet{Sullivan10} has the highest median redshift (0.65) of all the analyses which typically have a median redshift $\sim0.03$ or $\sim0.3$.

In summary, literature analyses either use the median of their host galaxy mass sample or use the break location from \citet{Sullivan10}; however, we fit for the location that maximized the size of the step.
We leave the discussion of what should be the best location for a break to future analyses.

\subsection{Are the NIR Residuals Adding More Information to the Optical Residuals?}
\begin{figure}
\plotone{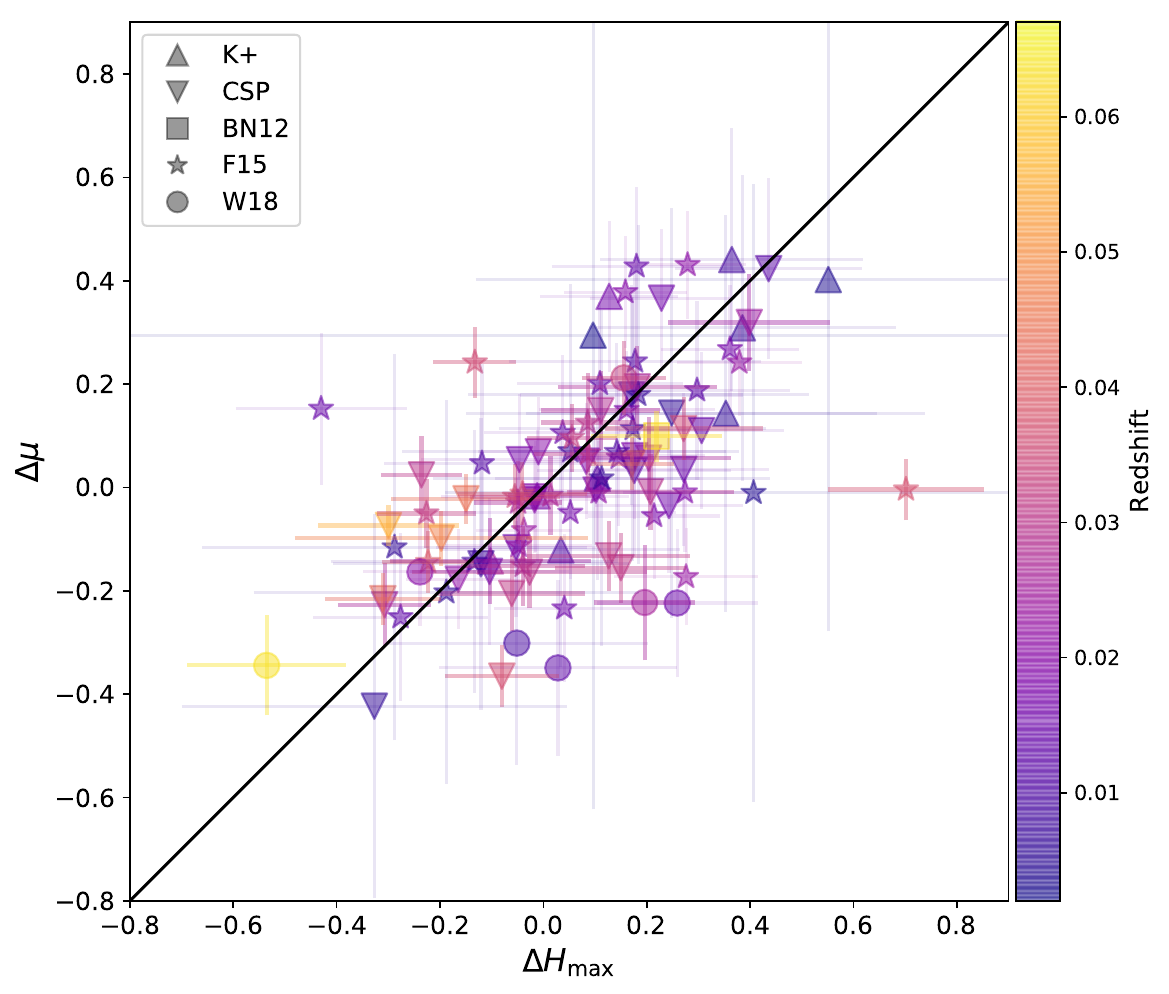}
\caption{
Optical Hubble residuals ($\mu$) vs. $H_{\rm max}$ residuals.
The errors from peculiar velocities \textbf{are} included in the error bars for both $H_{\rm max}$ and $\mu$ but not the additional intrinsic scatter (0.08~mag).
All objects in the smooth Hubble flow have thicker and less transparent error bars to improve readability.
 As in previous figures, the different shapes indicate the source of the SN~Ia lightcurve data.
 The color map indicates the redshift of the host galaxy.
 The black line illustrates a one-to-one relationship to guide the eye.
 The Pearson r-coefficient is 0.59.
}
\label{fig:h_v_dm_residuals}
\end{figure}

Is the NIR analysis an independent test of host galaxy correlations or degenerate with the tests done at optical wavelengths?
Figure~\ref{fig:h_v_dm_residuals} presents the $H_{\rm max}$ residuals plotted against the $\mu$ residuals for the 99 supernovae that have both NIR and optical data.

One source of potential correlation is a mis-estimate of the cosmological redshift for a supernova.
If we use the wrong cosmological redshift for an object, we would expect to see strong correlations between the optical and NIR brightnesses due to using the wrong cosmological redshift rather than due to any intrinsic physics about the supernova.
In particular, the lower redshift supernovae ($z < 0.02$) are affected by larger peculiar velocities.
In Table~\ref{tab:pcorrelation}, we present the mean, weighted mean, standard error on the mean, standard deviation and Pearson correlation coefficient ($r$) for the full sample and if the sample was split at $z = 0.02$.
The samples of $z < 0.02$ and $>0.02$ have offset weighted means but they are within $2~\sigma$ of each other.
Both samples have a strong correlation between optical and NIR supernovae indicating that peculiar velocities are not a driving their correlation.

The optical and $H$ Hubble residuals are clearly and strongly correlated.
This is not a surprise as effectively similar relationships are found when considering large sets of lightcurves in training lightcurve fitters.  But we here we have answered the question from a data-driven exploration with Hubble residuals directly from a significant selection of supernovae not involved in the construction of the SNooPy templates.

\begin{deluxetable}{ccccccc}
\tablewidth{0pt}
\tablecaption{Optical - $H$ Hubble Residuals \label{tab:pcorrelation}}
\tablehead{
\colhead{Sample} &  \colhead{\# SNeIa} &  \colhead{Mean}& \colhead{Wgt Mean}    & \colhead{SEM}   & \colhead{Std. Dev.} & \colhead{Pearson's $r$} \\ \colhead{} &  \colhead{} &  \colhead{mag}& \colhead{mag}    & \colhead{mag}   & \colhead{mag} & \colhead{}}
\startdata
All  & 99 & 0.05  & 0.02 & 0.02 & 0.19 & 0.59 \\
$z < 0.02$ & 51 & 0.06   & 0.06 & 0.03 & 0.18 & 0.61 \\
$z > 0.02$ & 48 & 0.04  & 0.00 & 0.03 & 0.18 & 0.54 \\
\enddata
\end{deluxetable}

\subsection{Comparison to the Carnegie Supernova Project}
\begin{figure}
\plotone{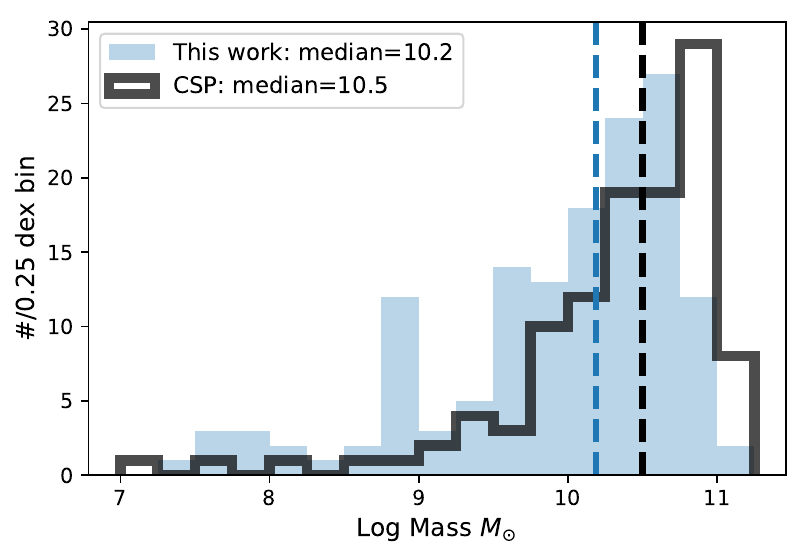}
\caption{
Mass distribution of this work compared to that of~\citet{Uddin20}.  This work has more lower-mass objects
}
\label{fig:csp_v_p20_mass_hist}
\end{figure}

We measured a $3.25$-$\sigma$ correlation between $H_{\rm max}$ residuals and host galaxy mass which dropped to $2$-$\sigma$ (at the best-fit step break) when removing the outliers.
These results are in agreement with U20.
A key difference between the CSP sample and our sample is that U20 had substantially fewer lower-mass galaxies, see Figure~\ref{fig:csp_v_p20_mass_hist}.
U20 used the ``max\_model'' fitting for the optical and NIR filters such that we can directly compare our $H_{\rm max}$ residuals but not the $\mu$ residuals since each optical filter was fit independently.
A constant location of the break for the step function was used in U20 which was the median of the sample at $10^{10.48} M_{\odot}$.
In contrast, the median of the sample in this paper is $10^{10.20} M_{\odot}$, but our models favor a step at a mass closer to the median from CSP.
We compared our results their ``All'' sample and find their $H$-band step magnitude in agreement with any of our subsamples.
Furthermore, examining all the optical filters, their measured step has an amplitude ranging from 0.147~mag to 0.074~mag with the trend ranging from 2.5 to $3.3~\sigma$.
This agrees with the recovered step magnitude for $\mu$ in this paper.

\subsection{Direction of the $H$ versus Optical Correlation}
The trend we observed in the $H$-band peak magnitudes is that larger galaxies, which are also more red and more likely to be ellipticals, host brighter supernova than lower mass galaxies, which are bluer and more likely spirals.
This correlation is the opposite of the trend observed in uncorrected optical brightness.
\citet{Hamuy95} first found that galaxies with a younger stellar population hosted brighter supernova.
Continued works such as \citet{Hamuy96} and \citet{Sullivan06} found correlations with the lightcurve shape parameters $\Delta m_B$ and stretch $s$, respectively, where faster declining (dimmer) SNeIa were hosted in elliptical and higher-mass galaxies whereas slower declining (brighter) SNeIa were hosted in spiral and lower-mass galaxies.
All subsequent analyses have found a correlation between host galaxy properties and the shape of the width of the lightcurve.
However, after width-luminosity standardization, the correlation with host galaxy properties for optical SNeIa switches such that larger mass galaxies have brighter residuals than lower mass galaxies.

\subsection{Physical Interpretation}

We interpret the results above as an indication that more massive galaxies host SNeIa that are brighter in the NIR than those hosted in less massive host galaxies.
This result is not strongly supported by all of the statistical tests implemented but it is supported at a $2$-$\sigma$ level.
Our interpretation of the data here is in agreement with the results from the U20 paper which is not affected by potential biases from multiple SNeIa surveys.

\citet{Brout20} claim that the host galaxy stellar mass dependence seen at optical wavelengths is due to the correlation of dust in different galaxies.
Because NIR is less sensitive to dust, we should not see any significant correlations between the NIR residuals and host galaxy mass if the correlation is driven by dust.
This analysis has found some evidence to support a correlation between the NIR and host galaxy stellar mass; however, the trend is only seen at a $2$-$\sigma$ level.
Therefore; we cannot conclude if the correlation is driven by dust alone.

\subsection{A Caveat on K-Corrections}

We note that the state of K-corrections in NIR SNeIa photometry remains in its beginning stages and we express concern that the K-corrections used here are not the final word.  The two significant previously explicitly published K-corrections are those of \citet{Krisciunas04b} and \citet{Hsiao07}.  The community has continued to gather NIR spectra, but these have not yet been compiled into a new set of spectral templates.
\citet{Stanishev15} presented their own K-correction methodology, but do not provide an updated set of spectral templates.
If SNeIa were all the same in the NIR, then the excellent NIR spectral series on SN~2011fe~\citep{Hsiao13} or SN~2014J~\citep{Marion15} would provide sufficient data for good K-corrections.
But while SNeIa NIR exhibit less scatter in the $H$-band than the optical, there is still clear evidence for some variation: single- vs. double-hump \citep[e.g., the dromedarian SN~2005hk detailed in][]{Phillips07}, and bridge objects such as iPTF13ebh \citep{Hsiao15}.
We remain of the opinion that a new effort in K-corrections for SNeIa in the NIR would be a worthwhile endeavor with a clear benefit to the community.

\section{Conclusion}\label{sec:conclusion}
We have collected and analyzed a data sample of \numFullHostSample\ SNeIa with observations in the restframe $H$-band.  We fit the lightcurves using SNooPy, and found \numAll\ of the SNeIa had lightcurve fits suitable for inclusion in a Hubble diagram.
We combined measurements from SDSS, DECaLS, PS1, 2MASS, and GALEX to determine photometric stellar masses for the host galaxies of \numLightHeavy\ of these \numAll\ SNeIa.

We explored possible correlations between $H_{\rm max}$ residuals from the SNooPy fitter and host galaxy properties.
Though we only presented the results from host galaxy stellar mass in the main text, further studies are presented in Appendix~\ref{sec:other}.
Using the \numLightHeavy\ SNeIa with host galaxy stellar mass measurements, we report a $0.13 \pm 0.04$~mag step at $10^{10.43} M_{\odot}$ in agreement with the step seen at optical wavelengths.
However, the AIC$_c$ and BIC only mildly prefer this step function over a constant model.
By further investigating the sample, we have shown that the correlation with $H$-band brightnesses is partially driven by outliers and removing these from the sample lowers the significance of the step to $2~\sigma$ at $10^{10.64} M_{\odot}$, but the ICs mildly prefer a constant model.
We showed that a GMM is strongly favored compared to a single Gaussian when the outlier population is present, but the ICs are split with them removed.

The apparent outlier population of SNeIa in the $H$-band is located within the smooth Hubble flow ($0.03 < z < 0.09$) except for one dim, low redshift object whose residual is within peculiar velocity uncertainties.
The set of 3 bright outliers from the W18 sample have a residual brightness $\leq -0.5$~mag after correction and are hosted in massive ($M > 10^{10} M_{\odot}$), bright ($M_r > -21.5$~mag), and red ($0.6 < g-r < 0.8$~mag), elliptical galaxies.
However, the host galaxy for the bright outlier SN~2005eu is a low mass, dim, and blue galaxy but it is small in angular size and possibly an elliptical galaxy.
SN~2008fr is similarly located in a low mass, dim, and blue galaxy but it is dimmer than expected unlike the other outliers.
These objects are not $5$-$\sigma$ outliers but without a clear reason for their large offsets, they should not be used for a cosmological analysis.

Using the optical lightcurves corresponding to the sample of NIR lightcurves, we measured a host galaxy mass step of $\sim$0.1~mag around $10^{10.43} M_{\odot}$ and 0.14~mag at the best fit step of $10^{10.65} M_{\odot}$.
This measurement not affected by the removal of the corresponding $H$-band outliers.
Interestingly, a GMM is more preferred in the optical once the outliers have been removed.
Since the ICs are split in preference between the GMM and single Gaussian, we cannot make a definitive conclusion.
Showing this trend using a third lightcurve fitter provides further evidence of either a physical phenomenon or that there is some intrinsic property that is not well understood in optical wavelengths.

The correlation found between $H$-band residuals and host galaxy mass is the opposite correlation seen with optical residuals.
In the NIR, high-mass galaxies host brighter SNeIa than low-mass galaxies, but in optical wavelengths the \textit{uncorrected} residuals show that brighter SNeIa are hosted in lower-mass galaxies.
But after correction for stretch and color, the correlation of optical Hubble residuals goes the other way, with more negative Hubble residuals in more massive galaxies.

If the cause of the host galaxy mass trend is dust, then we would expect to see no correlation in the NIR since
SNeIa in the NIR are less sensitive to dust.
As we find inconclusive evidence of a correlation, our results cannot distinguish between the possible drivers of the host galaxy mass correlation.

Based on our analysis, we conclude that SNeIa in $H$ show some evidence of correlations with host galaxy mass, but this is only a $2$-$\sigma$ signal without the outliers.
With more data  imminent (CSP~II, SIRAH, SweetSpot), we will be able to increase the sample size to test for the correlations again and to determine if there is a correlation or if there is a persistent outlier population.
Now is the time to examine these relationships in low redshift NIR lightcurve data to improve our NIR models in preparation for the $\sim$2,500 high-redshift NIR SNeIa that will be observed by Nancy Grace Roman Space Telescope~\citep{WFIRST15}.

\acknowledgments

K.A.P., M.W.-V., and L.G. were supported in part by the US National Science Foundation under Grant AST-1311862.
K.A.P. additionally acknowledges support from PITT PACC.
K.A.P. was also supported in part by the Berkeley Center for Cosmological Physics and the Director, Office of Science, Office of High Energy Physics of the U.S. Department of Energy under Contract No. DE-AC02-05CH11231 and
U.S. Department of Energy Office of Science under Contract No.DE-AC02-76SF00515.
L.G. was additionally funded in part by the European Union's Horizon 2020 research and innovation programme under the Marie Sk\l{}odowska-Curie grant agreement No. 839090.

We thank the referee whose comments have improved this paper and Saurabh Jha, Kyle Boone, and Ravi Gupta for useful conversations.

This research has made use of the The NASA/IPAC Extragalactic Database (NED) which is funded by the National Aeronautics and Space Administration and operated by the California Institute of Technology.

Funding for the Sloan Digital Sky Survey IV has been provided by
the Alfred P. Sloan Foundation, the U.S. Department of Energy Office of
Science, and the Participating Institutions. SDSS-IV acknowledges
support and resources from the Center for High-Performance Computing at
the University of Utah. The SDSS web site is www.sdss.org.

SDSS-IV is managed by the Astrophysical Research Consortium for the 
Participating Institutions of the SDSS Collaboration including the 
Brazilian Participation Group, the Carnegie Institution for Science, 
Carnegie Mellon University, the Chilean Participation Group, the French Participation Group, Harvard-Smithsonian Center for Astrophysics, 
Instituto de Astrof\'isica de Canarias, The Johns Hopkins University, 
Kavli Institute for the Physics and Mathematics of the Universe (IPMU) / 
University of Tokyo, Lawrence Berkeley National Laboratory, 
Leibniz Institut f\"ur Astrophysik Potsdam (AIP),  
Max-Planck-Institut f\"ur Astronomie (MPIA Heidelberg), 
Max-Planck-Institut f\"ur Astrophysik (MPA Garching), 
Max-Planck-Institut f\"ur Extraterrestrische Physik (MPE), 
National Astronomical Observatories of China, New Mexico State University, 
New York University, University of Notre Dame, 
Observat\'ario Nacional / MCTI, The Ohio State University, 
Pennsylvania State University, Shanghai Astronomical Observatory, 
United Kingdom Participation Group,
Universidad Nacional Aut\'onoma de M\'exico, University of Arizona, 
University of Colorado Boulder, University of Oxford, University of Portsmouth, 
University of Utah, University of Virginia, University of Washington, University of Wisconsin, 
Vanderbilt University, and Yale University.

This research uses services or data provided by the Astro Data Lab at NSF's National Optical-Infrared Astronomy Research Laboratory. NOIRLab is operated by the Association of Universities for Research in Astronomy (AURA), Inc. under a cooperative agreement with the National Science Foundation.

The Legacy Surveys consist of three individual and complementary projects: the Dark Energy Camera Legacy Survey (DECaLS; Proposal ID \#2014B-0404; PIs: David Schlegel and Arjun Dey), the Beijing-Arizona Sky Survey (BASS; NOAO Prop. ID \#2015A-0801; PIs: Zhou Xu and Xiaohui Fan), and the Mayall z-band Legacy Survey (MzLS; Prop. ID \#2016A-0453; PI: Arjun Dey). DECaLS, BASS and MzLS together include data obtained, respectively, at the Blanco telescope, Cerro Tololo Inter-American Observatory, NSF's NOIRLab; the Bok telescope, Steward Observatory, University of Arizona; and the Mayall telescope, Kitt Peak National Observatory, NOIRLab. The Legacy Surveys project is honored to be permitted to conduct astronomical research on Iolkam Du'ag (Kitt Peak), a mountain with particular significance to the Tohono O'odham Nation.

This project used data obtained with the Dark Energy Camera (DECam), which was constructed by the Dark Energy Survey (DES) collaboration. Funding for the DES Projects has been provided by the U.S. Department of Energy, the U.S. National Science Foundation, the Ministry of Science and Education of Spain, the Science and Technology Facilities Council of the United Kingdom, the Higher Education Funding Council for England, the National Center for Supercomputing Applications at the University of Illinois at Urbana-Champaign, the Kavli Institute of Cosmological Physics at the University of Chicago, Center for Cosmology and Astro-Particle Physics at the Ohio State University, the Mitchell Institute for Fundamental Physics and Astronomy at Texas A\&M University, Financiadora de Estudos e Projetos, Fundacao Carlos Chagas Filho de Amparo, Financiadora de Estudos e Projetos, Fundacao Carlos Chagas Filho de Amparo a Pesquisa do Estado do Rio de Janeiro, Conselho Nacional de Desenvolvimento Cientifico e Tecnologico and the Ministerio da Ciencia, Tecnologia e Inovacao, the Deutsche Forschungsgemeinschaft and the Collaborating Institutions in the Dark Energy Survey. The Collaborating Institutions are Argonne National Laboratory, the University of California at Santa Cruz, the University of Cambridge, Centro de Investigaciones Energeticas, Medioambientales y Tecnologicas-Madrid, the University of Chicago, University College London, the DES-Brazil Consortium, the University of Edinburgh, the Eidgenossische Technische Hochschule (ETH) Zurich, Fermi National Accelerator Laboratory, the University of Illinois at Urbana-Champaign, the Institut de Ciencies de l'Espai (IEEC/CSIC), the Institut de Fisica d'Altes Energies, Lawrence Berkeley National Laboratory, the Ludwig Maximilians Universitat Munchen and the associated Excellence Cluster Universe, the University of Michigan, NSF's NOIRLab, the University of Nottingham, the Ohio State University, the University of Pennsylvania, the University of Portsmouth, SLAC National Accelerator Laboratory, Stanford University, the University of Sussex, and Texas A\&M University.

The Legacy Survey team makes use of data products from the Near-Earth Object Wide-field Infrared Survey Explorer (NEOWISE), which is a project of the Jet Propulsion Laboratory/California Institute of Technology. NEOWISE is funded by the National Aeronautics and Space Administration.

The Legacy Surveys imaging of the DESI footprint is supported by the Director, Office of Science, Office of High Energy Physics of the U.S. Department of Energy under Contract No. DE-AC02-05CH1123, by the National Energy Research Scientific Computing Center, a DOE Office of Science User Facility under the same contract; and by the U.S. National Science Foundation, Division of Astronomical Sciences under Contract No. AST-0950945 to NOAO.

This publication makes use of data products from the Two Micron All Sky Survey, which is a joint project of the University of Massachusetts and the Infrared Processing and Analysis Center/California Institute of Technology, funded by the National Aeronautics and Space Administration and the National Science Foundation.

This research has made use of the NASA/IPAC Infrared Science Archive, which is operated by the Jet Propulsion Laboratory, California Institute of Technology, under contract with the National Aeronautics and Space Administration.

Some of the data presented in this paper were obtained from the Mikulski Archive for Space Telescopes (MAST). STScI is operated by the Association of Universities for Research in Astronomy, Inc., under NASA contract NAS5-26555. Support for MAST for non-HST data is provided by the NASA Office of Space Science via grant NNX09AF08G and by other grants and contracts.

\software{
Python\footnote{\url{http://python.org}},
NumPy\footnote{\url{http://www.numpy.org}},
SciPy\footnote{\url{http://www.scipy.org}},
matplotlib~\citep{Hunter07}\footnote{\url{http://matplotlib.org}},
AstroPy~\citep{Astropy13}\footnote{\url{http://www.astropy.org}},
IDL\footnote{\url{http://www.harrisgeospatial.com/ProductsandSolutions/GeospatialProducts/IDL.aspx}}
Stan~\citep{Carpenter17},
PyStan~\citep{Riddell18}
}

\appendix
\restartappendixnumbering

\section{Other Explored Correlations}\label{sec:other}

We ran the same analysis from Section~\ref{sec:distributions} on many different host properties including absolute $r$-band magnitude ($M_r$), $g-r$ color, galaxy morphology, NUV colors, smooth Hubble flow, and distance from the host galaxy.
The $M_r$ and $g-r$ color properties exhibited $\sim 1$--$2$-$\sigma$ correlations with the Hubble residuals, so we continued those two properties through every analysis step.
However, no significant correlation was found with these host galaxy properties, but we include summaries of the findings here for completeness.

\subsection{Absolute $r$-band Magnitude}
The restframe absolute $r$-band magnitude shows a similar correlation as with host galaxy mass for the distributions.
We define a ``bright'' and ``dim'' population with a threshold at between them at  $r=-21.0$~mag, which was chosen to correspond with the typical brightness of a galaxy with mass $\sim 10^{10} M_{\odot}$.
The outlier population found in mostly high mass and red galaxies is present here as well in the bright population.
The weighted means of the distributions for NIR result in $\sim2~\sigma$ detection and the optical residuals result in $\sim 1~\sigma$ detection.

We used 63 $H_{\rm max}$ and 70 $\mu$ residuals to test the functional form of correlations with the restframe, absolute $r$-band magnitude, $M_r$.
When fitting the step function with a floating break, we limited the range to $-21.96 < M_r < -19.4$~mag to ensure each bin had at least $20\%$ of the total SNeIa.
The model that most favors a correlation with the $H_{\rm max}$ residuals is the best-fit step function with a break at $-21.46$~mag.
The size of the best-fit step is $0.10 \pm 0.04$~mag, a $2.5~\sigma$ detection, but the ICs have no preference between a constant model or step function.
The distance modulus residuals prefer a break at $-21.76$~mag with an amplitude of  $0.10 \pm 0.03$~mag.
The AIC is around -4 and the BIC around 0.5 showing some preference for this step but not a large one.

We then ran the same sample changes presented in Section~\ref{sec:discussion}.
If the outlier population is removed, $H_{\rm max}$ correlations are degraded with the best-fit step function size stays the same but the break is moved to $M_r = -21.96$~mag.
Without the outliers, the $\mu$ residuals still prefer a step at $-21.76$~mag with a slight degradation of the significance of the step down to $3~\sigma$.
For the joint sample, the $H_{\rm max}$ residuals moves up to $-21.96$~mag again with a $\sim2~\sigma$ step as in the host mass analysis.
The joint sample for optical residuals showed a $2.75~\sigma$ step function correlation with a break at $-21.76$~mag, and the AIC$_c$ prefers a step function but the BIC showed no model preference.
These results match the mass results fairly well in significance and step size.
This evidence for a correlation between the $M_r$ of the host galaxy and optical brightnesses is unsurprising since our galaxies show a linear relationship between the log of the galaxy mass and absolute brightness.

\subsection{Other Correlations}
Other correlations that we tested are:
\begin{itemize}
\item \textbf{Restframe $g-r$ Color:} No significant correlation was found in either the NIR or optical using 132 and 94 objects, respectively.
Our initial study of $g-r$ color returned a $\sim1~\sigma$ correlation when comparing the distribution parameters from Section~\ref{sec:distributions}.
We found no correlations when running the analysis from Section~\ref{sec:functionalform}.
\item \textbf{Smooth Hubble Flow:} We tested the effects of using SNe with $z > 0.02$ corresponding to the smooth Hubble Flow.
This cut reduced our sample size by half and produced the same results as the full sample.
Table~\ref{tab:stddev} includes the results of using only Hubble flow SNeIa for three different host galaxy properties, and they are all labeled starting with ``Hubble''.
The distributions of residuals of SNeIa with $z<0.02$ in mass, color, and $M_r$ are the same as the distributions of SNeIa residuals with $z>0.02$.
Therefore, we find no evidence for evolution with redshift.
The one exception is the outlier group not from peculiar velocity of 5 SNeIa only appear at $z > 0.03$, but there are so few of them it is unclear if this is a real trend or a coincidence of small sample size.
\item \textbf{NUV colors:}
By using $NUV-H$, we are picking out young, blue stars versus old, red stars, which should act as a tracer for recent star formation.
We found that the SN~Ia Hubble residuals versus $NUV-H$ color histograms are mostly identical in scatter with a negligible offset.
$NUV-g$ exhibits the same distribution.
\item \textbf{Distance from host galaxy:} We found no discernible correlation in projected distances of supernovae from their host galaxies, except for the W18 outlier population which are all very separated from their host galaxy.
\end{itemize}

\section{Host Galaxy Photometry Table}
Here is the table of host galaxy photometry and derived masses.

\begin{longrotatetable}
\begin{deluxetable}{lllhlhlhlhlhlhlhlhlhlhlhrc}
\tablewidth{0pt}
\tabletypesize{\scriptsize}
\tablecaption{Host Galaxy Summary Table \label{tab:snia_gal}}
\tablehead{\colhead{Name} & \colhead{Host Galaxy} & \colhead{$F$} & \nocolhead{$F_{\rm err}$} & \colhead{$N$} & \nocolhead{$N_{\rm err}$} & \colhead{$u$} & \nocolhead{$u_{\rm err}$} & \colhead{$g$} & \nocolhead{$g_{\rm err}$} & \colhead{$r$} & \nocolhead{$r_{\rm err}$} & \colhead{$i$} & \nocolhead{$i_{\rm err}$} & \colhead{$z$} & \nocolhead{$z_{\rm err}$}  & \colhead{$y$} & \nocolhead{$y_{\rm err}$} & \colhead{$J$} & \nocolhead{$J_{\rm err}$} & \colhead{$H$} & \nocolhead{$H_{\rm err}$} & \colhead{$K$} & \nocolhead{$K_{\rm err}$} & \colhead{Mass\tablenotemark{a}} & \colhead{LC}}
\startdata
CSS121006:232854+085452 & SDSS J232854.54+085452.3 & -- & -- & -- & -- & -- & -- & 19.87 & 0.02 & 19.72 & 0.02 & -- & -- & 19.59 & 0.03 & -- & -- & -- & -- & -- & -- & -- & -- & 8.52 & N \\
CSS121009:011101-172841 & 2MASX J01105973-1728527 & -- & -- & 20.77 & 0.23 & -- & -- & 16.56 & 0.02 & 15.55 & 0.02 & -- & -- & 14.67 & 0.02 & -- & -- & 14.19 & 0.06 & 13.77 & 0.06 & 13.87 & 0.08 & 10.39 & N \\
CSS121114:090202+101800 & SDSS J090202.19+101759.7 & 19.59 & 0.04 & 19.35 & 0.03 & 18.77 & 0.07 & 17.87 & 0.02 & 17.57 & 0.02 & 17.37 & 0.02 & 17.24 & 0.05 & -- & -- & -- & -- & -- & -- & -- & -- & 8.76 & Y \\
LSQ12fhs & 2MASX J22522347-2036373 & 18.13 & 0.07 & 17.67 & 0.04 & -- & -- & 16.13 & 0.02 & 15.75 & 0.02 & 15.56 & 0.02 & 15.53 & 0.02 & 15.35 & 0.02 & 15.27 & 0.12 & 15.46 & 0.2 & 15.31 & 0.23 & 9.54 & Y \\
LSQ12fmx & 2MASX J031252.93-001223.8 & 18.96 & 0.05 & 18.58 & 0.03 & 17.9 & 0.06 & 16.77 & 0.02 & 16.24 & 0.02 & 15.93 & 0.02 & 15.72 & 0.03 & -- & -- & 15.41 & 0.11 & 15.28 & 0.18 & 15.28 & 0.16 & 10.14 & Y \\
LSQ13cmt & ESO 541- G 013 & -- & -- & -- & -- & -- & -- & 14.37 & 0.02 & 13.45 & 0.02 & -- & -- & 12.76 & 0.02 & -- & -- & 12.54 & 0.04 & 12.28 & 0.06 & 12.36 & 0.06 & 11.27 & Y \\
LSQ13cwp & 2MASX J04035024-0239275 & -- & -- & -- & -- & -- & -- & 15.41 & 0.02 & 14.51 & 0.02 & -- & -- & 13.85 & 0.02 & -- & -- & 13.33 & 0.04 & 13.13 & 0.04 & 13.43 & 0.07 & 11.0 & Y \\
PS1-10w & CGCG 290-062 & 18.33 & 0.06 & 17.96 & 0.04 & 16.92 & 0.05 & 15.72 & 0.02 & 15.13 & 0.02 & 14.83 & 0.02 & 14.64 & 0.03 & -- & -- & 14.7 & 0.09 & 14.65 & 0.18 & 14.98 & 0.14 & 9.91 & Y \\
\enddata
\tablenotetext{a}{Units of $\log M_{\odot}$. No errors are reported for this parameter.}
\tablecomments{The errors on the photometry are hidden here to fit the page but are available in the full machine-readable table.}
\end{deluxetable}
\end{longrotatetable}

\section{SNeIa SNooPy Fits Table}
Here is the table of SNooPy fits for the H-band ``max\_model'' and Optical ``EBV\_model2''.

\clearpage
\begin{longrotatetable}
\begin{deluxetable}{lrlllllrlllllllll}\tablewidth{0pt}
\tabletypesize{\scriptsize}
\tablecaption{SN H-band ``max\_model'' and Optical ``EBV\_model2'' $\mu$ fits \label{tab:sn_hubble}}
\tablehead{\colhead{Name} & \colhead{$t_{\rm max}$} & \colhead{$t^{\rm err}_{\rm max}$} & \colhead{$z_{\rm cmb}$} & \colhead{$z^{\rm err}_{\rm cmb}$} & \colhead{$H_{\rm max}$} & \colhead{$H^{\rm err}_{\rm max}$} & \colhead{$N_{H}$} & \colhead{$s_{BV, H}$} & \colhead{$s^{\rm err}_{BV, H}$} & \colhead{$\mu$} & \colhead{$\mu^{\rm err}$} & \colhead{$N_{\mu}$} & \colhead{$s_{BV, \mu}$} & \colhead{$s^{\rm err}_{BV, \mu}$} & \colhead{Ref.\tablenotemark{a}} & \colhead{Samp.\tablenotemark{b}}\\ \colhead{ } & \colhead{MJD} & \colhead{MJD} & \colhead{$\mathrm{}$} & \colhead{$\mathrm{}$} & \colhead{$\mathrm{mag}$} & \colhead{$\mathrm{mag}$} & \colhead{ } & \colhead{ } & \colhead{ } & \colhead{$\mathrm{mag}$} & \colhead{$\mathrm{mag}$} & \colhead{ } & \colhead{ } & \colhead{ } & \colhead{ } & \colhead{ }}
\startdata
CSS121006:232854+085452 & 56207.0 & 0.07 & 0.0799 & 0.0008 & 19.945 & 0.168 & 2 & 0.955 & 0.0 &   &   &   &   &   & W18 & W18 \\
CSS121009:011101-172841 & 56220.0 & 0.46 & 0.06 & 0.0006 & 18.162 & 2.079 & 2 & 0.666 & 0.949 &   &   &   &   &   & W18 & W18 \\
CSS121114:090202+101800 & 56252.0 & 0.46 & 0.0419 & 0.0004 & 17.949 & 0.093 & 3 & 1.174 & 0.109 &   &   &   &   &   & W18 & W18 \\
LSQ12fhs & 56206.0 & 0.07 & 0.0329 & 0.0003 & 17.448 & 0.135 & 4 & 0.918 & 0.048 &   &   &   &   &   & W18 & W18 \\
LSQ12fmx & 56227.0 & 0.46 & 0.0669 & 0.0006 & 19.143 & 0.121 & 4 & 0.880 & 0.050 &   &   &   &   &   & W18 & W18 \\
LSQ13cmt & 56582.0 & 0.0 & 0.0569 & 0.0005 & 18.002 & 0.161 & 3 & 0.684 & 0.200 &   &   &   &   &   & W18 & W18 \\
LSQ13cwp & 56606.0 & 0.07 & 0.0669 & 0.0006 & 18.413 & 0.150 & 3 & 0.911 & 0.048 & 37.053 & 0.091 & 17 & 1.003 & 0.060 & W18 & W18 \\
PS1-10w & 55248.0 & 0.0 & 0.0322 & 0.0001 & 17.340 & 0.096 & 10 & 0.915 & 0.013 &   &   &   &   &   & F15 & F15 \\
\enddata

\tablenotetext{a}{Reference codes  J99: \citet{Jha99}; H00: \citet{Hernandez00}; K00: \citet{Krisciunas00}; K04a: \citet{Krisciunas04a}; K04b: \citet{Krisciunas04b}; Ph06: \citet{Phillips06}; Pa07a: \citet{Pastorello07a}; Pa07b: \citet{Pastorello07b};St07: \citet{Stanishev07};  F15: \citet{Wood-Vasey08}; C10: \citet{Contreras10}; S11: \citet{Stritzinger11}; BN12: \citet{Barone-Nugent12}; W14: \citet{Weyant14}; F15: \citet{Friedman15}; W18: \citet{Weyant18}}
\tablenotetext{b}{Sample name used for the divisions in the analysis.  Some SNe~Ia were observed by multiple projects.
We assign each SNe~Ia to a single sample for the purposes of quoting dispersions and distributions in the analysis.}
\tablenotetext{c}{Redshifts from the PISCO survey~\citep{PISCO}.}
\end{deluxetable}
\end{longrotatetable}

\bibliographystyle{aasjournal}
\bibliography{refs}

\end{document}